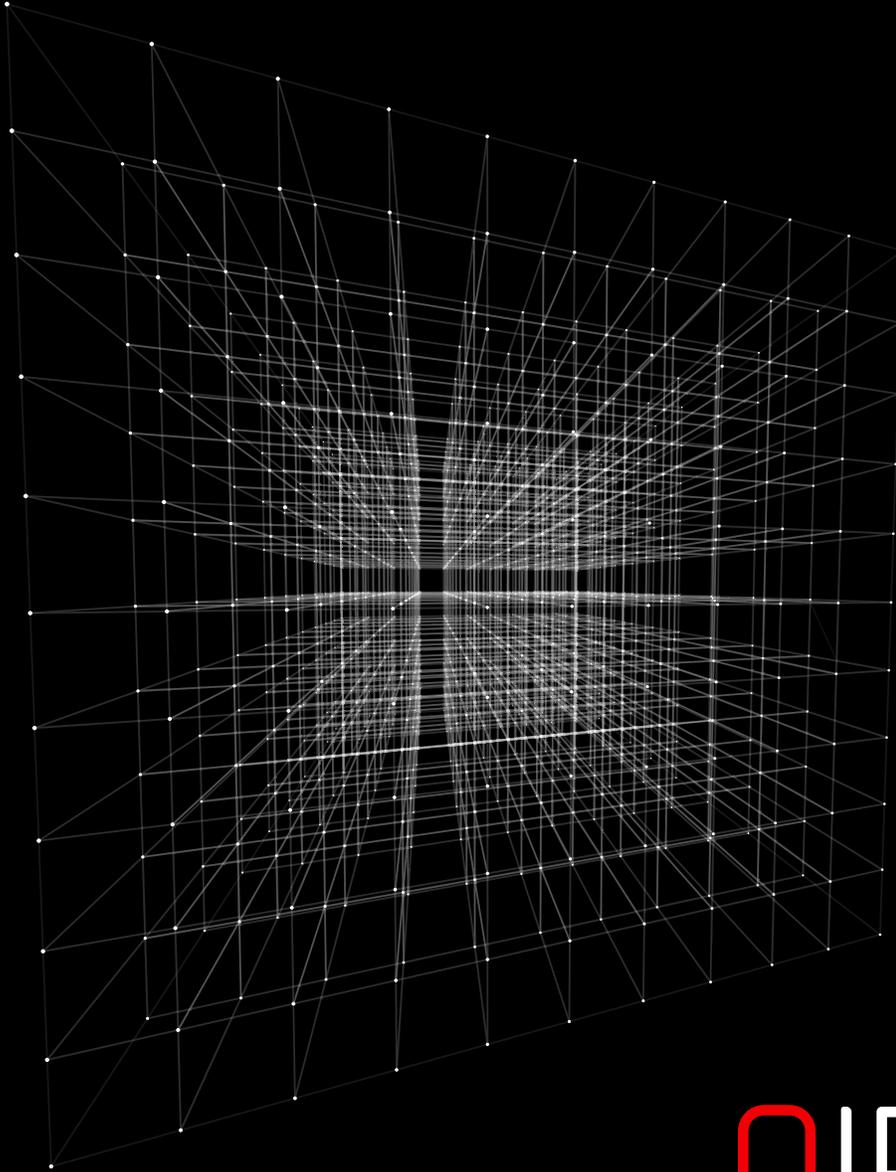

# QIR

Quantum
Index Report
2025

MIT INITIATIVE ON THE DIGITAL ECONOMY

### ▸ How to cite this report

Ruane, J., Kiesow, E., Galatsanos, J., Dukatz, C., Blomquist, E., Shukla, P., "The Quantum Index Report 2025", MIT Initiative on the Digital Economy, Massachusetts Institute of Technology, Cambridge, MA, May 2025.

The Quantum Index Report 2025 by Massachusetts Institute of Technology is licensed under **CC BY-ND 4.0** Attribution-NoDerivatives 4.0 International.

### ▸ Interactive website and public data

The Quantum Index Report 2025 is accompanied with interactive tools available on our website (qir.mit.edu) and we share our raw data with the community available to download from our website (qir.mit.edu/data).

*In memory of Shawneric Hachey, whose unique talent and dedication shaped the way this project is presented today.*

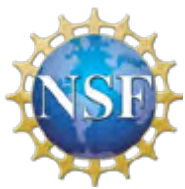
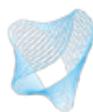

This work was supported by the Engineering Research Centers Program of the National Science Foundation under NSF Cooperative Agreement No. 1941583. Any opinions, findings and conclusions or recommendations expressed in this material are those of the authors and do not necessarily reflect those of the National Science Foundation.

## ▸ Team

**Jonathan Ruane**, Principal Investigator and Editor-in-Chief
MIT Sloan School of Management
MIT Initiative on the Digital Economy

**Elif Kiesow**, Senior Researcher and Project Manager
MIT Initiative on the Digital Economy

**Johannes Galatsanos**, Researcher
MIT Initiative on the Digital Economy

**Carl Dukatz**
Accenture

**Edward Blomquist**
Accenture

**Prashant Shukla**
Accenture

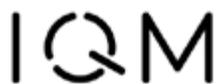 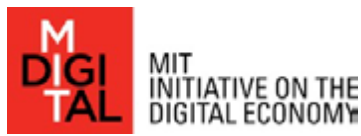 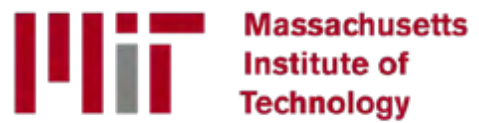

# Introduction
2025 Quantum Index Report

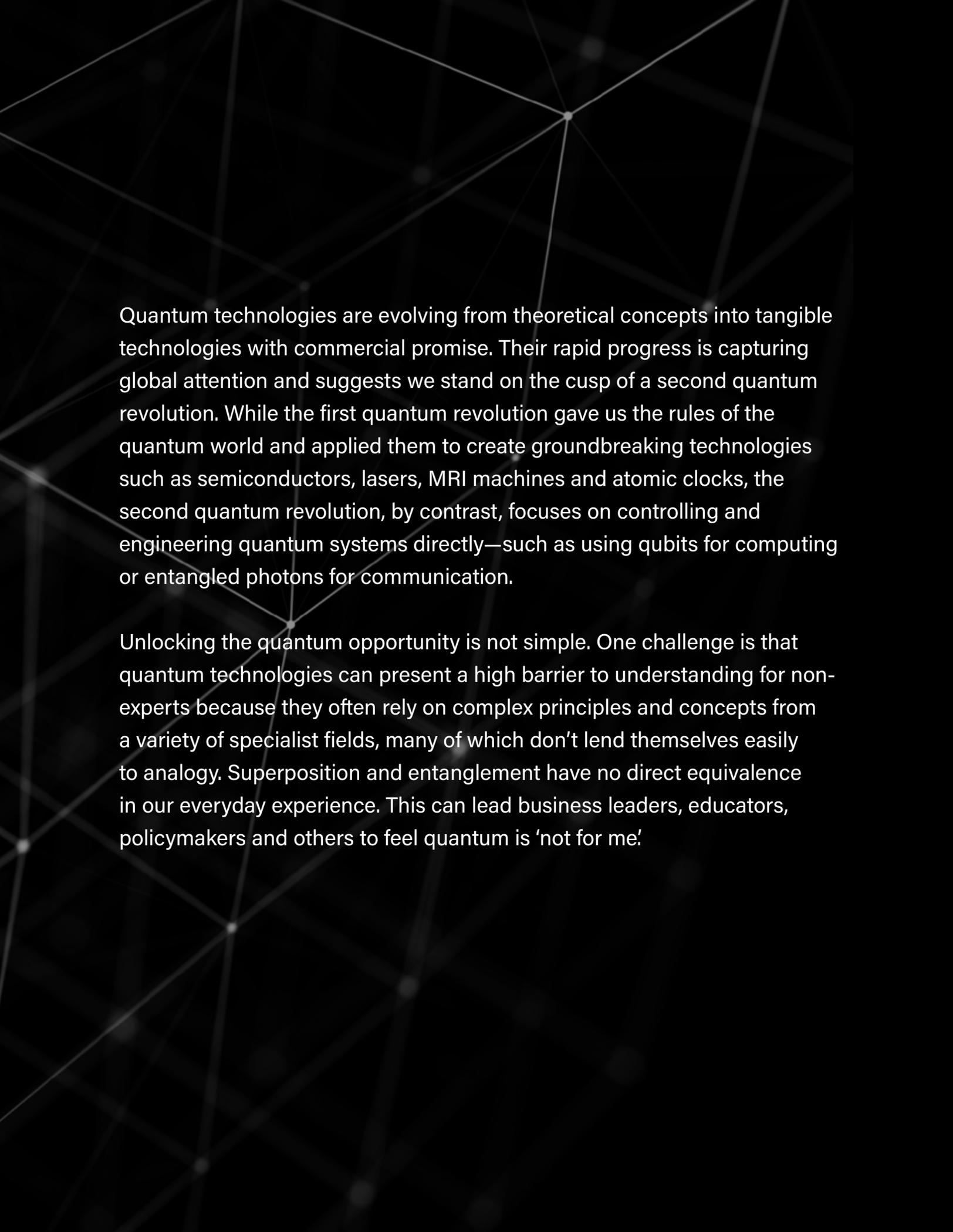

Quantum technologies are evolving from theoretical concepts into tangible technologies with commercial promise. Their rapid progress is capturing global attention and suggests we stand on the cusp of a second quantum revolution. While the first quantum revolution gave us the rules of the quantum world and applied them to create groundbreaking technologies such as semiconductors, lasers, MRI machines and atomic clocks, the second quantum revolution, by contrast, focuses on controlling and engineering quantum systems directly—such as using qubits for computing or entangled photons for communication.

Unlocking the quantum opportunity is not simple. One challenge is that quantum technologies can present a high barrier to understanding for non-experts because they often rely on complex principles and concepts from a variety of specialist fields, many of which don't lend themselves easily to analogy. Superposition and entanglement have no direct equivalence in our everyday experience. This can lead business leaders, educators, policymakers and others to feel quantum is 'not for me.'

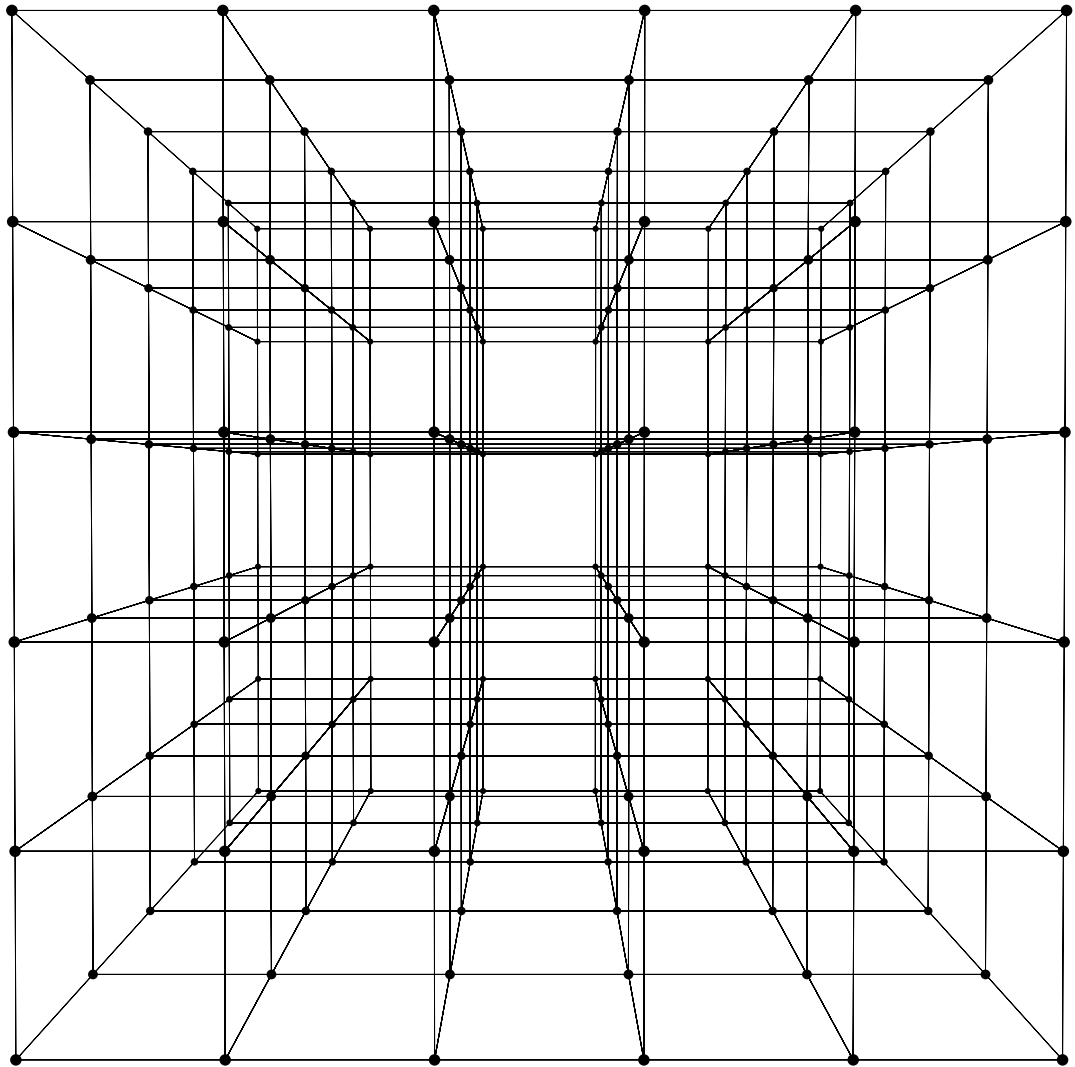

Our vision for the Quantum Index Report is to create a comprehensive, data-driven assessment of the state of quantum technologies.

## ▸ Vision

The Quantum Index Report aims to reduce the complexity and make it possible for a wider audience to have a deeper understanding of the quantum landscape. Like most transformative technologies, the success of quantum will depend not only on inventors, physicists and engineers, but also on entrepreneurs, investors, designers, teachers, and decision-makers who can help shape how the technology is developed, commercialized, and governed. By making the field more accessible and inclusive, we stand a better chance of realizing its full potential—for science, industry, and society at large.

Our vision for the Quantum Index Report is to create a comprehensive, data-driven assessment of the state of quantum technologies. For this inaugural edition we have focused on quantum computing and networking. The report tracks, measures, and visualizes trends across research, development, education and public acceptance. It aggregates data from academia, industry and policy sources and aims to provide nonpartisan insights. Where possible, the underlying data behind this report is available online where you will also find additional data and visualizations (www.qir.mit.edu).

## ▸ Community

We look at activity in the quantum landscape through a broad range of perspectives. We have aggregated publicly available data, contributed original data, and extracted new metrics by combining data series. However, the challenges are substantial, the field remains nascent and data is oftentimes sparse, difficult to gather, invisible to us or non-existent. We acknowledge there are many limitations and biases, such as our US focus in this edition. To achieve the broader goals of this project we need the support of a global community, and invite you to participate in any way you can. We welcome datasets, analysis, commentary or descriptions of what else you would like to see included. Please connect via the Get-Involved section of our website (www.qir.mit.edu/get-involved) or directly by email (contact@qir.mit.edu).

MIT's motto is *mens et manus*, translated as "mind and hand". This motto reflects the ideals of the institute which promotes education and research for practical application. The Quantum Index Report hopes to serve the quantum community with this same ethos as we present the 2025 report with a commitment to bridge science, commercialization, entrepreneurship and societal needs.

Jonathan Ruane

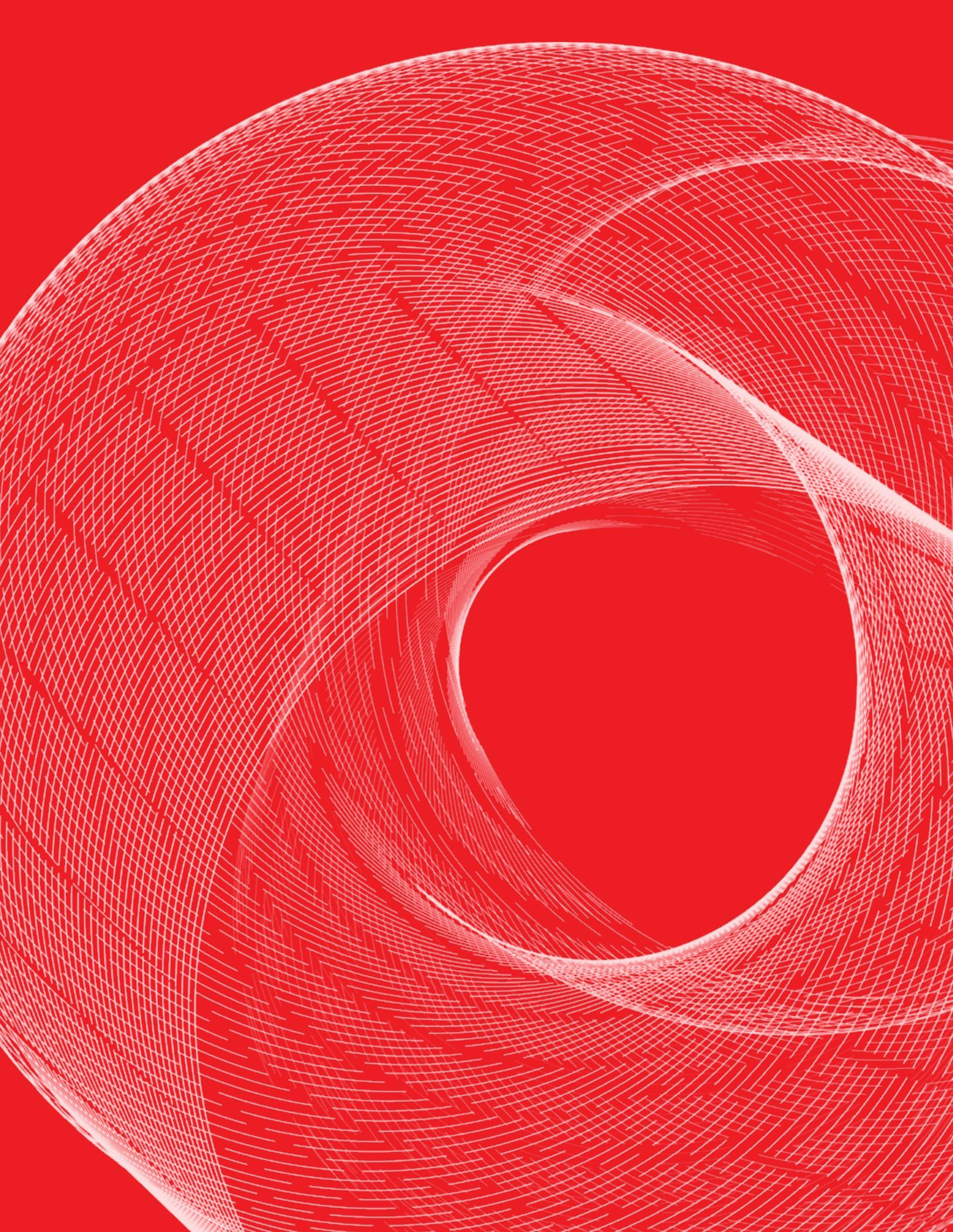

# Table of Contents | 2025 Quantum Index Report

# Key Insights | 2025 Quantum Index Report

▸ **Patents**

Corporations and universities lead innovation efforts, accounting for 91% of quantum computing patents.

▸ **Academic Research**

While China produces more papers overall in quantum computing, American research tends to have greater impact and influence.

▸ **Venture Funding**

Quantum computing firms lead the sector, securing $1.6 billion in publicly announced investments during 2024, followed by quantum software companies with $621 million.

▸ **Quantum in Corporate Communications**

In corporate communications, there has been a marked increase in the discussion of quantum computing over the last two years.

▸ **Policy**

While countries maintain unique approaches to quantum governance, they face common challenges in balancing innovation promotion with security concerns, leading to emerging hybrid governance models.

▸ **Workforce**

The US labor market shows strong growth, with quantum skills demand almost tripling since 2018.

▸ **Education**

In higher education, Germany leads globally with master's programs that include "quantum" in the degree name, followed by the UK and the US. These three nations represent 45% of all quantum master's degree programs worldwide.

▸ **Public Opinion**

Public views on governance show strong support for private sector involvement in quantum technology development, while expressing skepticism about government oversight.

▸ **Quantum Networking**

Quantum Networking Testbeds play a crucial role in the development of quantum networking and, by extension, the quantum internet. Currently, our data identifies 28 quantum networking testbeds in the US and Europe.

▸ **Quantum Processor Global Landscape**

Two dozen manufacturers are commercially offering more than 40 quantum processing units (QPUs) today. The United States leads the field, both in terms of the number and diversity of QPUs, followed by China.

▸ **Quantum Processor Benchmarks**

Overall, quantum processing units (QPUs) are making impressive progress in performance, but they remain far from meeting the requirements for running large-scale commercial applications such as chemical simulations or cryptanalysis.

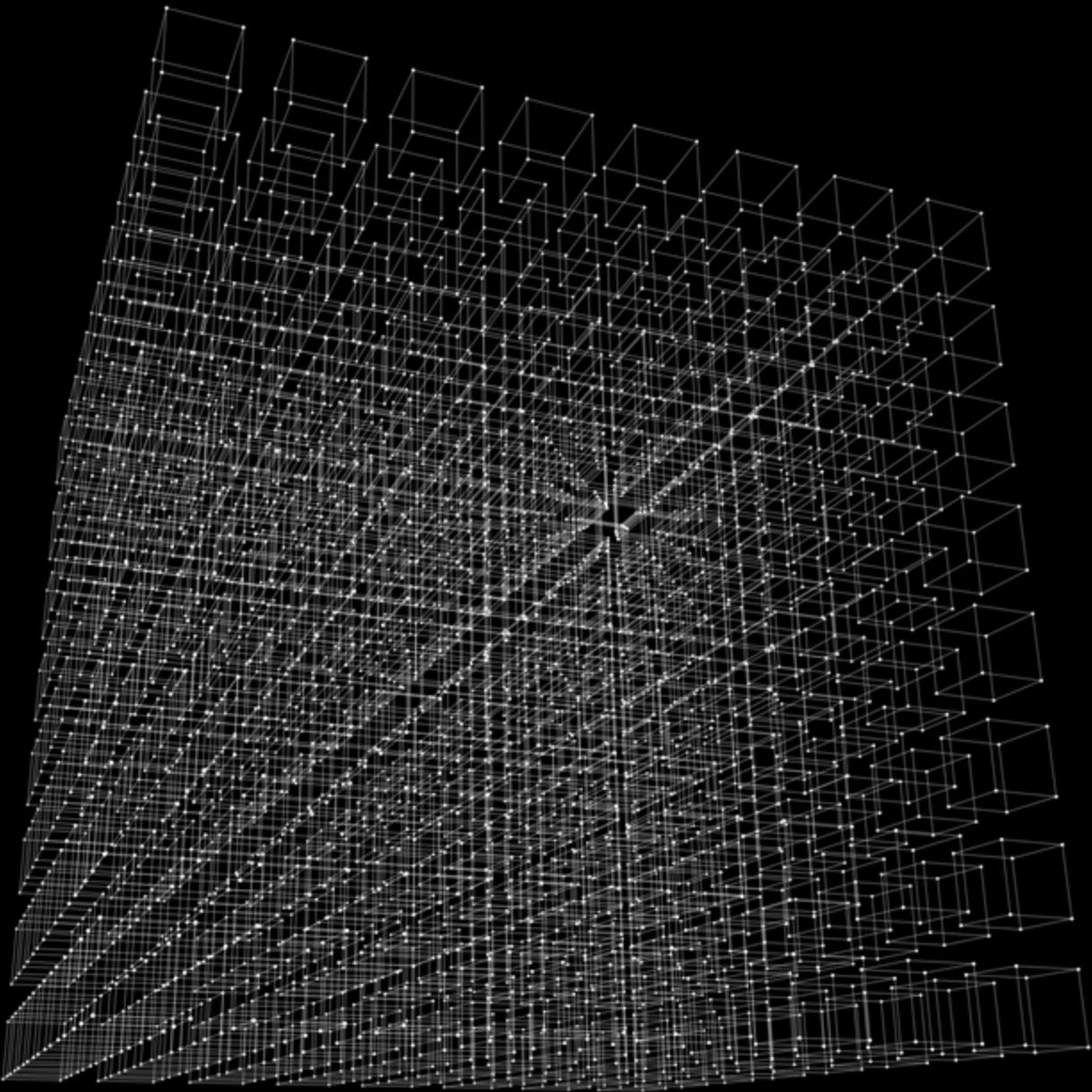

# Executive Summary
## 2025 Quantum Index Report

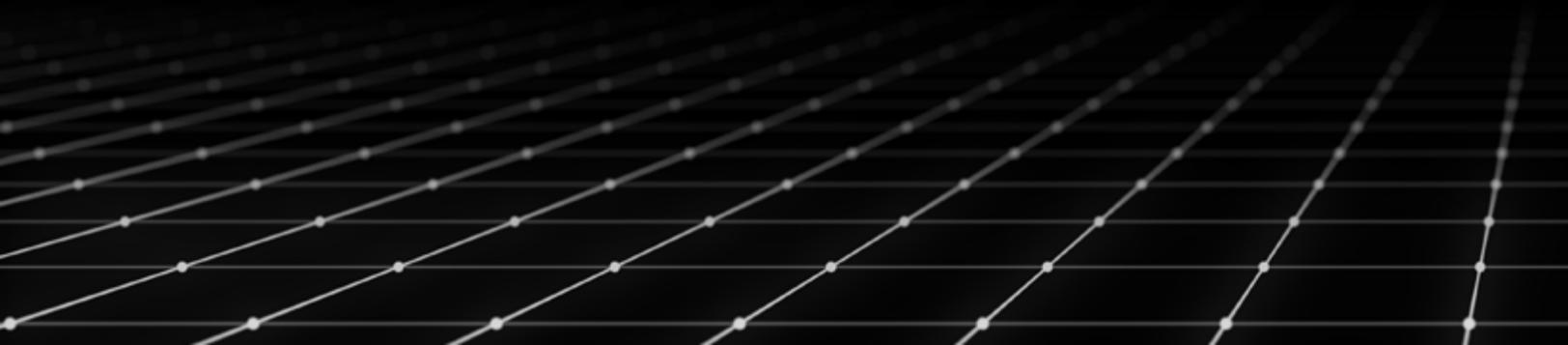

### ▸ Patents

The quantum technology patent landscape has shown remarkable growth and concentration over recent years. Between 2016-2021, quantum computing patent family filings increased by over 300%, while total quantum technology patents grew five-fold from 2014 to 2024. Corporations and universities lead innovation efforts, accounting for 91% of quantum computing patents, with corporations holding 54% and universities 37% of total filings. Geographically, China leads with 60% of patents as of 2024, followed by the United States and Japan. The sector has evolved through distinct phases, from early development between 1999-2004 to rapid expansion between 2013-2019. Recent trends from 2020 to 2023 show universities reached a peak in total patent filings in 2023 while corporate patents showed a decline as of 2023, suggesting potential market adjustments.

### ▸ Academic Research

The United States holds a leading position in quantum computing research output, particularly in terms of research quality. In contrast, China has established itself as the clear leader in quantum communications, with the United States following at a distance. The research quality metrics also reveal interesting insights: while China produces more papers overall in quantum computing, American research tends to have greater impact and influence. These differences suggest strategic specialization, with the US focusing on quantum computing and China prioritizing quantum communications, particularly evident in China's development of extensive satellite quantum communication capabilities. European nations maintain significant research presence across both areas, though typically trailing behind the two leaders in publication volume while demonstrating strong research quality.

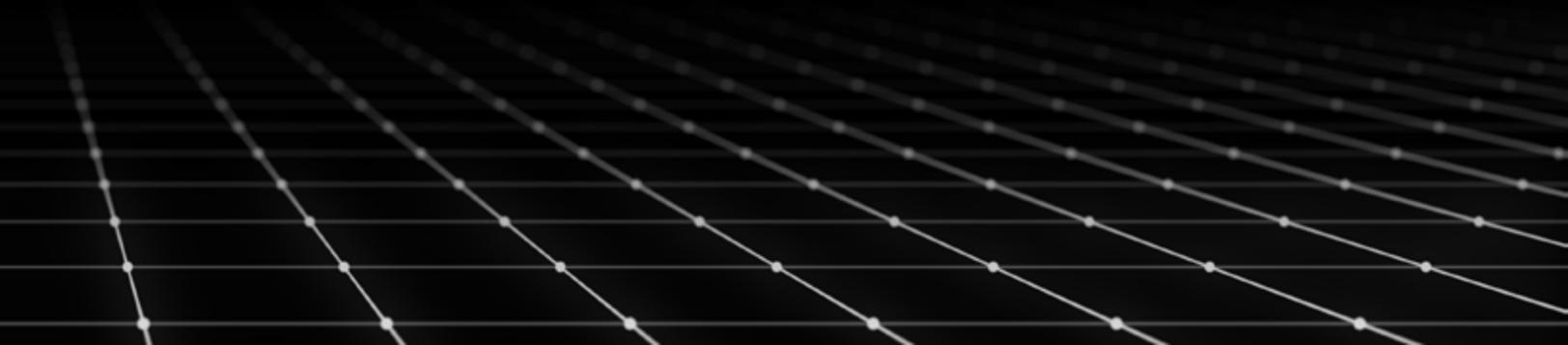

### ▸ Venture Funding

The quantum technology funding landscape has shown remarkable evolution and growth in recent years, consistently surpassing previous milestones. 2024 was a new high-water mark for the sector, although it is worth noting quantum represents less than 1% of total venture capital funding worldwide. Quantum computing firms lead the sector, securing $1.6 billion in publicly announced investments during 2024, followed by quantum software companies with $621 million. The United States and United Kingdom lead global investment with a combined share exceeding 60% across 2012 to 2024. Recent notable investments include Australian firm PsiQuantum securing $620 million in 2024. The structure of this particular deal highlights the increasing role of public-private co-funding arrangements. While established powers such as the US continue to invest, other players such as Canada and the Netherlands show impressive commitment to the sector, indicating accelerated expansion strategies and commercialization success.

### ▸ Quantum in Corporate Communications

Our data tracks mentions of quantum technology across more than 50,000 corporate communications such as press releases and earnings calls. There has been a marked increase in the discussion of quantum computing over the last two years. This trend spans multiple document types, including news articles, and earnings calls, where quantum computing references have shown significant growth. Industry leaders are driving this dynamic, particularly IBM and NVIDIA. The increasing frequency of quantum computing discussions in corporate communications suggests a growing presence of the technology in mainstream business discourse, as companies increasingly recognize its potential impact on future operations and competitive advantage.

### ▸ Policy

The global quantum technology landscape reveals a complex interplay between national sovereignty and international cooperation, with countries developing distinct approaches while acknowledging the need for coordinated governance frameworks. Major powers such as China (claimed $15 billion investment), the United States (National Quantum Initiative), and the European Union (Quantum Flagship program) have established comprehensive strategies, though priorities vary.

While countries maintain unique approaches to quantum governance, they face common challenges in balancing innovation promotion with security concerns, leading to emerging hybrid governance models. The future of quantum technology policy appears to be moving toward increasingly sophisticated international frameworks, with success depending on developing flexible structures that can adapt to rapid technological advancements while maintaining trust among participating nations.

## ▸ Workforce

The quantum technology sector is experiencing significant workforce development change amid sustained demand growth, with major nations implementing comprehensive strategies to address these needs. The US National Quantum Initiative, places strong emphasis on workforce development, while Canada and Australia have launched similar national quantum strategies focusing on labor capacity expansion. The US labor market shows strong growth, with quantum skills demand almost tripling since 2018, though stabilizing into a more moderate upward trend. Key developments include the establishment of quantum hubs at universities, specialized training programs connecting business managers with researchers, and the emergence of a "quantum-as-a-service model" which aids experiential learning. Despite initial rapid acceleration from 2018-2020, recent years show more stable growth patterns, suggesting a leveling of demand.

## ▸ Education

Global quantum technology education is experiencing rapid expansion across all educational levels, with significant developments in K-12 programs and higher education. At the primary and secondary level, initiatives such as the National Q-12 Education Partnership in the US, industry partnerships in China, and the EU's Quantum Flagship project are introducing quantum concepts to younger students. In higher education, Germany leads globally with master programs with "quantum" in the degree name, followed by the UK and the US. These three nations represent 45% of all quantum master's programs worldwide. Bachelor degree enrollment trends in the US for QIST related disciplines shows strong growth especially in related topics such as Computer Science while Electrical Engineering and Physics enrollments remained stable. Some commentators suggest the field faces significant workforce challenges in the future, highlighting the need for expanded domestic talent pipelines while maintaining international recruitment capabilities.

▸ **Public Opinion**

Our survey of 1,375 US residents conducted in October 2024 reveals distinct public perceptions about quantum computing and networking, showing a split between those with domain knowledge and those who remain unfamiliar. Public awareness tends to cluster at opposite ends of the spectrum, with either minimal exposure or significant understanding of quantum computing. Emotional responses vary considerably across different applications, with practical uses like materials discovery generating the strongest enthusiasm, while security-related applications raise more concerns due to their dual nature of potentially breaking current encryption methods while enabling new security solutions. Public views on governance show strong support for private sector involvement in quantum technology development, while expressing skepticism about government oversight. Throughout the survey, consistent neutral responses suggest widespread recognition that quantum computing represents a complex technology whose ultimate societal impact remains uncertain for the general public.

▸ **Quantum Networking**

Quantum internet and quantum networking are emerging frontiers in quantum information science. Quantum networking is the field of study and development focused on enabling that quantum internet. Quantum networks make the transmission of quantum information possible between devices and they allow the distribution of quantum entanglement. Quantum networks will not replace classical communications or the classical internet however they have potential to offer novel functionalities such as more secure communication and the ability to connect quantum computers for enhanced computing power. Quantum Networking Testbeds play a crucial role in the development of quantum networking and, by extension, the quantum internet. Currently, our data identifies 28 quantum networking testbeds in the US and Europe. Testbeds are essential for advancing quantum networking because they provide realistic environments in which to explore the performance, interoperability, and scalability of quantum components. Investments in testbeds are not merely about testing hardware, they also represent a commitment to advancing the foundational science and engineering needed for a transformative quantum era. Beyond technical development, testbeds also play a critical role in workforce training and industry engagement.

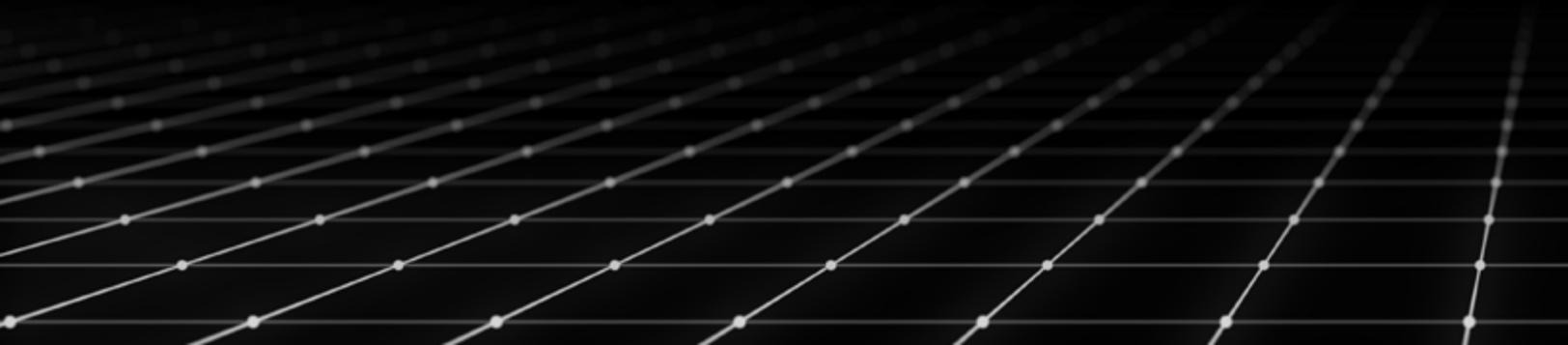

▸ **Quantum Processor Global Landscape**

Two dozen manufacturers are commercially offering more than 40 quantum processing units (QPUs) today. The United States leads the field, both in terms of the number and diversity of QPUs, followed by China. However, China's commercially available QPUs tend to be smaller and have lower performance compared to those from the US and Europe. Within Europe, the UK, Netherlands, France, and Finland each have 4-6 commercial QPUs. In total, over 160 QPUs are currently in the prototype, planning, or commercial stages, developed by close to 80 manufacturers across 17 countries. Among the different QPU modalities, superconducting systems dominate the commercial market, representing over 40% of available QPUs. However other modalities, such as photonics, trapped ions, and especially neutral atoms and electron spins, are gaining momentum; their share is expected to grow in the coming years.

▸ **Quantum Processor Benchmarks**

Overall, quantum processing units (QPUs) are making impressive progress in performance, but they remain far from meeting the requirements for running large-scale commercial applications such as chemical simulations or cryptanalysis. To evaluate the maturity of different QPU offerings and modalities, multiple benchmarks must be considered. One such metric is the number of qubits, which historically followed an almost exponential growth trend. However, in recent years, especially within leading modalities like superconducting and trapped-ion systems, this growth has slowed. The industry focus has shifted toward building higher-performance machines by improving error correction, gate and readout fidelity, and gate speed rather than merely increasing qubit counts. Another important metric is fidelity, or the amount of errors produced by a QPU. In particular trapped-ion systems have demonstrated the highest fidelity operations and qubit connectivity and have set ambitious goals for further improvements. However, they continue to face challenges with low qubit counts and relatively slow gate speeds. Neutral atom platforms, a more recent entrant, have shown promising qubit scalability while maintaining reasonable fidelities. Photonic systems, still in the early stages, suggest the potential for high qubit counts, albeit with trade-offs in fidelity and scaling costs. Across all current and planned QPU technologies, no single modality or manufacturer has yet emerged as a clear leader. Each platform presents a distinct set of strengths and limitations, and the race toward useful, scalable quantum computing remains wide open.

▸ **Recent Highlights**

**December 2024** Google announces its Willow chip with error correction below the surface code threshold

**February 2025** QuEra raises $230 million financing to accelerate development of large-scale fault-tolerant quantum computers

**February 2025** Quantum Machines announces that it has raised $170 million in Series C funding

**January 2025** UN International Year of Quantum officially starts

**February 2025** Microsoft introduces the Majorana 1 quantum processor

**February 2025** Amazon Web Services unveils its Ocelot quantum chip

▸ **Recent Highlights**

**March 2025** IonQ raises over $370 million in addition to its 2025 acquisition of ID Quantique and Qubitekk that strengthen its quantum networking capabilities

**March 2025** NVIDIA announces plan to build Quantum Computing Research Center in partnership with labs at Harvard, MIT, and Boston Quantum firms

**April 2025** Spain launches its first National Quantum Strategy backed by €800 million

**March 2025** Quantum Internet Alliance announces the first operating system designed for quantum networks

**April 2025** DARPA announces cooperation with nearly 20 quantum companies for its Quantum Benchmarking Initiative

**April 2025** IBM announces its plans to invest more than $30 billion in R&D to enhance IBM's American manufacturing of mainframe and quantum computers

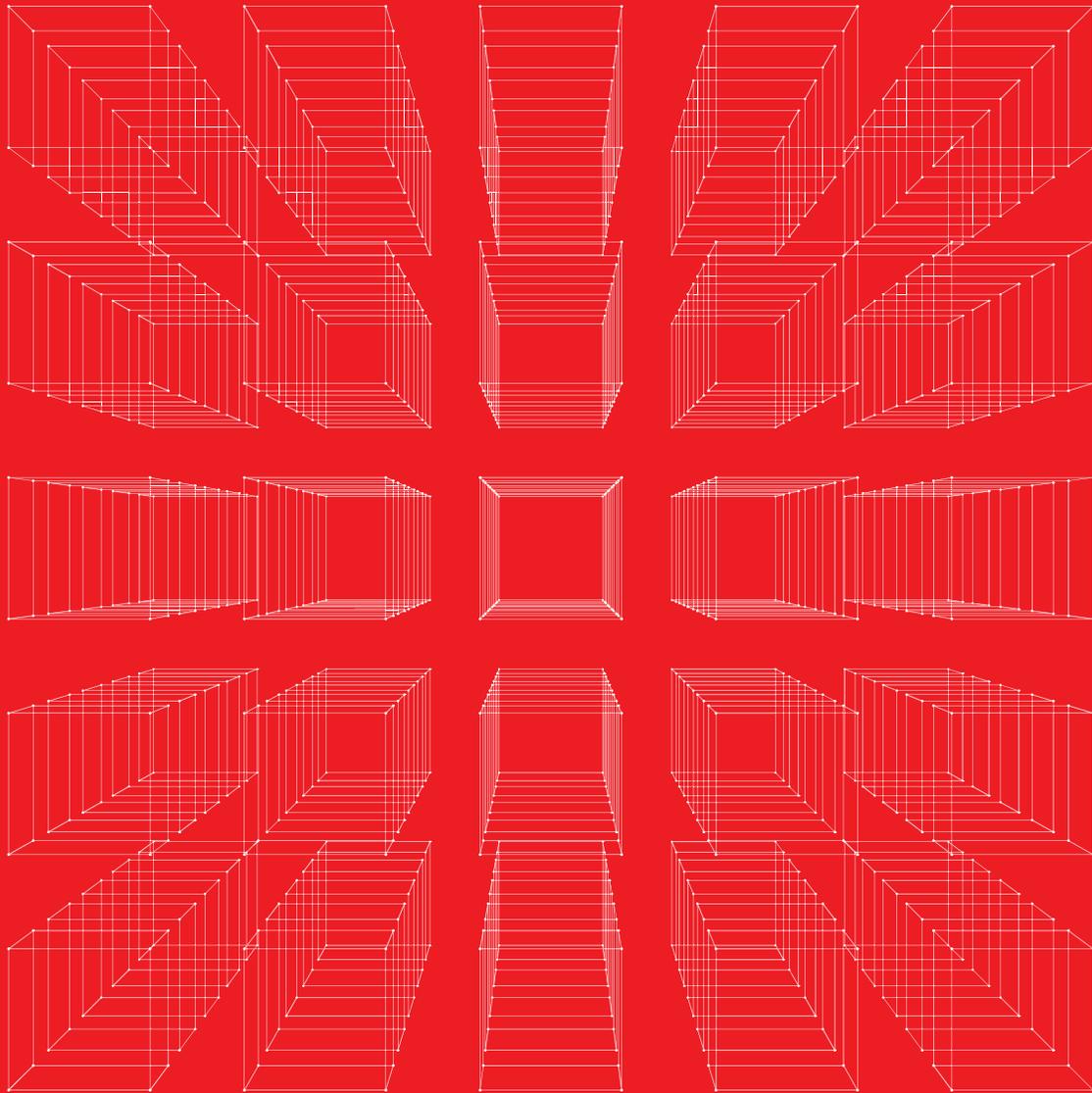

# 1 | Patents

As quantum technologies transition from the lab to the marketplace, patents and other forms of Intellectual Property (IP) are becoming increasingly important strategic assets in the race for quantum leadership. As well as serving as key indicators of general innovation activity, the growth in volume of quantum-related patent filings reflects both the maturing of research efforts and the intensifying competition among companies, institutions, and nations.

The quantum IP landscape is being shaped not only by traditional hardware systems and foundational qubit architectures but also by new frontiers such as quantum error correction[1], hybrid classical-quantum algorithms[2], and novel materials and qubit fabrication processes.

This report tracks patent data based on patent families. A patent family consists of multiple patent applications covering the same fundamental invention, filed in different countries. Therefore patent families serve as a better metric for analyzing new technology developments because they provide a comprehensive view of innovation scope and global market intentions. Unlike individual patents, patent families account for variations in filing requirements across jurisdictions. This interconnected structure allows researchers to track how inventors protect their IP across international borders, revealing both the breadth of innovation and possible geographic expansion plans.

Furthermore, patent families help normalize comparisons between regions with different patent systems and requirements, offering a more accurate picture of global innovation trends.

The data for this chapter was provided by Accenture Research in cooperation with The Quantum Insider.

## 1.1 | Patents by entity

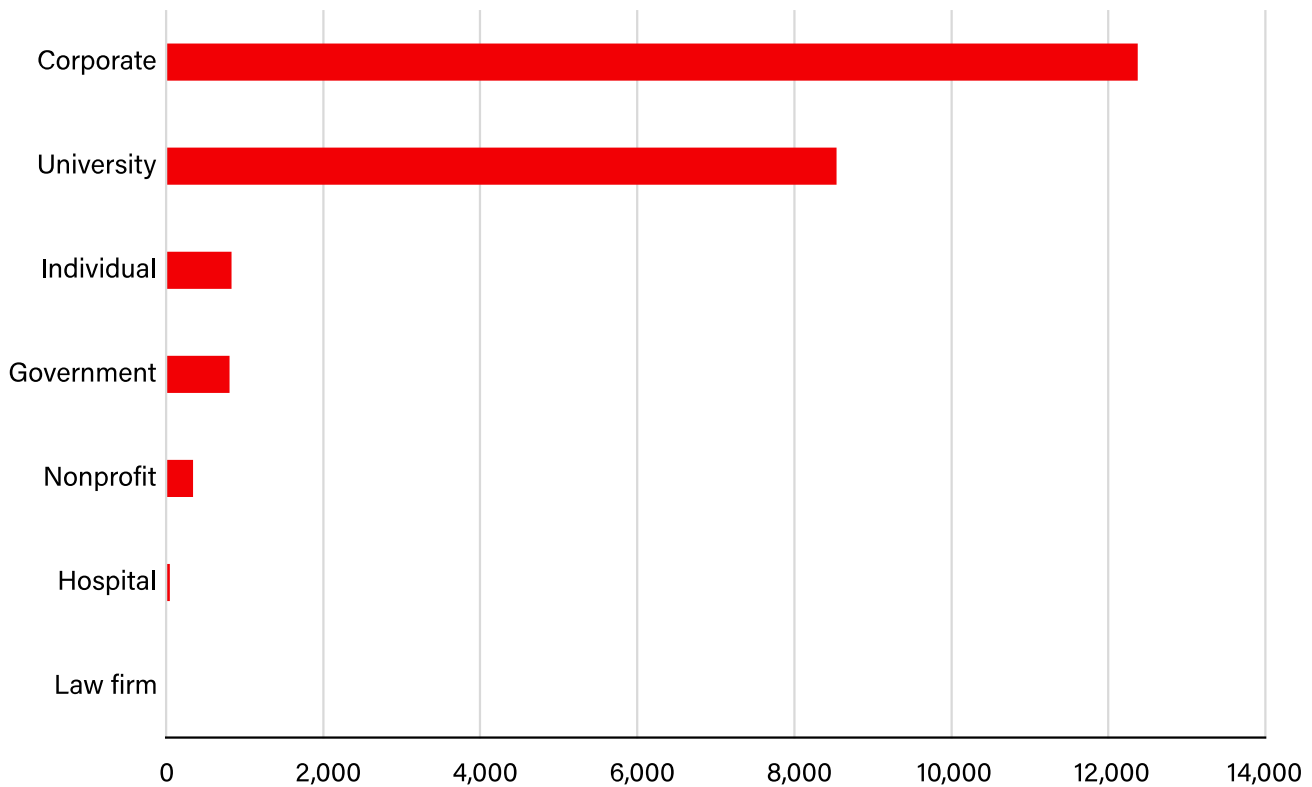

Quantum computing patent families by origin, 1999-2023

Corporations have emerged as the dominant force in quantum computing patent development, demonstrating significant investment in intellectual property protection. Recent data shows that global players such as IBM, Google, Microsoft, Intel, and Baidu are among the top patent filers.[3]

Corporate entities lead the landscape with 54% of total patents, followed by universities with 37%. Together, these two sectors account for 91% of all quantum computing patents, demonstrating a high concentration around these two categories. Individual inventors hold the third position with 3.6%, closely followed by government institutions at 3.5%. Nonprofit organizations contribute 1.5%, while hospitals and law firms show minimal participation.

During the early development period from 1999 to 2004, initial patents came primarily from corporate and university sectors, while government entities entered the patent landscape in 2002. This slow-growth period saw annual totals remain

▸ ```
Over the 2016–2021 period, quantum computing patent
family filings increased by over 300%.
```



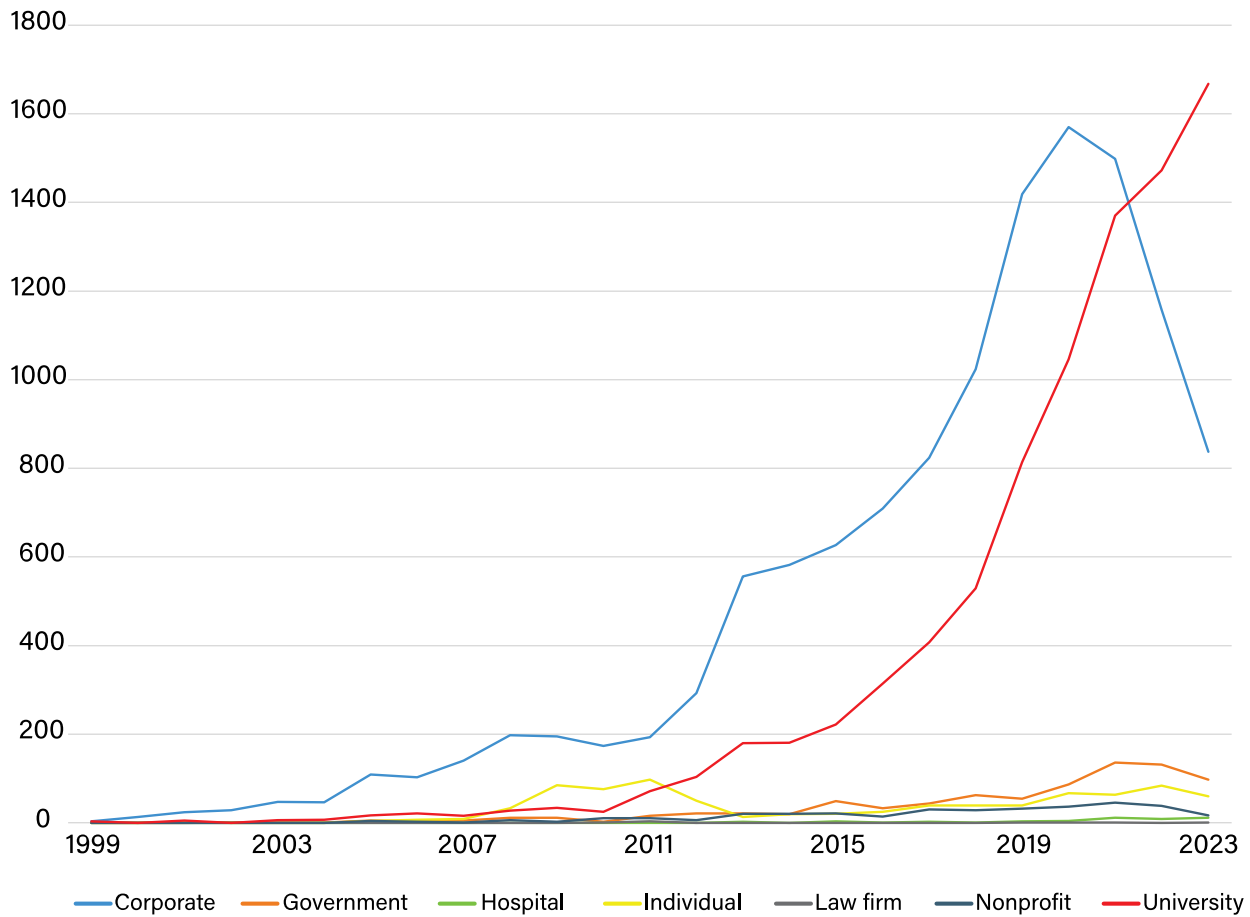

Quantum computing patent families by origin, 1999-2023

under 150 patent families, indicating the challenges associated with building research capabilities in the nascent field.

The transition period from 2005 to 2012 marked a significant shift in patent activity. Beginning in 2005, corporate patents jumped notably. Individual inventors began making more substantial contributions after 2008.

A period of rapid expansion occurred from 2013 to 2019 driven almost entirely by corporations and universities. Recent trends from 2020 to 2023 show continued evolution in patent activity. The corporate sector reached its maximum of 1,570 patents in 2020, while universities continued strong growth to reach 1,668 patents in 2023. Government participation accelerated notably after 2019. 2023 also marked the first significant decline across most categories except for universities, suggesting potential market adjustments. Throughout this entire period, universities and corporations consistently led patent development efforts, maintaining their positions as primary drivers of quantum computing innovation.

▸ In 2023, 837 patent family filings were made by corporations while 1668 were made by universities — indicating substantial commitment to quantum technology development by public and private institutions.



## 1.2 | Patents by country

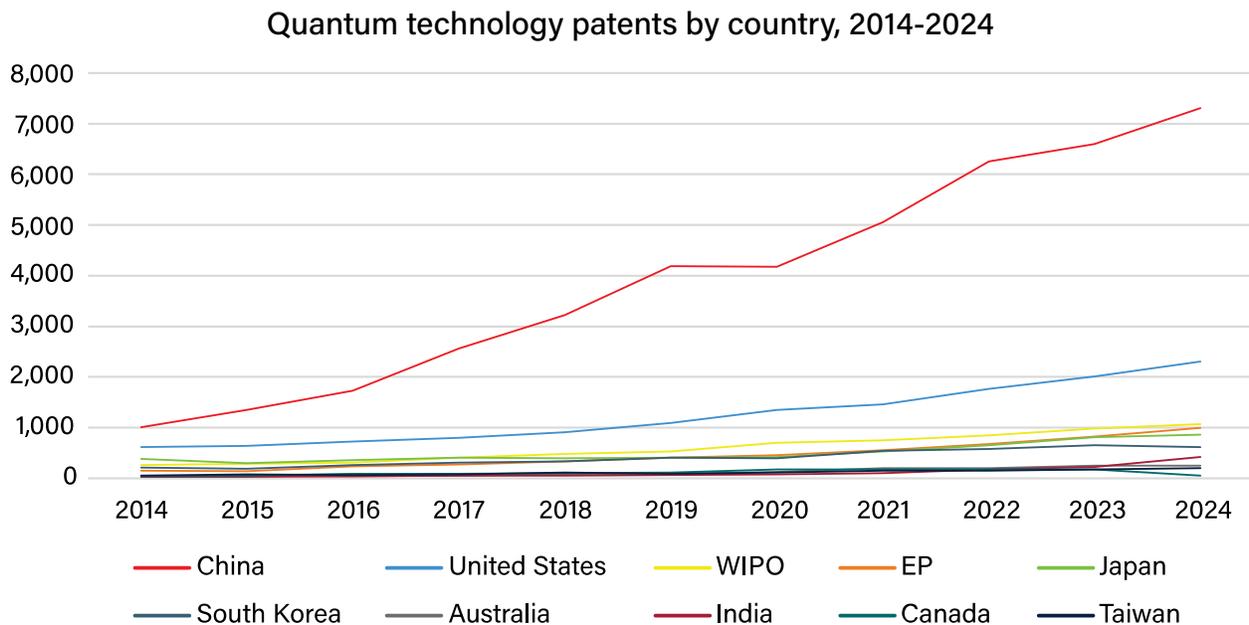

Quantum technology patents by country, 2014-2024

In the period 2014-2024, the total number of quantum technology patent filings grew significantly, representing a five-fold increase over this period. The growth has been particularly pronounced since 2020.

China emerged as the dominant location for quantum technology patent filing, growing from 1,011 patents in 2014 to 7,308 in 2024. The United States maintained second position throughout the period, increasing from 613 to 2,301 patents, while the World Patent Office secured third place, growing from 265 to 1,072 patents.

Analyzing growth patterns during the 2014-2016 period, the total number of patents grew moderately. This period saw relatively balanced growth across regions, with China holding a lead over the United States. A significant acceleration occurred in 2017, marking the beginning of a more rapid growth phase. This surge was driven by China with an expansion from 1,726 to 2,560 patents, accompanied by increases in the United States and WPO.

The period from 2018 to 2020 saw sustained growth momentum, including the emergence of India as a new player.

The most recent years (2021-2024) have witnessed continued strong growth rates. China's leadership became increasingly pronounced, while the United States maintained strong growth, and India demonstrated accelerated expansion.

Throughout 2014-2024, the geographic distribution of patent activity has become increasingly concentrated. China's market share expanded from 42% in 2014 to 60% in 2024, while the United States maintained the second

21 | MIT INITIATIVE ON THE DIGITAL ECONOMY

place with a relatively stable share around 19%, and the World Patent Office held steady at approximately 9%. Together, these three entities controlled 88% of all quantum computing patents in 2024, indicating a highly concentrated intellectual property landscape in this technology sector. It is important to note that, in terms of individual countries, Japan is placed as the third in total patent filing numbers after China and the US across this period.

According to recent patent research reports by QuIC[4] and by QEDC[5], China has established itself as the global leader in quantum communications patents. The country's strong emphasis on quantum communications research likely resulted in significant patent activity, with Chinese institutions leading the field. Organizations such as QuantumCTek, Ruban Quantum Technology, and Beijing University of Posts & Telecom are among the major patent holders in this domain.[6]

`QED-C research on patents states that the US Patent and Trademark Office (USPTO) has issued more quantum computing patents than any other country's office, and that the Chinese patent office has issued the most quantum communications patents.`

The disparity in patent numbers between China and the US highlights the competitive dynamics in quantum technology development, with each nation pursuing different aspects of the quantum technology ecosystem with different prioritization levels.



## 1.3 | Future research

We intend to provide the community with ongoing monitoring of the rapidly evolving quantum technology patent landscape. We aim to investigate geographic and market concentration evolution, tracking emerging patent hubs and their technological specializations. For future iterations of this section, we are interested in breaking down patents by quantum technology subfields and analyzing patent families across different technical classifications. Stakeholders interested in sharing data regarding these aspects of the quantum technology patent landscape are encouraged to contact us.

You can reach us at contact@qir.mit.edu.

---

### ▸ Footnotes

[1] Matt Swayne, 'US Leads in Steady Rise of Patents Covering Key Quantum Performance Measures' (The Quantum Insider, 30 April 2024) <https://thequantuminsider.com/2024/04/30/us-leads-in-steady-rise-of-patents-covering-key-quantum-performance-measures/> accessed 3 April 2025.

[2] Yudong Cao, Jonathan P Olson and Eric R Anschuetz, 'Hybrid Quantum-Classical Computer System and Method for Performing Function Inversion' <https://patents.google.com/patent/US20200394547A1/en> accessed 3 April 2025.

[3] A Portrait of the Global Patent Landscape in Quantum Technologies' (QuIC 2024) <https://www.euroquic.org/wp-content/uploads/2024/03/QuIC-White-Paper-IPT-January-2024.pdf> accessed 4 February 2025.

[4] ibid.

[5] Elliott J Mason QED-C, 'State of Quantum Industry Innovation – What Patents Tell Us' (11 December 2024) <https://quantumconsortium.org/blog/state-of-quantum-industry-innovation-what-patents-tell-us/> accessed 27 March 2025.

[6] ibid.



# 2 | Academic Research

Academic publications serve as an early indicator of scientific progress, typically preceding commercial applications in Quantum Technologies. Citation patterns and collaborations in journal papers can highlight research quality, interdisciplinary connections, and international partnerships shaping the field. Through comprehensive analysis of publication metrics, researchers can identify emerging trends, assess knowledge transfer mechanisms, and understand the evolution of quantum research ecosystems globally. This chapter presents data on academic publications on quantum computing and quantum communications based on research by the ASPI.[1]

Quantitative assessment of research output provides valuable insights into national capabilities in quantum technologies. Total publication counts offer a broad perspective on research activity levels across nations, reflecting the scale of investment in quantum research infrastructure and the size of research communities. When combined with H-index[2] measurements and citation data[3], which account for both productivity and citation impact, these metrics reveal the depth and influence of research programs. This dual perspective helps distinguish between quantity and quality of research output, allowing for a nuanced understanding of each nation's contribution to the rapidly developing quantum knowledge base.



## 2.1 | Quantum computing

### 2.1.1. Rankings by published research

▶ The US and China represent nearly half of all academic publications in quantum computing. The European Union accounts for 22% of global output. India, Japan and the United Kingdom produce a large volume of research but significantly trail the two global leaders.

The global landscape of quantum computing publications reveals a highly concentrated distribution pattern, with China and the United States collectively accounting for nearly half of all published research. China leads with 23% of publications, followed closely by the US at 22%, while India emerges as the third major contributor with 5%. This top-heavy distribution creates a clear tier structure in global quantum computing research output.

The middle tier of contributors demonstrates significant diversity, with Germany and Japan each contributing 5% and 4% respectively, followed by the United Kingdom at 4%. Canada, Italy, and Russia each contribute 3% of global publications, forming a secondary cluster of substantial contributors. This middle tier represents a crucial segment of global quantum computing research, bridging the gap between the dominant players and smaller contributors.

The lower tier of the distribution reveals a broad base of international participation, with several countries each contributing 2% of global publications, including France, Spain, Australia, South Korea, the Netherlands, and Switzerland, showing widespread engagement across multiple regions staying active in quantum computing research. The remaining countries, including Iran, Poland, Brazil, Austria, Singapore, Taiwan, Israel, Saudi Arabia, and Denmark, each contribute 1%, forming a diverse foundation of global participation in this field.



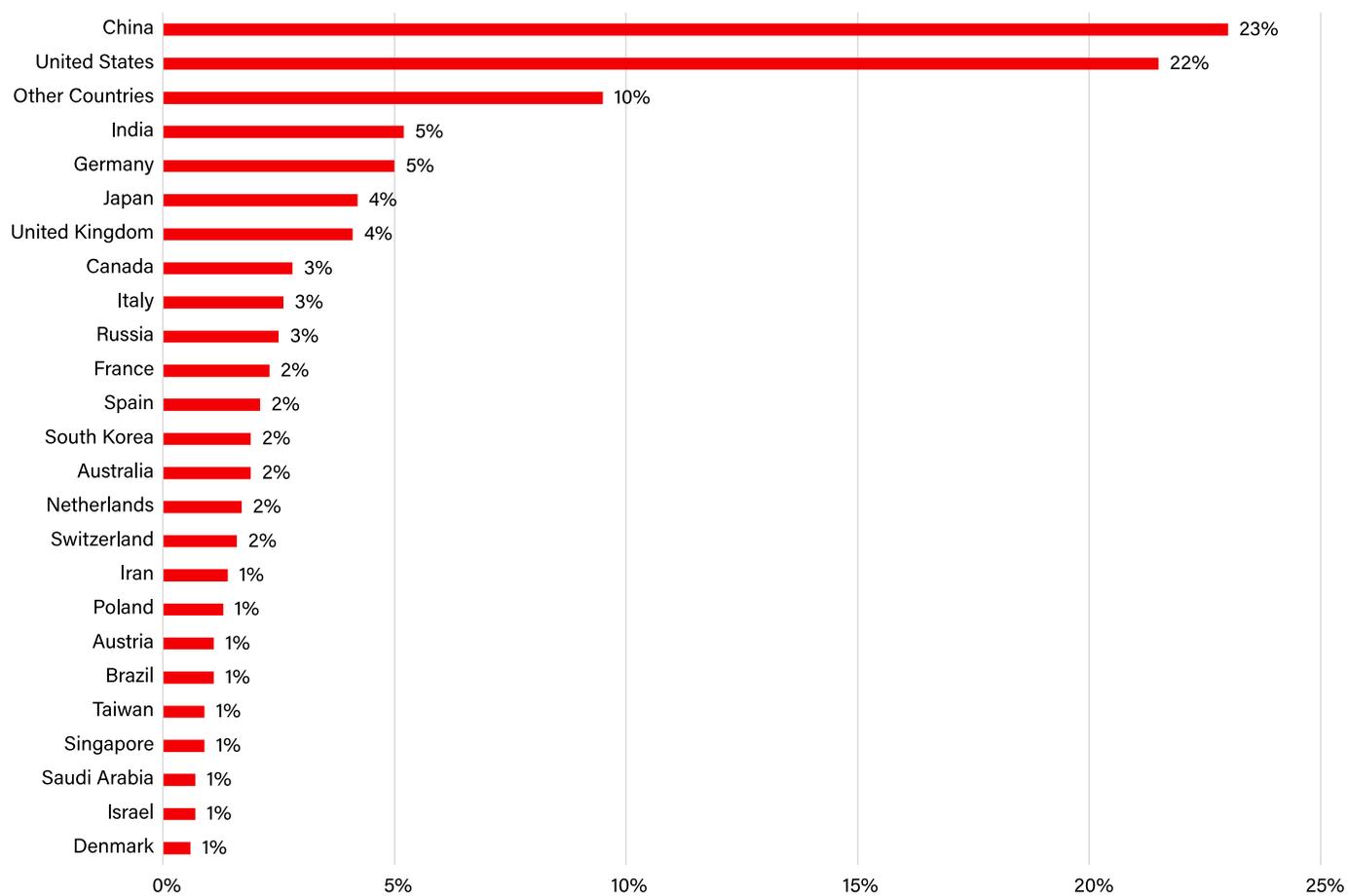

National share of quantum computing published research, 2019-2023



## 2.1.2. Rankings by H-index

▸ US quantum computing research quality is ranked highest in the world, significantly ahead of other countries. China ranks second with an H-index of 61, followed by the United Kingdom with 46.

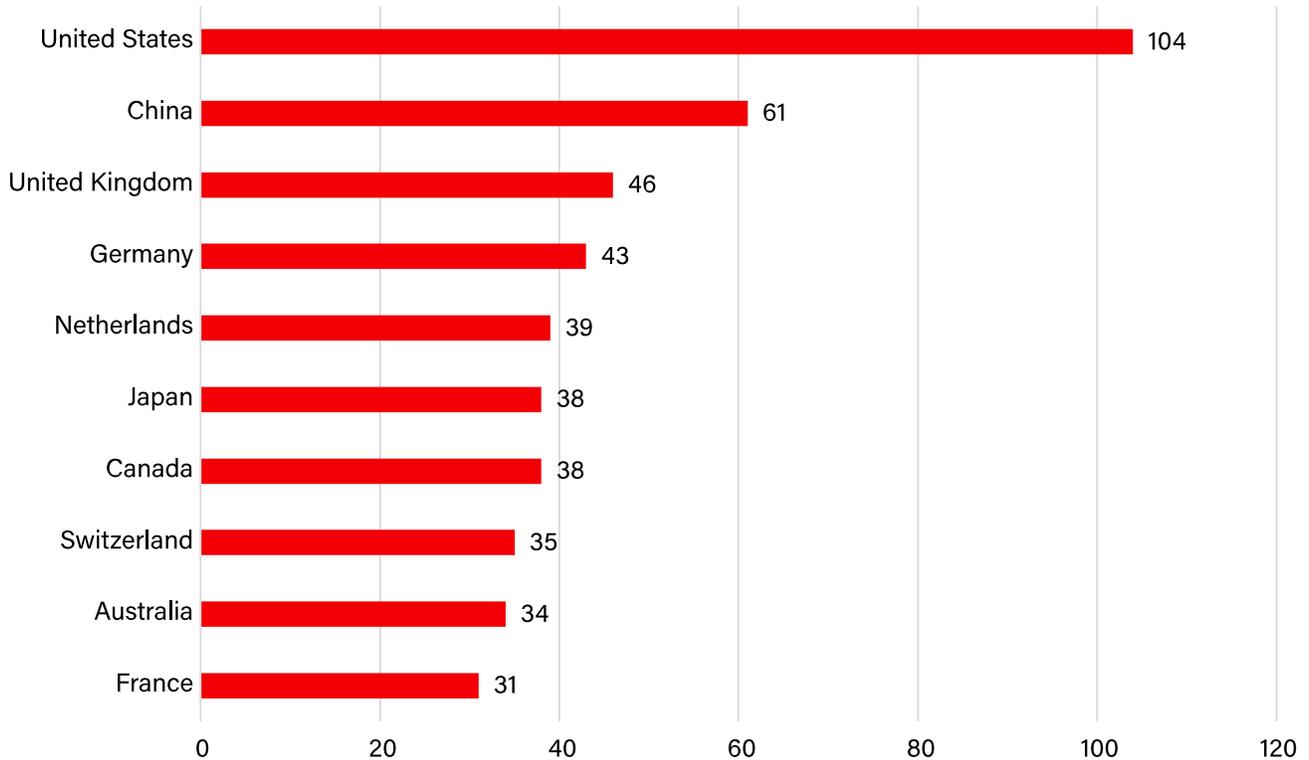

National quantum computing published research ranked by H-index, 2019-2023



In the landscape of quantum computing research quality measured by H-index, the United States stands as the clear leader with an H-index of 104, demonstrating exceptional research productivity and citation influence. This score reinforces why American institutions are at the forefront of quantum computing advancement, significantly outperforming all other nations.

Following the United States, China emerges in second position with an H-index of 61, representing a substantial research presence. The top two nations highlight the current state of global competition in quantum computing research.

The United Kingdom rounds out the top three positions with an H-index of 46, demonstrating Europe's significant contribution to quantum computing research excellence.

The middle tier of the distribution reveals intense activity among nations. Germany and the Netherlands lead this group with H-index scores of 43 and 39 respectively, followed closely by Canada and Japan, which tie at 38. Switzerland maintains a strong research presence with an H-index of 35, and Australia contributes meaningfully at 34. France completes this tier with an H-index of 31, demonstrating impressive research productivity despite being slightly lower than its European counterparts.



## 2.1.3. Rankings by most highly cited papers

▶ Among the top 10% of the most highly cited papers in quantum computing, the United States has the largest proportion of publications with 34%. China follows with the second highest proportion, 16%. The European Union accounts for an estimated 17% of the global total.

By analyzing the country of origin of the top 10% most highly cited quantum computing papers, we gain insight into where the field's most influential ideas are emerging from.

The United States leads this group with 34% of the most highly cited quantum computing publications, demonstrating exceptional research impact and influence.

China emerges as a strong second with 16% of highly cited publications, reinforcing its position as the primary challenger to US research leadership. This position is noteworthy as it represents a substantial gap between China and the next tier of countries. The United Kingdom and Germany tie for third place with 6% each, illustrating Europe's strong presence in high-impact quantum computing research.

The middle tier of the distribution shows interesting patterns of research excellence. Japan and Canada each contribute 4% of highly cited publications, while Switzerland, the Netherlands, India, and Australia form a closely grouped cluster at 3% each. This relatively small spread among these countries suggests a competitive landscape where institutions are achieving similar levels of citation impact despite their geographical and institutional differences.



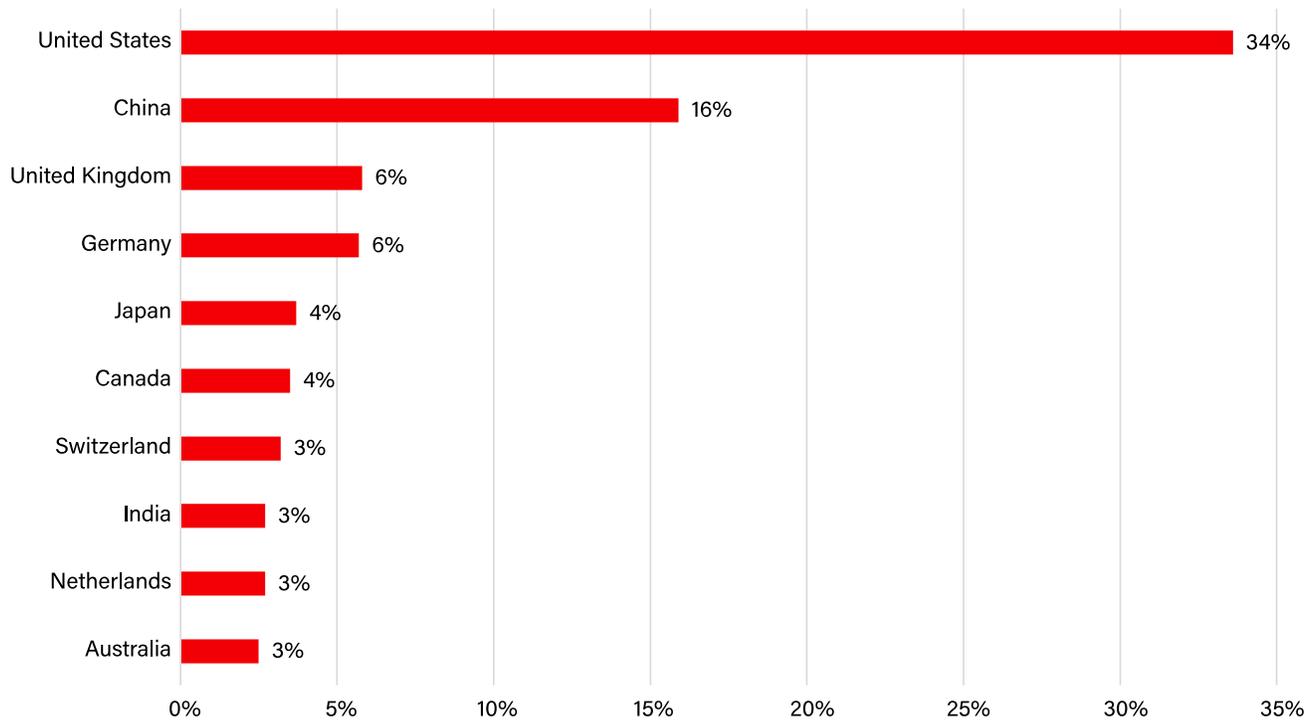

National share of top 10% most highly cited quantum computing publications, 2019-2023

This distribution of highly cited publications reveals several key characteristics of quantum computing research excellence. First, it shows a clear hierarchical structure with the United States holding a dominant position, followed by China in a secondary tier, and then a cluster of countries achieving similar levels of impact. Second, it demonstrates that research excellence isn't solely determined by absolute size or resources, as evidenced by smaller nations such as the Netherlands maintaining a strong position. Finally, it highlights the internationally diverse nature of high-impact quantum computing research, with representation from North America, Europe, and Asia-Pacific regions. This suggests a globalized research ecosystem, albeit one with a concentration of leadership from developed nations.



## 2.1.4. Rankings overall

The United States demonstrates remarkable consistency across all three metrics, placed among leaders in total publications (22%), and in top position in the H-index (104), and highly cited publications (34%). This strong leadership position suggests not only high research productivity but also exceptional research quality and impact.

China presents an intriguing case of varying performance across metrics. While leading in total publications (23%), it ranks second in highly cited publications (16%) and drops to second place in H-index (61). This pattern suggests that while Chinese institutions produce the highest volume of research, they haven't yet achieved the same level of citation impact as US institutions.

European nations show distinct patterns across the metrics. The United Kingdom, for instance, ranks third in H-index (46) but wasn't among the top contributors in total publications, indicating high-quality research despite lower publication volumes. Conversely, Germany ranked fourth in H-index (43) while maintaining fifth place in publication count (5%), showing strong consistency across both metrics. The Netherlands demonstrates exceptional efficiency, ranking fifth in H-index (39) despite being absent from the top publication counts, suggesting highly impactful research despite moderate publication volume.

Japan offers another compelling case study, appearing in the middle tier of both rankings (tied for sixth in H-index at 38 and seventh in publications at 4%). This consistency suggests a balanced approach to research quality and quantity. Canada maintains similar positioning in both metrics (seventh in H-index at 38 and eighth in publications at 3%), demonstrating steady performance across both dimensions.



## Quantum computing research rankings overview

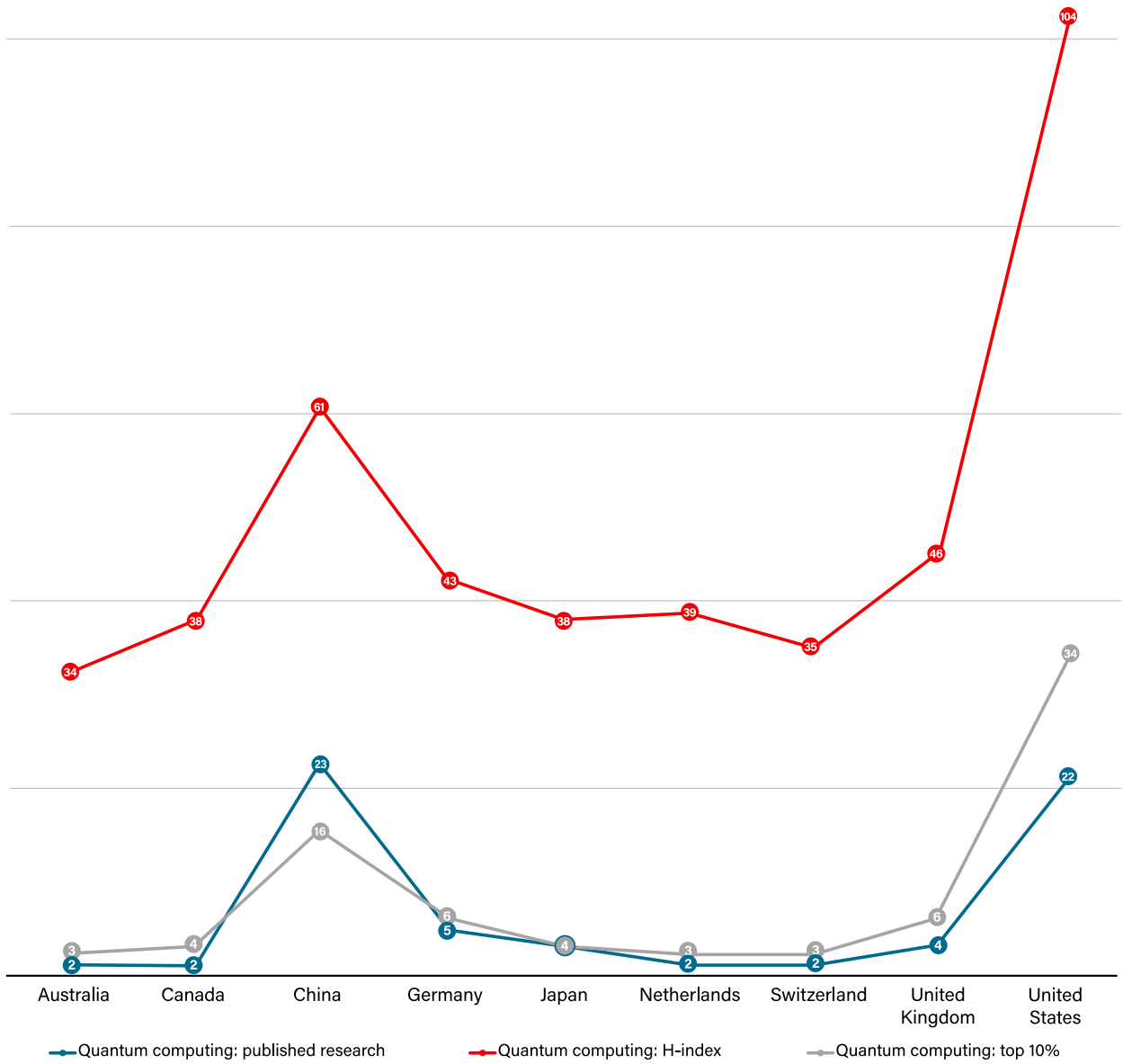



## 2.2 | Quantum communications

### 2.2.1. Rankings by published research

▸ China publishes over one-third of all quantum communications research whereas the United States follows as a distant second. The European Union accounts for 21% of the global total publications. UK, Germany and India are the only others generating 5% of the global output.

China has emerged as the dominant force in quantum communications research output with an impressive 39% of total publications. This substantial lead is particularly notable given that quantum communications represents a specialized area within quantum technology, suggesting deliberate and targeted research efforts and significant institutional capacity. This research effort has likely aided China's success in deploying space-based quantum communications that utilize satellites and long distance QKD networks.

The United States follows as a distant second with 12% of publications, while the United Kingdom, Germany, and India form a closely grouped cluster, each contributing 5% of total publications, demonstrating balanced research output across these major scientific powers.

The middle tier of the distribution shows interesting patterns of research engagement. Russia, Japan, Italy, and Canada each contribute 3% of publications, forming a secondary cluster of substantial contributors. South Korea, Spain, and Austria follow with 2% each, while France, Australia, Switzerland, Singapore, the Netherlands, Poland, Iran, and Denmark complete the distribution with 1% each. This broad international participation suggests a vibrant global research ecosystem in quantum communications, though with clear tiers of research intensity.



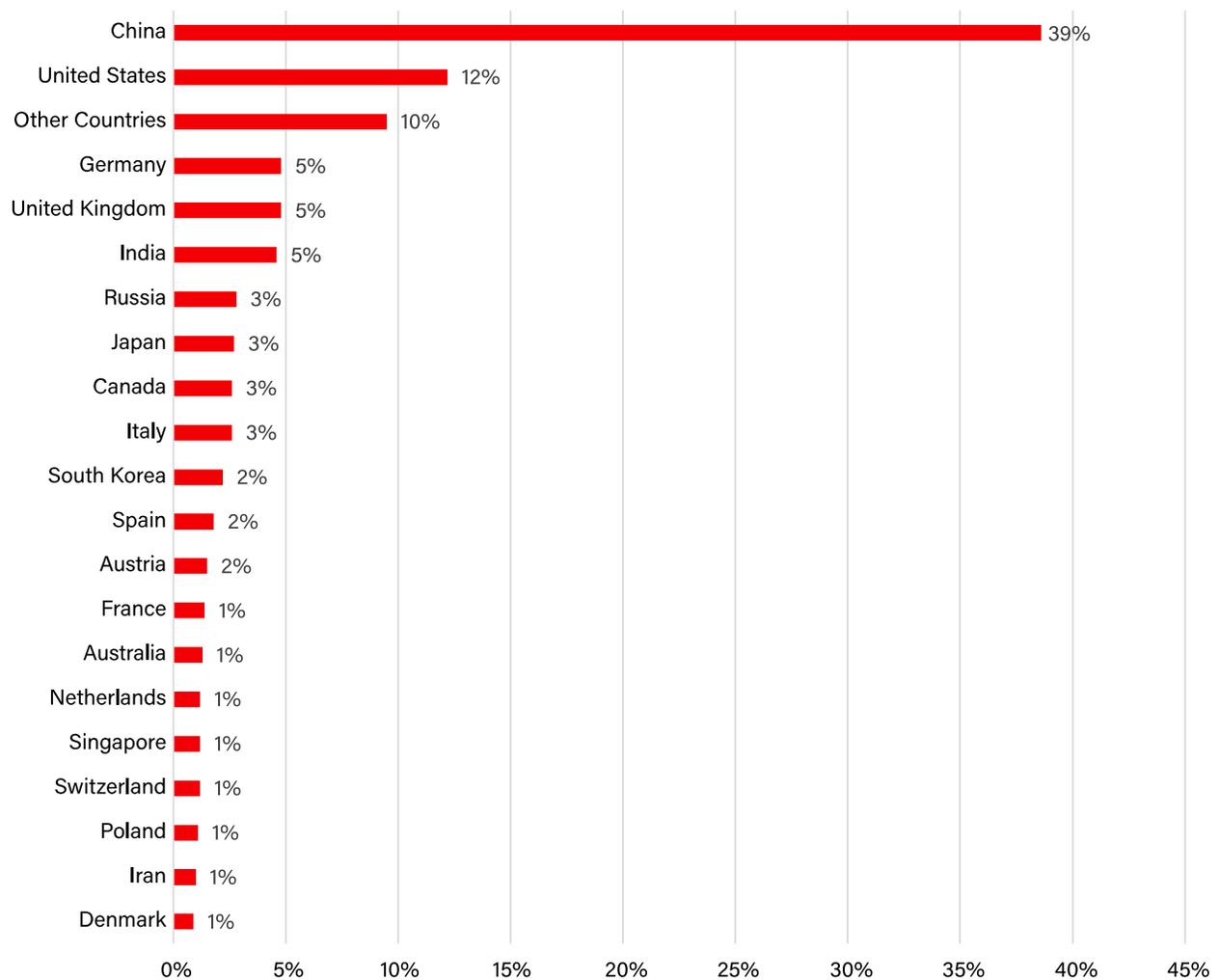

National share of quantum communications published research, 2019-2023



## 2.2.2. Rankings by H-index

▸ In quantum communications, China leads in terms of research quality with an H-index of 51, ahead of the US which takes the second place with 39. Germany and the United Kingdom follow with H-index of 27 and 26, respectively.

National quantum communications published research ranked by H-index, 2019-2023

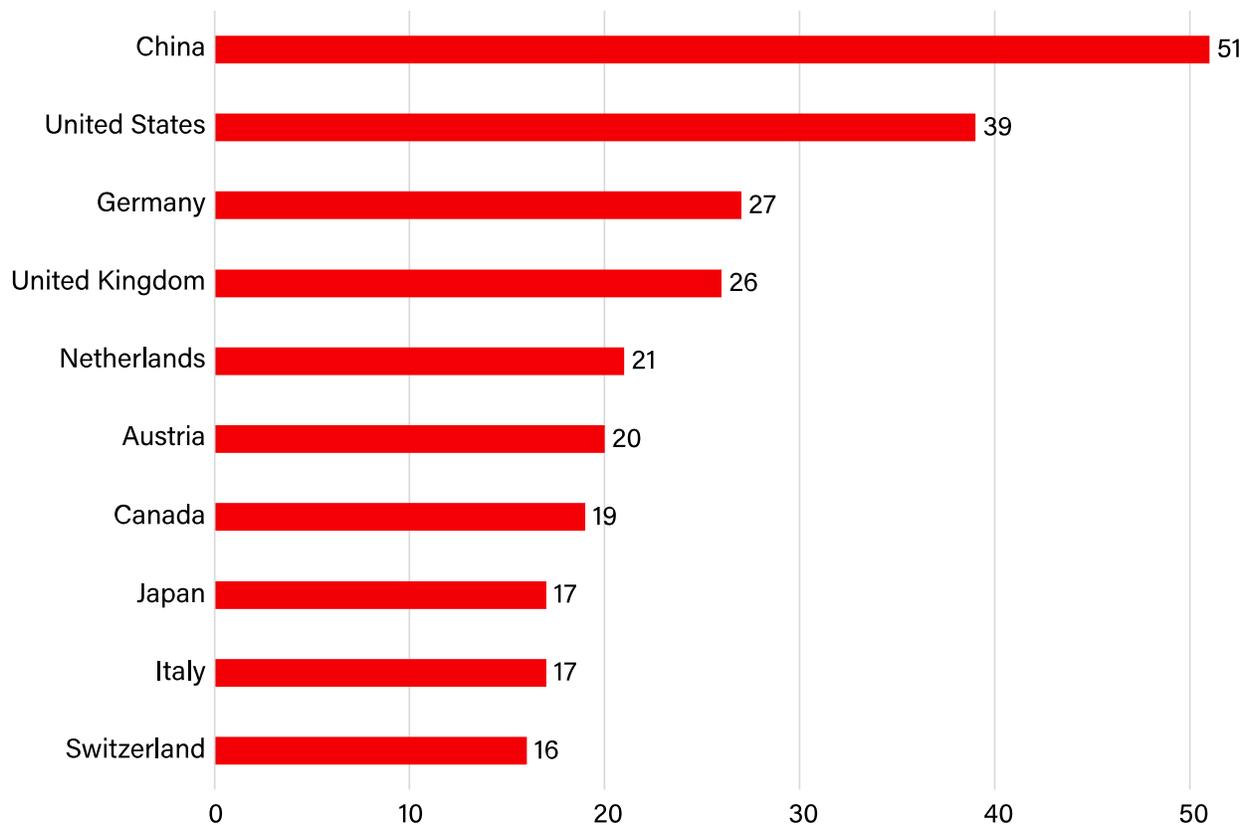



China stands as the leader in the field, with an H-index of 51, demonstrating exceptional research productivity and citation influence. This score places Chinese institutions such as the University of Science and Technology (USTC) firmly at the forefront of quantum communications advancement, significantly outperforming all other nations.

Following China's lead, the United States emerges as a strong second with an H-index of 39, representing a substantial research presence despite being notably lower than China's figure. The top two nations highlight the current state of global competition in quantum computing research.

The middle tier of the distribution reveals intense competition among European nations and Japan. Germany leads this group with an H-index of 27, followed closely by the United Kingdom at 26. The Netherlands maintains a strong research presence with an H-index of 21, while Austria contributes meaningfully at 20. Canada demonstrates an H-index of 19, while Italy and Japan tie at 17, suggesting balanced research ecosystems outside the leading powers. Switzerland completes this tier with an H-index of 16, demonstrating consistent research productivity despite being slightly lower than European counterparts.



### 2.2.3. Rankings by most highly cited papers

▸ China has a third of the top 10% of the most highly cited
  quantum communications publications. The United States
  follows with 17%. The combined European nations account for
  28% of the global total.

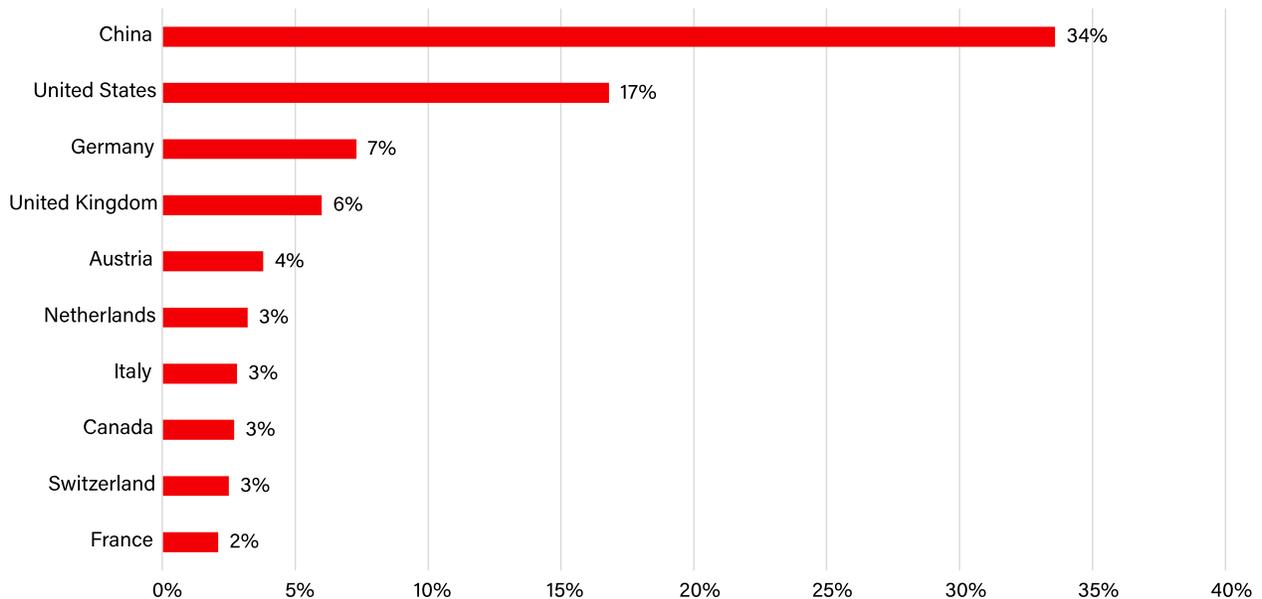

National share of top 10% most highly cited quantum communications publications, 2019-2023

| Country | Share |
|---|---|
| China | 34% |
| United States | 17% |
| Germany | 7% |
| United Kingdom | 6% |
| Austria | 4% |
| Netherlands | 3% |
| Italy | 3% |
| Canada | 3% |
| Switzerland | 3% |
| France | 2% |

Regarding the pattern of global research leadership based on the top 10% most cited publications, China stands prominently at the forefront, accounting for 34% of this field. The United States follows as a strong second, contributing 17% of these influential works, while Germany rounds out the top three with 7%. Together, these three nations dominate the landscape of quantum communications research excellence, collectively producing 58% of the field's most impactful publications.



## 2.2.4. Rankings overall insights

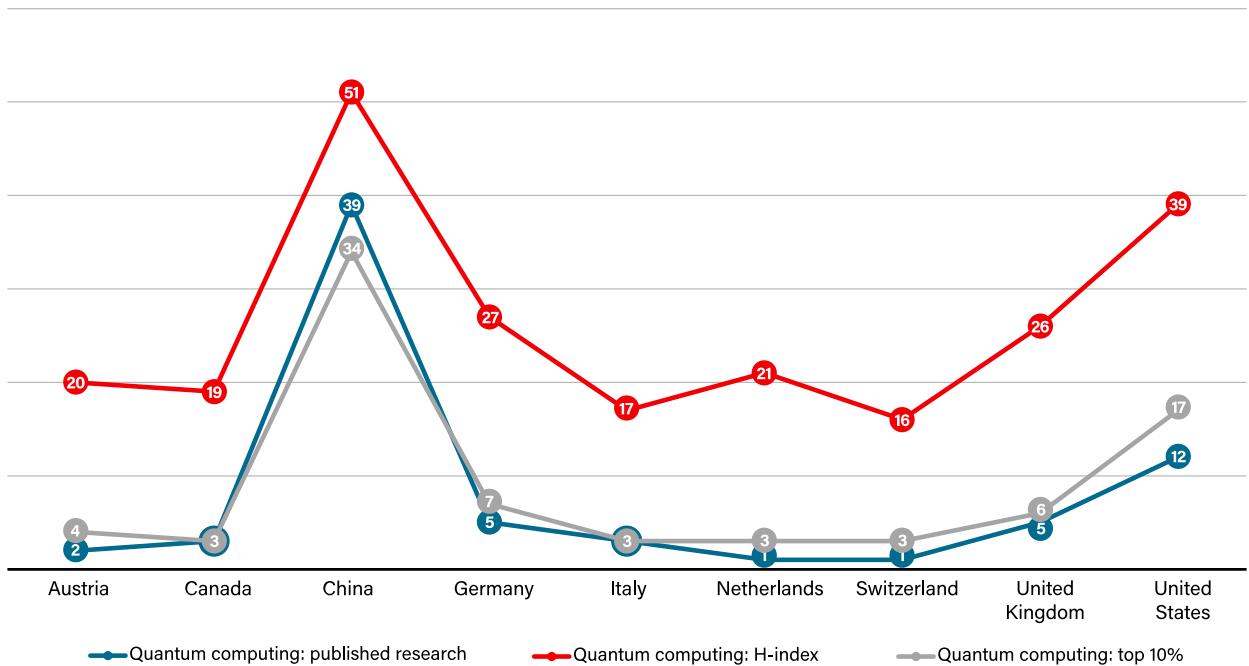

Quantum communications research rankings overview

China dominates the quantum communications field with 39% of total publications, significantly outpacing the next closest contributor, the United States, which accounts for 12% of global output. This substantial gap in publication volume is matched by differences in research quality metrics, where China achieves an H-index of 51 and places 34% of its publications in the top 10% most-cited works. The United States exemplifies high-quality research with an H-index of 39 and 17% of publications in the top 10%, despite lower publication volume compared to China.

European countries show varying levels of performance in quantum communications research. Germany leads as the strongest European contributor with 5% of global publications and an H-index of 27, followed by Austria and the United Kingdom, each contributing 5%. Notably, smaller European nations demonstrate efficiency in research quality despite lower publication volumes: Austria achieves 4% of publications in the top 10%, while the Netherlands maintains an impressive H-index of 21 despite contributing only 1% of global publications.



## 2.3 | Quantum computing vs. quantum communications research

▸ The US maintains leadership in quantum computing research while China leads in quantum communications.

Quantum computing published research by region

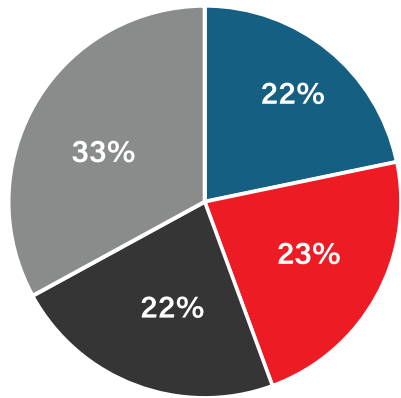

Quantum communications published research by region

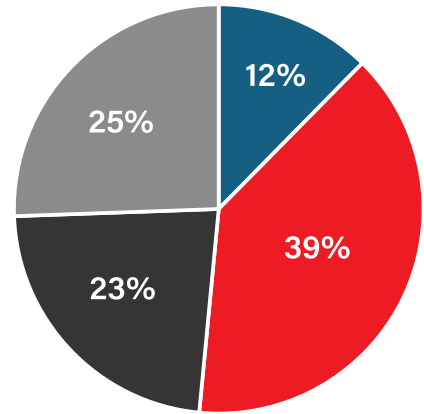

■ United States   ■ China   ■ European Union   ■ Other Countries

China demonstrates leadership in quantum communications with 39% of publications, while maintaining a lower, but still significant position in quantum computing at 23%. This difference of 16 percentage points suggests a strategic focus on quantum communications research in China, likely achieved by specific national priorities and infrastructure investment.

The United States shows an interesting inverse pattern. While ranking second in quantum communications with 12% of publications, it is among the leaders in quantum computing at 22%. This reversal might indicate different strategic priorities between the two nations, with the US maintaining stronger leadership in quantum computing while China focuses more intensively on quantum communications.

European nations present distinct patterns across both fields. Germany and the United Kingdom maintain consistent performance with 5% in both areas, suggesting a balanced approach across the quantum subspecialties.



# 2.4 | Future research

We are interested in extending our analysis by looking into quantum computing and quantum communications subfields, examining how research productivity and quality vary across specialized subdomains. By mapping publication patterns and quality metrics onto these specialized areas, we will better understand the complex interplay between theoretical foundations and practical applications, potentially revealing emerging trends and opportunities for innovation that might be obscured at the broader field level. We invite contributions from the quantum research community to future editions of this report.

You can reach us at contact@qir.mit.edu.

## ▸ Footnotes

[1] Gaida, J., Wong-Leung, J., & Robin, S. (2023). Critical technology tracker. Who Is Leading the Critical Technology Race. A Project by the Australian Strategic Policy Institute. https://techtracker.aspi.org.au

[2] Ibid, H-index (Hirsch index) is an established performance metric used for analyzing the impact of scholarly output. It's a combined measure of quantity and impact. To calculate H-index, a set of papers (e.g. all those on quantum computing from a particular country over a certain time period) has an H-index of N if the relevant authors have published N papers that have N or more citations each. The H-index is based on Times Cited data from the Web of Science database. It will not include citations from non-indexed resources.

[3] The top 10% of the most highly cited papers were analyzed to generate insights into which countries are publishing the greatest share of high-quality, innovative and high-impact research. Credit for each publication was divided among authors and their affiliations and not assigned only to the first author (for example, if there were two authors, they would each be assigned half the allocation). Fractional allocation of credit was used for all metrics.



# 3 | Venture Funding

Financial investment is a critical resource for the development of early stage quantum firms. The embryonic profile of the technology, the field's inherent complexity, and the long-term nature of its development generally make it more suitable for specialist, patient investors.

Our data, compiled in collaboration with Accenture and The Quantum Insider, focuses mainly on publicly available funding announcements from open media sources (press releases, articles, etc). Not all entities fully disclose their funding details, and challenges remain in terms of data gathering and classification (e.g. "Other" category in 2024 data on page 42). Investment levels within large companies such as Google, IBM, Microsoft, or Amazon are not known—and these are some of the largest scale actors in the space.

Within our dataset, total funding for quantum technologies first peaked in 2021. Although there was a decline of approximately 40% in 2023, the sector quickly recovered and reached a new peak in 2024. Quantum computing firms have generated the highest share of overall funding compared to other quantum technologies such as quantum communications and security firms (e.g. quantum networking) and software firms (e.g. quantum algorithms). Despite the growth in recent years, quantum technology investment still represents only a tiny fraction of total venture funding (<1%).

---

Despite the growth in recent years, quantum technology investment still represents only a tiny fraction of total venture funding (<1%).



## 3.1 | Quantum technology funding landscape by round

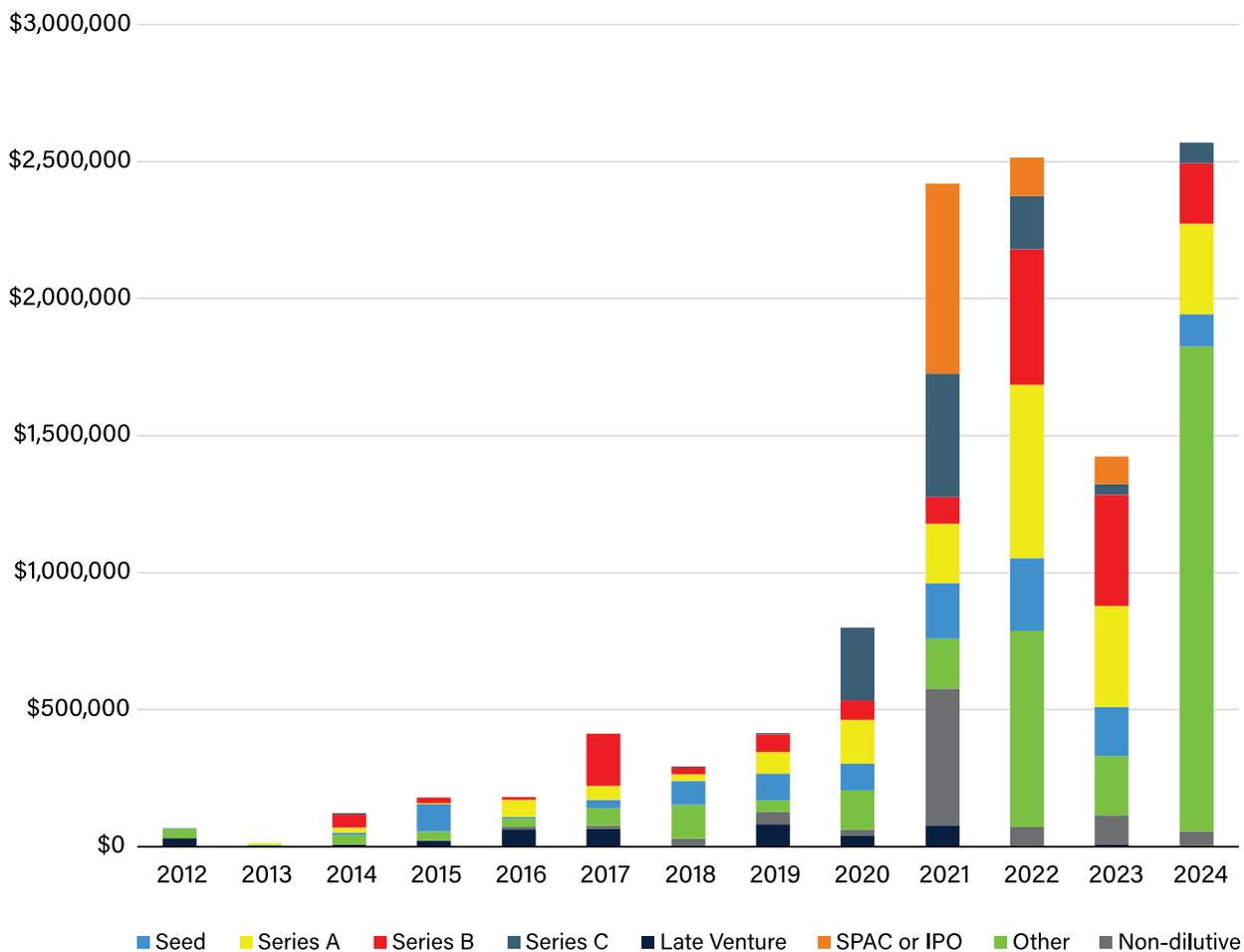

Quantum technology funding landscape by round, 2012-2024

* The 'Other' funding category encompasses a wide variety of investments that either did not fit discreetly into the standard classification groups or was not precisely reported. Individual investments are sometimes complex and opaque as companies may secure funding from multiple sources simultaneously which have different terms and structures. There are also a wide variety of sources including government, institutional, traditional venture capital, grant aid, or quasi-debt. Due to the evolving nature of quantum funding structures in 2024, a number of transactions that might otherwise be categorized under 'Late Venture' have been included in 'Other' to maintain consistency with earlier data and emerging patterns.

The quantum technology funding landscape has undergone multiple transformations over the past decade, marked by periods of explosive growth punctuated by strategic corrections. The journey began with modest investments in 2012, followed by rapid acceleration in 2014, establishing the foundation for future growth.



After a brief correction in 2018, the sector entered a new phase. 2019 marked a recovery. The significant inflection point in the market occurred in 2020 when total funding approximately doubled compared to the previous year. This continued in 2021 when it roughly tripled again. Spikes in 2021 and 2022 were somewhat driven by quantum computing companies going public in the form of Special Purpose Acquisition Companies (SPACs). These are atypical funding vehicles that are unlikely to occur on a regular basis going forward. Significant SPAC examples included IonQ and Rigetti in 2021 and D-Wave in 2022.

Quantum technology funding has shown remarkable growth and diversification across various investment categories from 2012 to 2024. The total quantum technology investment landscape is dominated by "Other" funding sources, which account for 30% of all investments, followed by Series A funding at 17% and Series B at 14%. Seed investments represent 10% of the total funding, while Series C rounds contribute 9%, and SPAC/IPO activities account for 8%. Non-dilutive funding and Late Venture investments make up smaller portions at 7% and 4%, respectively.

Non-dilutive funding has grown substantially over time, increasing from minimal amounts in early years to reach significant levels, with its highest point at $500 million in 2021. SPAC/IPO activities, while less frequent, represent major funding milestones, with investments reaching $693 million in 2021.

This evolution might reflect the maturation of quantum technology companies, transitioning from early-stage venture funding to later stage capital structures. The sector has demonstrated substantial resilience, with each temporary downturn serving as a stepping stone for a rebound and further growth.

## 3.2 | Quantum technology venture funding by category

The quantum technology sector has experienced significant shifts in investment patterns over the past decade. Quantum computers have generated the highest share of overall funding compared to other technologies such as quantum networks and software.

The data shows strong growth in 2024, particularly amongst Quantum computing firms which received $1.59 billion in investments across the year. Quantum software investments also showed remarkable growth in 2024, reaching $621 million.

Several major investments during 2021-2022 marked significant milestones in the sector's development. In January 2021, Quantinuum secured a $300 million equity



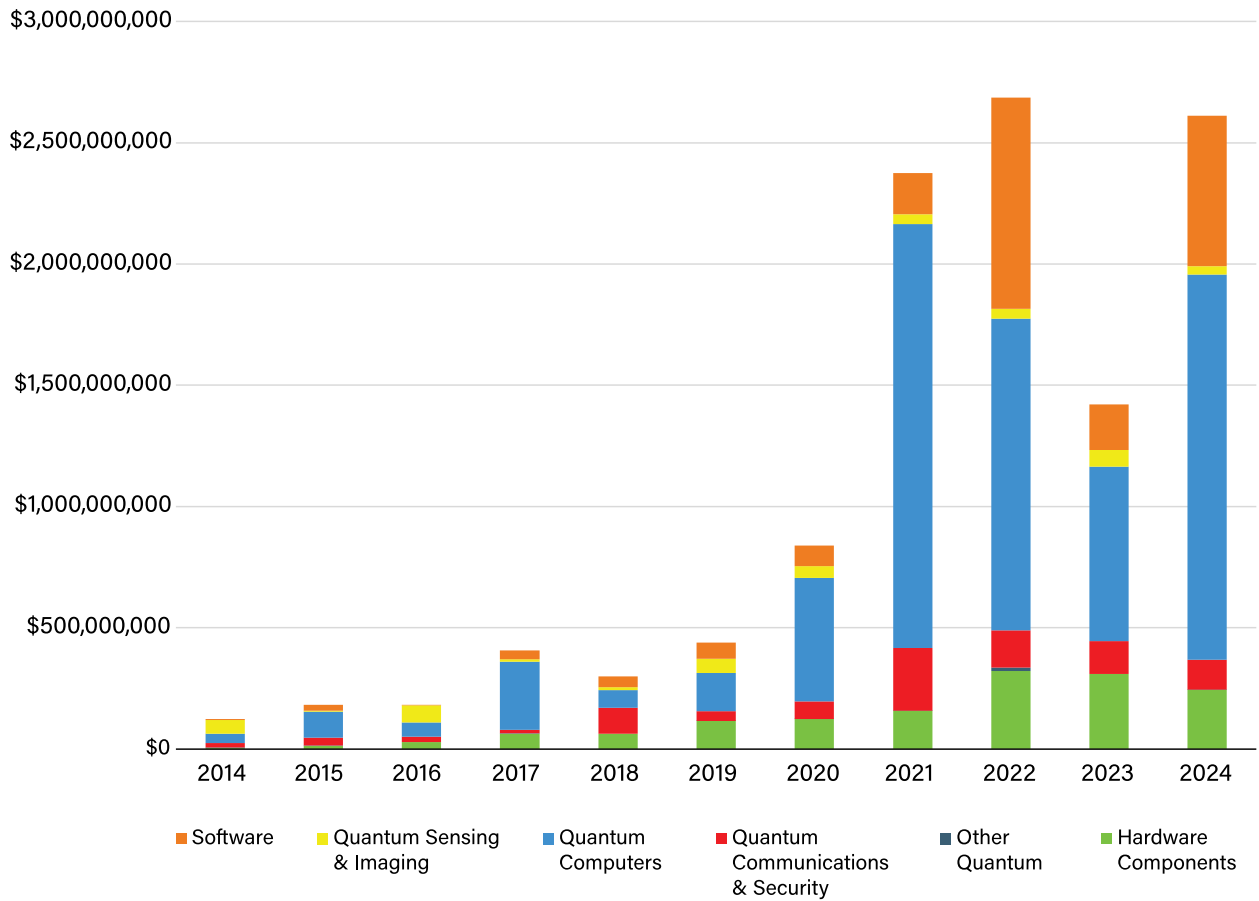

Quantum technology funding landscape by category, 2014-2024

investment, achieving a pre-money valuation of $5 billion.[1] This round drew participation from prominent investors including JPMorgan Chase, Mitsui & Co., Amgen, and Honeywell. Later that same year, PsiQuantum attracted $450 million in funding from Temasek, BlackRock, Microsoft, and other strategic partners.[2]

In 2022, Finland-based IQM Quantum Computers raised €128 million including a venture loan from the European Investment Bank and participation from World Fund, Bayern Kapital, and Tencent.[3] IQM provides full-stack quantum computing systems to supercomputing data centres, research institutes, universities, and enterprises.

More recently, in April 2024, the Australian Commonwealth and Queensland Governments made a substantial commitment to the sector, investing $620 million in PsiQuantum through a financial package consisting of equity, grants, and loans to support the development of a utility-scale quantum computer.



## 3.3 | Quantum technology funding landscape by countries

The global quantum technology funding landscape has evolved into a significant international competition, with multiple countries making substantial investments to secure their position in this emerging field.

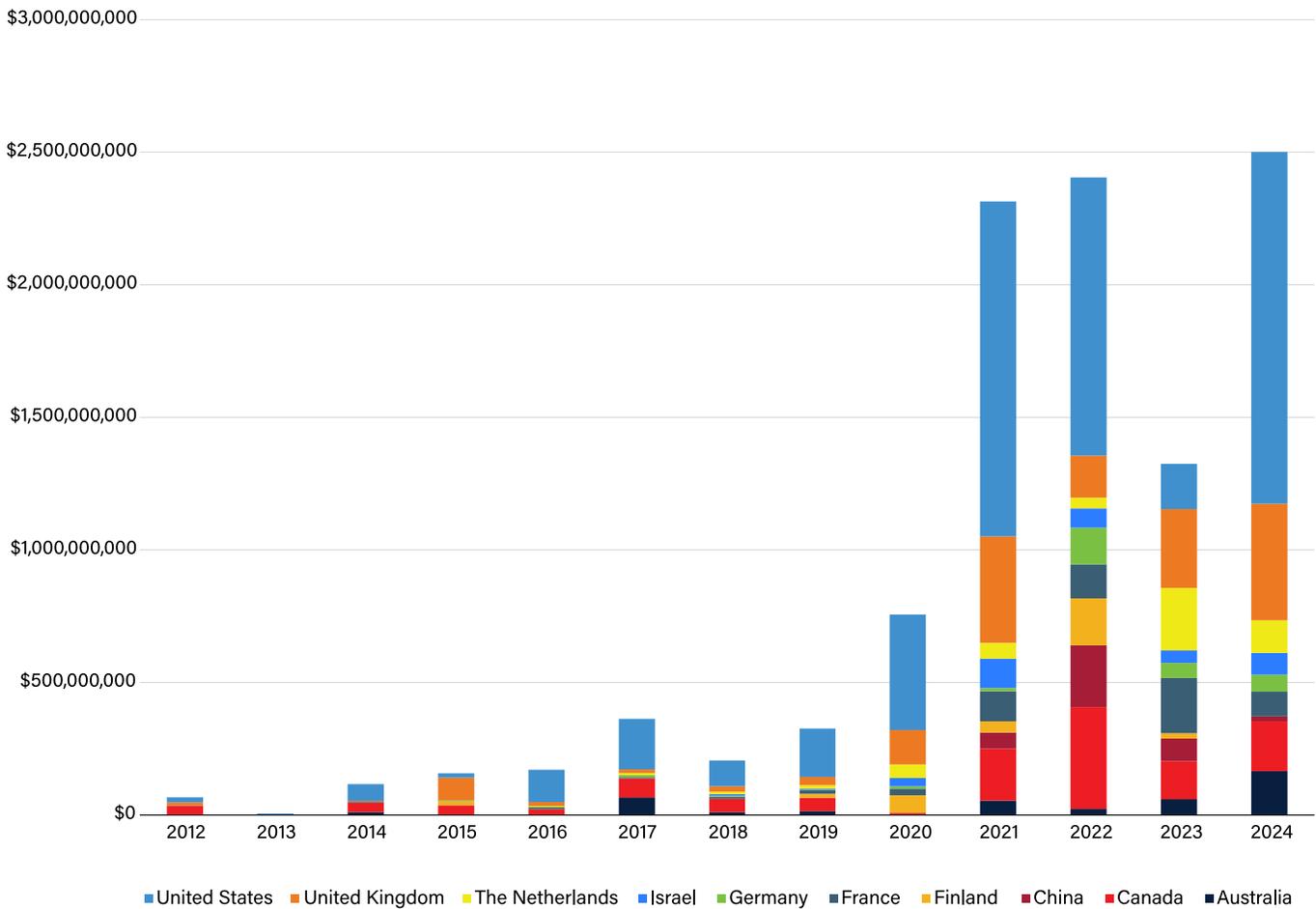

Quantum technology funding landscape by top 10 countries, 2012-2024

The United States and United Kingdom lead the field with a combined share of more than 60% of total funding across 2012 to 2024. The US headquartered firms secured $4.94 billion, followed by UK ventures at $1.6 billion. Canada ranks third with $1.2 billion. At the next level down, France ($606 million), the Netherlands ($540 million), Australia ($412 million), and China ($398 million) form a middle tier, while Israel ($352 million), Finland ($334 million) and Germany ($303 million) are clustered just behind.

The funding history over time shows that most countries maintained very modest but relatively stable investment levels until 2016, after which they adopted either steady growth trajectories or pursued clear bursts of increased funding.



The recent growth patterns reveal a different dynamic between established and emerging players. While the major nations maintain substantial absolute funding levels, several smaller nations demonstrate remarkable growth trajectories. Australia leads with the highest growth rate, reflecting an assertive investment strategy with substantial government support. In contrast, despite their larger absolute investments, the US and UK demonstrate more moderate growth in recent years. This suggests that while established powers maintain their position through sustained investment, smaller nations are dedicated to closing the gap through accelerated funding increases.

The consistent upward trend in investment across a wide range of nations in recent years indicates growing global recognition of quantum technology's strategic importance as well as the desire to build firms that can translate scientific research efforts into commercial success.

## 3.4 | Future research

Given the relatively small annual volume of quantum venture deals internationally, there will be gaps in reporting. Improving data gathering is an opportunity for the community in general. If you have data that would aid this endeavor and enrich future Quantum Index Reports please contact the QIR team.

Building on our analysis, several important research areas emerge for future investigation, including cross-country comparative studies to investigate the relationship between basic science investment, workforce development and startup emergence in quantum (linkages between research, talent and capital), investment impact analysis to quantify relationships between funding patterns and delivery of technological milestones, and the role played by corporate and strategic investors.

You can reach us at contact@qir.mit.edu.

---

### ▸ Footnotes

[1] 'Honeywell Announces the Closing of $300 Million Equity Investment Round for Quantinuum at $5B Pre-Money Valuation' <https://www.quantinuum.com/press-releases/honeywell-announces-the-closing-of-300-million-equity-investment-round-for-quantinuum-at-5b-pre-money-valuation> accessed 28 March 2025.

[2] 'PsiQuantum Raises $450 Million to Build Its Quantum Computer' (PsiQuantum) <https://www.psiquantum.com/news-import/psiquantum-raises-450-million-to-build-its-quantum-computer> accessed 3 April 2025.

[3] 'European Quantum Computing Leader IQM Raises €128m Led by World Fund to Help Combat the Climate Crisis | Press Releases IQM' <https://www.meetiqm.com/newsroom/press-releases/european-leader-in-quantum-computing-iqm-raises-128m-led-by-world-fund> accessed 3 April 2025.



# 4 | Quantum in Corporate Communications



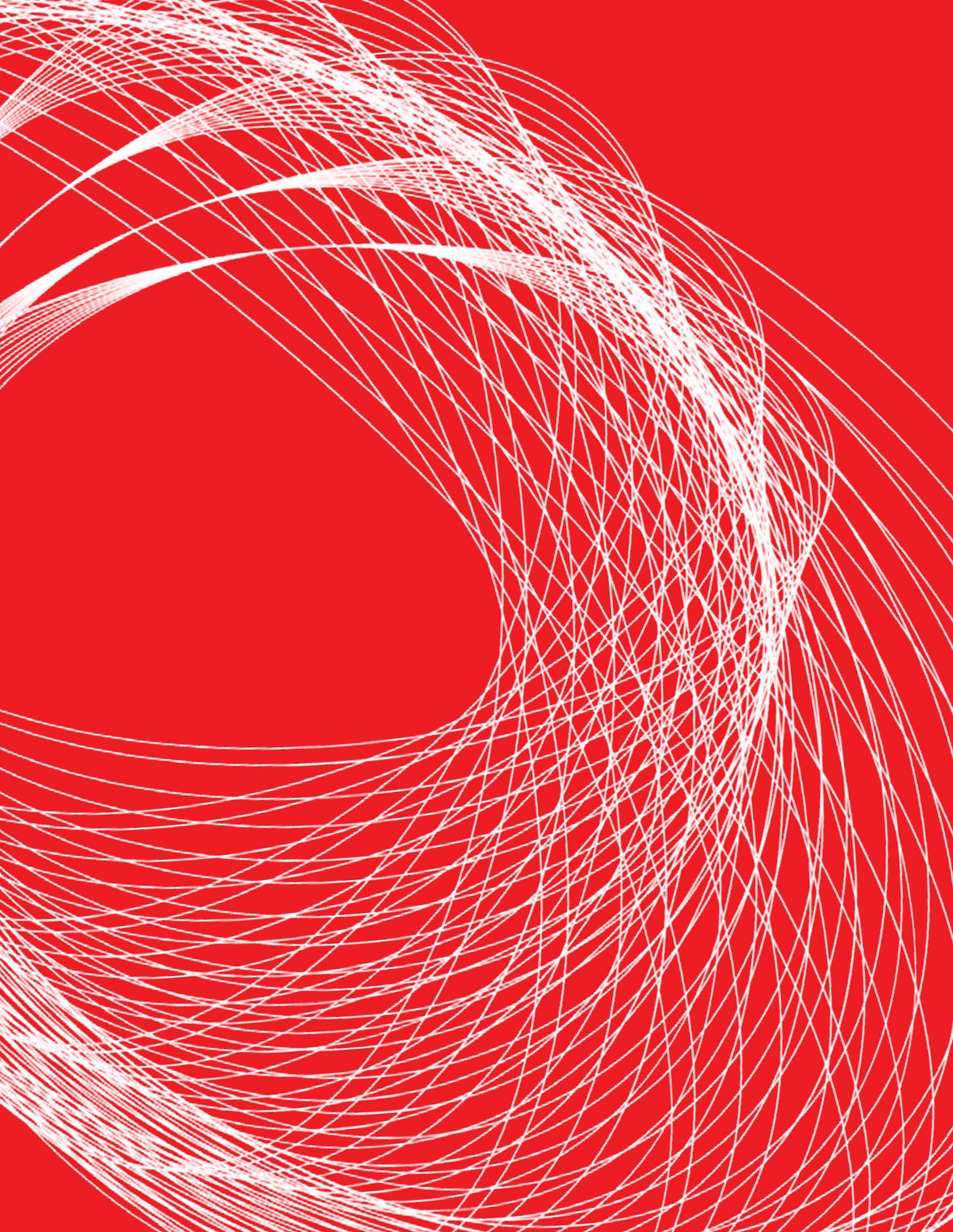

## 4.1 | Mentions in corporate communications

Over the past two years, there has been a significant surge in mentions of quantum computing across corporate communications, with news sources showing particularly pronounced increases. This trend might be reflecting growing business interest in quantum technology. The data for this section was compiled in collaboration with Accenture.

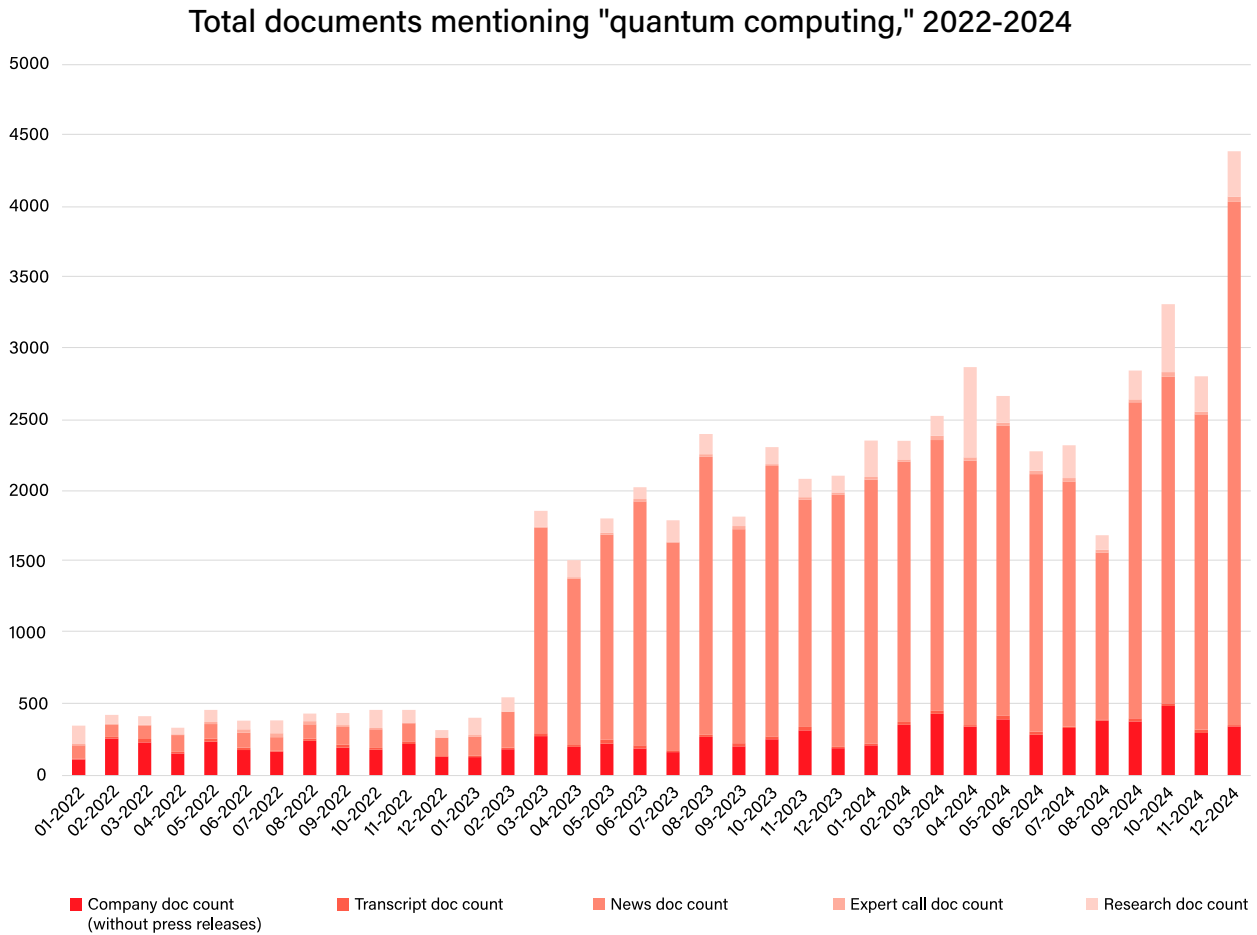

Total documents mentioning "quantum computing," 2022-2024

▶ According to the analysis of 58,070 company documents between 2022 and 2024, the frequency of quantum computing references has shown consistent upward momentum, suggesting an expanding awareness and importance of quantum concepts into mainstream business discourse.

The increase in quantum computing mentions spans multiple types of corporate communications, with especially notable growth in news-related content. Expert calls and research documents demonstrate heightened engagement with quantum technology discussions. This broad-based increase suggests that quantum computing



is becoming progressively integrated into various aspects of corporate communication, moving beyond purely technical contexts to enter mainstream business dialogue.

The quarterly mean analysis of the company documents reveals distinct patterns across the three-year period. In 2022, the year began with a quarterly mean of 203 mentions, characterized by significant monthly fluctuations between 114 and 262 mentions. The second quarter maintained a similar level at 192 mentions, showing more consistent monthly values ranging from 157 to 239 mentions. The third quarter mean slightly increased to 203 mentions, with monthly values between 170 and 244 mentions, while the fourth quarter showed a slight decline to 178.67 mentions, ranging from 132 to 222 mentions.

The pattern shifted notably in 2023, with the first quarter showing 198 mentions, followed by a modest increase to 206 mentions in the second quarter. The third quarter demonstrated stronger growth, reaching 215 mentions, while the fourth quarter showed the highest quarterly mean of the year at 256 mentions. This upward trend in 2023's quarterly means indicates a gradual but consistent increase in quantum computing mentions across company documents.

The more significant changes occurred in 2024, with each quarter showing substantial increases. The first quarter began at 335 mentions, followed by a slight increase to 342 mentions in the second quarter. The third quarter reached 367 mentions, and the fourth quarter achieved the highest quarterly mean of the entire period at 379 mentions. This represents a significant acceleration in the frequency of quantum computing mentions, with 2024's quarterly means consistently exceeding those of previous years by a substantial margin.

The overall trend shows a clear progression from relatively stable quarterly means in 2022 (ranging from 179 to 203 mentions) to moderate growth in 2023 (198 to 256 mentions) and finally to substantial increases in 2024 (335 to 379 mentions). This pattern suggests an accelerating adoption of quantum computing discussions in company documents, with the most pronounced growth occurring in the latter half of the period.

The steady rise in quantum computing mentions across different document types might be indicating a shift in how businesses approach and discuss quantum technology. While research documents naturally maintain high levels of quantum computing references, the significant increase in mentions across other document categories, news articles in particular, might be suggesting a maturing of the technology's presence in corporate communications.

▸ ```
This trend aligns with the growing commercial interest in
quantum computing, as companies increasingly recognize its
potential impact on future business operations.
```

## 4.1.1. Mentions in earnings calls

The growth in quantum computing mentions during earnings calls represents a significant trend in corporate communications within the period 2011-2024. After an insignificant number of mentions in the first years, and starting from just 4 mentions in Q1 2016, the number has grown substantially to reach 25 mentions in Q1 2024, demonstrating a six-fold increase over this eight-year period.

This steady rise in mentions aligns with the broader pattern of increasing quantum computing discussion across corporate communications, where news sources have shown particularly pronounced increases. The trend is supported by concrete evidence of major corporations actively incorporating quantum computing into their strategic discussions, with companies like IBM establishing dedicated quantum facilities and launching new quantum-focused initiatives.[1]

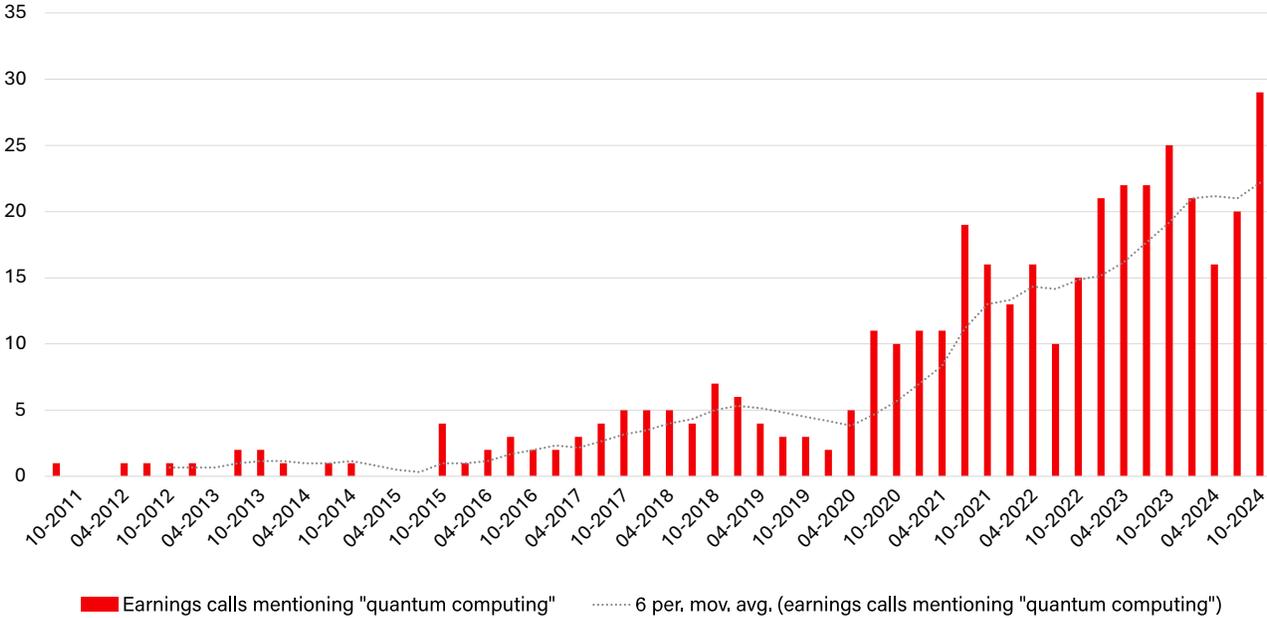

Total earnings calls mentioning "quantum computing," 2011-2024

The growth in earnings call mentions reflects an increasing and broader commercial interest in quantum computing. It has been accompanied by significant market expansion projections; for example, a 2024 report from Technavio estimated the global quantum computing market will grow by $17.34 billion (USD) from 2024 to 2028, a compound annual growth rate of 26%.[2] The market has also seen sustained and substantial startup investment, for example a $300 million equity injection for Quantinuum in 2024.[3]

## 4.1.2. Quantum computing mentions by company

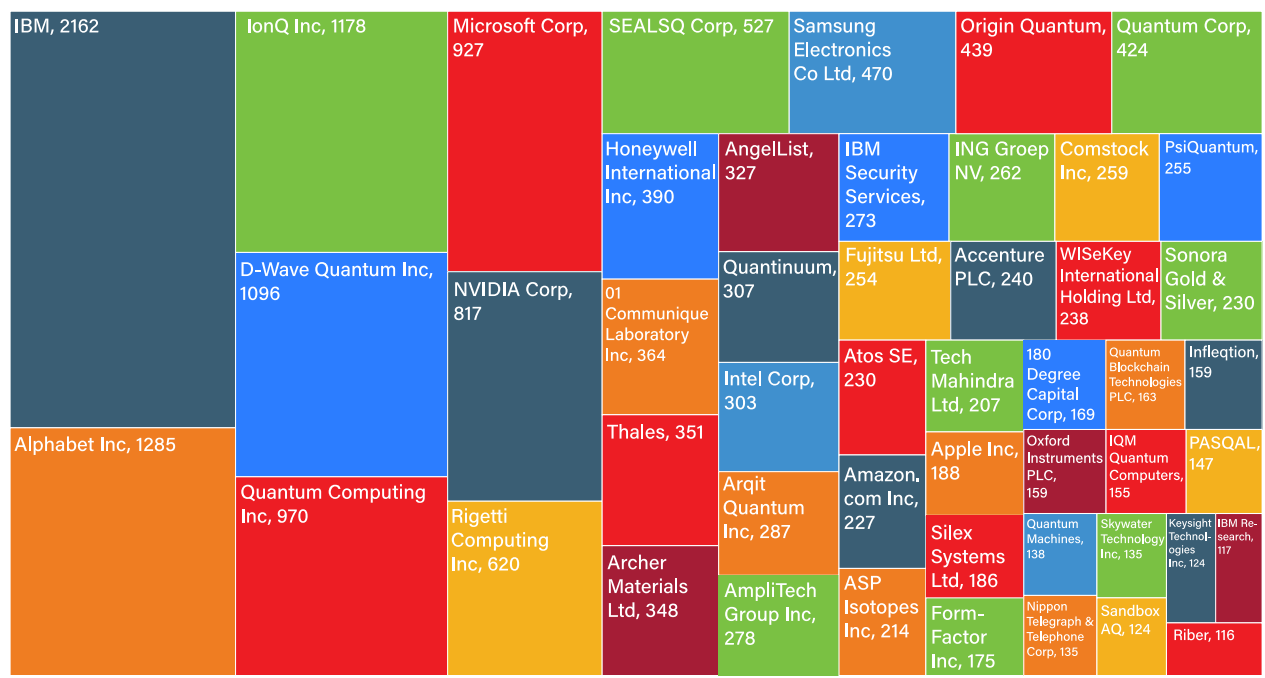

Total document count company (top 50)

IBM and Alphabet stand out as the leading firms in terms of producing external communications, as measured by number of documents (e.g. press releases or earnings calls) mentioning quantum computing. They are followed by IonQ in the third place, D-Wave in the fourth place, and Quantum Computing Inc. in the fifth place completing the top five. Microsoft, NVIDIA, Rigetti, SEALSQ, and Samsung complete the top ten.

## 4.2 | Future research

We look forward to expanding this research in the future editions of this report. We are interested in analyzing industry-specific patterns in quantum technology adoption messaging, examining cross-industry collaboration patterns in quantum technology communications and exploring other relevant trends that might align with mentions in corporate communications.

You can reach us at contact@qir.mit.edu.

▸ **Footnotes**

[1] 'Embracing the Quantum Economy: A Pathway for Business Leaders' (World Economic Forum 2025) Insight Report <https://reports.weforum.org/docs/WEF_Embracing_the_Quantum_Economy_2024.pdf> accessed 2 February 2025.

[2] Technavio https://www.technavio.com, 'Quantum Computing Market Growth Analysis Research Report - Historical & Forecast 2024 - 2028' <https://www.technavio.com/report/quantum-computing-market-industry-analysis> accessed 28 March 2025.

[3] 'Honeywell Announces the Closing of $300 Million Equity Investment Round for Quantinuum at $5B Pre-Money Valuation' <https://www.quantinuum.com/press-releases/honeywell-announces-the-closing-of-300-million-equity-investment-round-for-quantinuum-at-5b-pre-money-valuation> accessed 28 March 2025.

# 5 | Policy

National governments around the world are increasingly recognizing quantum technologies as a domain of strategic importance—one that intersects with economic competitiveness, national security, and scientific leadership. As a result, quantum policy is no longer confined to academic funding or isolated research programs. It is becoming a centerpiece of industrial strategy, with countries racing to define and implement national quantum initiatives, invest in infrastructure, and shape global standards. This policy momentum reflects a broader geopolitical dynamic, where early movers aim to secure technological sovereignty and influence the trajectory of the quantum future.

Policy frameworks worldwide face common challenges in managing quantum technology development. One critical issue is the tension between promoting innovation and ensuring security. Nations must balance the need to protect sensitive quantum research with the requirement for international collaboration to advance the field. In 2024, several countries, including the US, Australia, UK, Canada and the Netherlands imposed aligned export controls on quantum technologies.[1]

Most countries have pursued largely independent approaches to their quantum plans. In contrast, the European Union's Quantum Flagship[2] program serves as a model for coordinated continental-level quantum research, pooling national resources while maintaining a shared framework for ethical oversight and societal impact. The future of quantum technology policy making will likely involve increasingly sophisticated international frameworks. Current trends suggest a move toward hybrid models that combine national sovereignty with international cooperation. This evolution in governance approaches reflects the unique nature of quantum technologies, which demand high levels of international cooperation especially at the research level, while simultaneously respecting legitimate national security concerns.

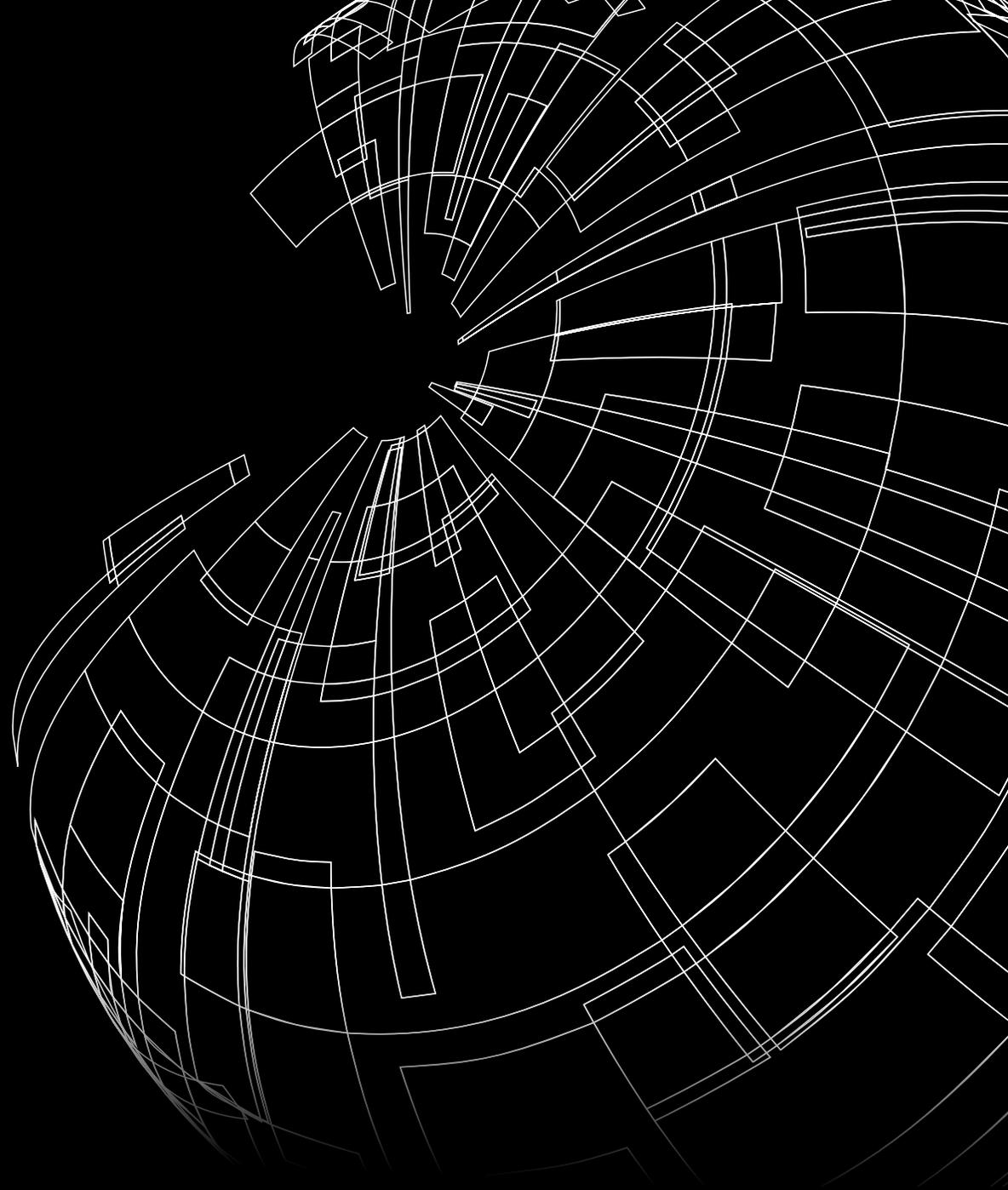

In order to provide a focused analysis of the global policy landscape, this chapter zooms in on policies of seven countries, two leading players in the quantum technology space, the US and China; three anglophone economies with comprehensive national policies in quantum technologies, the UK, Canada and Australia; and two important players serving as European technology hubs, the Netherlands and Ireland.

## 5.1 | United States

The United States has established a comprehensive framework for quantum technology policy development through the National Quantum Initiative (NQI), launched in 2018.[3] This whole-of-government approach coordinates contributions from across federal departments and agencies through either the National Science and Technology Council (NSTC) Subcommittee on Quantum Information Science or the NSTC Subcommittee on Economic and Security Implications of Quantum Science. Recent policy developments have strengthened this framework, notably through the CHIPS and Science Act of 2022, which authorized quantum networking infrastructure development and STEM education integration. The National Defense Authorization Act (NDAA) for fiscal year 2022 further expanded the initiative's scope by formalizing the NSTC Subcommittee on the Economic and Security Implications of Quantum Science.

Cybersecurity is a central pillar of US quantum technology policy, particularly in light of emerging threats to classical encryption systems. The President's National Security Memorandum 10, released in May 2022, established comprehensive policies for promoting quantum computing leadership while addressing cryptographic vulnerabilities.[4] The policy emphasizes transitioning to quantum-resistant cryptography and protecting sensitive technological information, with particular concern about adversaries potentially collecting encrypted data for future decryption. International cooperation has become increasingly important, with strategic documents highlighting the need for dedicated funding mechanisms and enhanced interagency coordination of international practices.[5]

Another crucial pillar of US quantum technology policy is international cooperation. The US government has signed bilateral quantum cooperation statements with Australia, Denmark, Finland, France, Germany, Japan, South Korea, the Netherlands, Sweden, Switzerland, and the United Kingdom.[6] These bilateral partnerships facilitate high-level dialogues between relevant government agencies and create opportunities for enhanced collaboration between research institutions, universities, and industry.

In 2024, the US issued new export controls relating to quantum technologies.[7] The restrictions apply to quantum computers and a broad range of associated items including "related equipment, components, materials, software, and technology that can be used in the development and maintenance of quantum computers."

Overall, the US approach aims to balance national interests with global collaboration—promoting mutual benefits while protecting intellectual capital and property.

## 5.2 | Australia

Australia has launched a comprehensive National Quantum Strategy that aims to transform the nation's future through technological advancement. The strategy was developed through extensive consultation with the quantum sector and wider community before its release in May 2023.[8] By 2030, Australia aims to be recognized as a leader of the global quantum industry, with quantum technologies becoming integral to a prosperous, fair, and inclusive Australia. The strategy identifies both opportunities and challenges, including the potential to capitalize on existing expertise, build sovereign capability, and benefit from economic growth through increased productivity, while addressing challenges in commercialization, capital attraction, infrastructure access, and skills development.

The strategy is built around five central themes that will guide actions over seven years: creating thriving research and development, securing essential quantum infrastructure and materials, building a skilled workforce, establishing supportive standards and frameworks that support national interests, and building a trusted, ethical ecosystem. Key initiatives include investing in quantum ecosystem growth, supporting commercialization, and establishing new programs to incentivize quantum use cases. The government has committed significant resources, including earmarking at least $1 billion from the $15 billion National Reconstruction Fund for critical technologies. Implementation will be collaborative, drawing on the strengths of industry, businesses, universities, states, territories, and trusted international partners to ensure Australia realizes its quantum opportunity.

In 2024, Australia imposed export controls on quantum by the expansion of its "Defence and Strategic Goods List" dual-use section Category 4 (Computers) to include quantum computers.[9]

## 5.3 | Canada

Canada has launched a comprehensive National Quantum Strategy backed by a $360 million investment over seven years, positioning the country to maintain its competitive position in quantum research and technology development.[10] This builds on the fact that Canada has a number of globally recognized academic institutions with strong quantum research efforts, such as the University of Waterloo and University of Toronto. It is also the home to quantum firms such as D-Wave and Xanadu.

The Canadian strategy is built on three interrelated pillars: research, talent, and commercialization, which support key missions that will guide Canada's quantum development. The initiative aims to strengthen Canada's existing quantum research capabilities while growing domestic quantum technologies, companies, and talent, with particular focus on making Canada a world leader in quantum computing hardware and software development.

The strategy's three missions focus on specific technological areas: (1) developing and deploying quantum computing hardware and software, (2) establishing a national secure quantum communications network with post-quantum cryptography capabilities, and (3) supporting the development and early adoption of quantum sensing technologies. Implementation will be supported through various programs, including the Natural Sciences and Engineering Research Council of Canada (NSERC) quantum streams, the National Research Council's Quantum Research and Development Initiative (QRDI), and Canada's Global Innovation Clusters. The strategy emphasizes collaboration between academia, industry, and government, with a focus on creating thousands of jobs and establishing Canada as a global leader in quantum technologies.

According to national strategy, Canada is also committed to strengthening country-to-country collaboration both bilaterally and multilaterally, with an emphasis on key allied countries. This should not only permit jointly advancing knowledge, but also position Canada to work towards ensuring the interoperability of these technologies. The strategy envisions Canada playing a central role in emerging supply chains, with attention paid to protecting sensitive technologies where deemed necessary.

In 2024, Canada implemented export controls on quantum technologies with an update to its export control list, adding quantum computers.[11]

## 5.4 | United Kingdom

The UK's National Quantum Strategy, published in March 2023, outlines a comprehensive 10-year vision to establish the UK as a leading quantum-enabled economy.[12] The strategy builds on the UK's existing quantum strengths, with the government dedicating £2.5 billion to its quantum research and innovation program over ten years. The first two years will see £25 million targeted at training skilled quantum workers through quantum-related fellowships and doctoral training.

The UK has already established itself as a significant player in quantum technology, with approximately 160 companies in the quantum sector and the second-highest percentage of private equity investment in quantum computing globally, second only to the US.[13]

The strategy focuses on five key missions:[14] developing UK-based quantum computers capable of running 1 trillion operations by 2035; deploying the world's most advanced quantum network at scale by 2035; implementing quantum sensing solutions in every National Health Service (NHS) Trust by 2030; deploying quantum navigation systems on aircraft by 2030; and lastly, implementing mobile, networked quantum sensors across critical infrastructure sectors (transport, telecoms, energy, and defense sectors) by 2030. These missions are supported by the National Quantum Technology Program (NQTP)[15], which connects government, academia, and industry to accelerate quantum technology development and commercialization. The strategy emphasizes collaboration between academia and industry, with partnerships involving prominent institutions

57 | MIT INITIATIVE ON THE DIGITAL ECONOMY

and companies while maintaining a strong focus on responsible development and international cooperation.

In 2024, the UK announced export controls on quantum technologies by including quantum computers under the section "Computers and related equipment, materials, software and technology" via amendments to its Export Control Order.[16]

## 5.5 | The Netherlands

The Netherlands has established a comprehensive quantum strategy through Quantum Delta NL, a National Growth Fund program focused on positioning the country as an internationally leading center for quantum technology research and development.[17,18] The program is structured around three catalyst (CAT) programs: (1) quantum computing and simulation, (2) a national quantum network, and (3) quantum sensing applications.

These catalyst programs aim to provide the resources for members to accelerate introduction of quantum to the market via easier access to quantum networks, computers, and simulators—and in doing so, the Netherlands intends to lower barriers to development and testing. Within each catalyst, Quantum Delta NL created four action lines centered around the following themes: research and innovation, quantum ecosystem, human capital, and societal impact.

Funded research initiatives fall within one or more of the six research lines as described in the National Agenda for Quantum Technology[19]: Quantum computing, quantum simulation, quantum communication, quantum sensing, quantum algorithms, and post-quantum cryptography. The program has already demonstrated significant impact, with 16 projects awarded funding in 2022 and 19 in 2023. In March 2025, the National Growth Fund advisory committee approved its updated programming.[20]

Despite its strong research foundation, the Netherlands faces significant challenges in attracting private investment to support its quantum ambitions, according to findings presented in the Invest-NL report "The role of the Netherlands in quantum technology".[21] The 18 existing or upcoming Dutch quantum companies require between €1 billion and €2 billion to reach profitability, with €150-300 million needed within 18 months.[22] While the government has allocated over €600 million through the National Growth Fund, private investors have contributed only €10-15 million in startup capital, significantly less than comparable investments in the United States. The government is actively working to address this funding gap, with Invest-NL prepared to invest part of its €250 million allocation for fundamental technologies in quantum companies.[23] The Netherlands has also strengthened its international position through strategic partnerships, notably signing a joint statement with the United States in February 2023 to enhance cooperation in quantum information science and technology.[24]

With regard to internationalization, the Netherlands advocates balancing different objectives. The country endorses the EU's ambition of building up its own strategic tech industry, which would help guard against undesired dependencies. At the same time, the Netherlands sees 'open markets' as the appropriate departure point and is willing to promote mutual trust between innovative clusters worldwide.[25]

In 2024, the Netherlands expanded the list of items subject to export control by including quantum computers under the category "Computers: Systems, equipment and components".[26]

## 5.6 | China

China has emerged as an ambitious possible global leader in quantum technology, with its strategy characterized by significant state investment and comprehensive national planning. It was announced (and frequently disputed) that the government has allocated an estimated $15 billion to quantum research and development, accounting for over 50% of global public investment in the field.[27] This investment has enabled China to achieve several notable milestones, including the launch of the world's first quantum satellite, Micius, in 2016,[28] and the development of the world's largest quantum communication network spanning 12,000 kilometers.[29] The strategy emphasizes both quantum computing and quantum communication, with particular success in the latter area, where China leads the world in patents and implementation.

China's quantum strategy is distinct from Western approaches, with a strong emphasis on state-led development and technological sovereignty.[30] The Ministry of Industry and Information Technology (MIIT) has identified quantum computing as a "future industry" within its broader industrial policy, focusing on fault-tolerant quantum computing technology and quantum software development.[31] The government has invested in an extensive quantum research facility in Hefei, Anhui Province, covering an area of 37 hectares, aiming for it to be an internationally leading research hub.

The country has implemented a systematic approach to drive and shape standards development.[32] In 2025, China launched its own initiative to develop quantum-resistant encryption standards,[33] paralleling the push by the US to create such standards in its effort organized by the US National Institute of Standards and Technology (NIST).[34]

While private investment in Chinese quantum companies appears to be limited when compared to the US, the state-led approach has enabled a rapid buildup of capabilities.[35] China's strategy also includes significant investment in quantum education and workforce development, with initiatives like the "Education Modernisation 2035 Plan" to prepare future generations for quantum technology development.[36]

## 5.7 | Ireland

Ireland launched its national quantum strategy in 2023, "Quantum 2030: A National Quantum Technologies Strategy for Ireland," which aims to establish the country as an internationally competitive hub for quantum technology by 2030.[37] The strategy recognizes Ireland's unique position as a global technology hub, with nine of the top ten global software companies and three of the top four internet companies maintaining significant operations in the country. While Ireland currently trails similarly sized European states in quantum technologies, the strategy outlines an ambitious plan to capitalize on the country's existing technology ecosystem and develop indigenous quantum capabilities.

The five pillar approach focuses on supporting excellent fundamental and applied quantum research, fostering top science and engineering talent, prioritizing national and international collaboration, stimulating innovation, entrepreneurship and economic competitiveness, and building awareness of quantum technologies and real-world benefits.

The strategy focuses on developing Ireland's quantum research capabilities and building a strong quantum workforce. The initiative emphasizes increasing training through research for scientists, engineers, mathematicians, and electrical engineers, with particular attention to developing "quantum engineers". The government is actively encouraging major technology companies with existing Irish operations to establish quantum technology research labs and recruit early stage Irish scientists. One early success is the hosting of IBM's sole European Research Lab[38] with a focus on quantum in Dublin.

While Ireland faces challenges in competing with countries like the Netherlands, Denmark, and Finland, the strategy represents a significant step toward establishing Ireland as a competitive player in the global quantum technology landscape. The strategy also sets out to build and maintain collaboration with international partners. For instance, Ireland intends to strengthen quantum-technologies research links with the EU, US, and UK. This could be supported by developing new funding mechanisms, as well as drawing on established sources of funding.

# 5.8 | Overview

| | United States | Australia | Canada | United Kingdom |
|---|---|---|---|---|
| **Main Initiative** | National Quantum Initiative Act | National Quantum Strategy | National Quantum Strategy | National Quantum Strategy |
| **Date** | December 2018 | May 2023 | January 2023 | March 2023 |
| **Funding** | $2.7 billion over 5 years (proposed reauthorization) | A$1 billion | Up to C$360 million | £2.5 billion over 10 years |
| **Focus Areas** | Quantum computing, sensing, communications | Research, commercialization, infrastructure, workforce, ethics, international partnerships | Quantum computers and software, quantum communications, quantum sensors | Quantum science, engineering, business support, regulatory framework |
| **Workforce Development** | Quantum education and workforce hub | Aim to be world's top destination for quantum talent | Talent development as a key pillar | New skills initiatives, doctoral training, fellowships |
| **Key Feature** | Shift from basic research to applications in reauthorization | Attention to responsible innovation | Missions-based approach | Aim to be quantum-enabled economy by 2033 via support for business and standards development |

|  | **Netherlands** | **China** | **Ireland** |
|---|---|---|---|
| **Main Initiative** | National Agenda on Quantum Technology | National Strategy for Quantum Science and Technology | Quantum 2030 |
| **Date** | September 2019 | Ongoing since 13th five-year plan (2016-2020) | November 2023 |
| **Funding** | €615 million (Quantum Delta NL program) | $15 billion (estimated) | Full budget unknown (IrelandQCI project: €10 million) |
| **Focus Areas** | Commercialization, education, ethical development, community building | Quantum communications, computing, sensing | Quantum computing, communications, sensing, international collaboration, ecosystem building |
| **Workforce Development** | Focus on training new talents | Centralized talent development at USTC | Develop quantum skills base, support researchers, SMEs, and innovators |
| **Key Feature** | Quantum Delta NL program implementation with focus on ecosystem building | Centralized, state-controlled approach with rapid development focus | Emphasis on coordination, talent development, and leveraging EU/UK/US partnerships |

## 5.9 | Future research

Building on this analysis of national quantum policies from select countries, our aim for future reports is to incorporate additional nations. We also intend to add more detailed comparative analysis with more precise assessments of different governance approaches in the coming years. The tracking framework established in this chapter would provide valuable longitudinal insights into policy evolution and effectiveness over time, allowing researchers to identify successful regulatory adaptations and best practices in balancing innovation with governance objectives. Please reach out to share additional data or insights that might contribute to this work. Our QIR website provides further up-to-date documentation of evolving policies, serving as a dynamic complement to this chapter.

You can reach us at contact@qir.mit.edu.

## ▸ Footnotes

[1] These states are parties to Wassenaar Arrangement on Export Controls for Conventional Arms and Dual-Use Goods and Technologies, however this particular export control implementation effort took place outside of the official Wassenaar Arrangement. This could potentially mean that export control limitations might apply to countries party to the Arrangement if they do not implement similar controls or until the Arrangement ends up covering quantum technologies.

[2] 'Quantum Technologies Flagship | Shaping Europe's Digital Future' <https://digital-strategy.ec.europa.eu/en/policies/quantum-technologies-flagship> accessed 3 April 2025.

[3] 115th Congress, 'H.R.6227 - National Quantum Initiative Act' (21 December 2018) <https://www.congress.gov/bill/115th-congress/house-bill/6227/text> accessed 3 April 2025.

[4] The White House, 'National Security Memorandum on Promoting United States Leadership in Quantum Computing While Mitigating Risks to Vulnerable Cryptographic Systems' (The White House, 4 May 2022) <https://bidenwhitehouse.archives.gov/briefing-room/statements-releases/2022/05/04/national-security-memorandum-on-promoting-united-states-leadership-in-quantum-computing-while-mitigating-risks-to-vulnerable-cryptographic-systems/> accessed 20 March 2025.

[5] National Science and Technology Council, 'Advancing International Cooperation In Quantum Information Science And Technology' (2024).

[6] ibid.

[7] National Quantum Coordination Office, 'Department of Commerce Releases Export Controls on Quantum Technologies' (National Quantum Initiative, 6 September 2024) <https://www.quantum.gov/department-of-commerce-releases-export-controls-on-quantum-technologies/> accessed 3 April 2025.

[8] Department of Industry Science and Resources, 'National Quantum Strategy' (2023) <https://www.industry.gov.au/publications/national-quantum-strategy> accessed 3 April 2025.

[9] Australian Government, 'Defence and Strategic Goods List 2024' <https://www.legislation.gov.au/F2024L01024/asmade> accessed 3 April 2025.

[10] Government of Canada, 'Canada's National Quantum Strategy' (Innovation, Science and Economic Development Canada 2023) <https://ised-isde.canada.ca/site/national-quantum-strategy/en/national-quantum-strategy-roadmap-quantum-computing> accessed 3 April 2025.

[11] Global Affairs Canada, 'Notice to Exporters No. 1129 – Amendment to the Export Control List: Quantum Computing and Advanced Semiconductors' (GAC, 19 June 2024) <https://www.international.gc.ca/trade-commerce/controls-controles/notices-avis/1129.aspx?lang=eng> accessed 3 April 2025.

[12] UK Department for Science, Innovation and Technology, 'National Quantum Strategy' (GOV.UK, March 2023) <https://www.gov.uk/government/publications/national-quantum-strategy> accessed 3 April 2025.

[13] UK Department for Science, Innovation and Technology, 'National Quantum Strategy Additional Evidence' (GOV.UK, December 2023) <https://www.gov.uk/government/publications/national-quantum-strategy> accessed 3 April 2025.

# 6 | Workforce

The quantum technology sector faces a critical challenge in developing and maintaining a qualified workforce to support its continued advancement. Occupations often require a combination of theoretical knowledge and practical expertise, making it challenging for employers to find candidates with the right mix of skills and acumen.

Major nations understand that establishing and nurturing a critical mass of quantum talent is a priority and have responded by developing comprehensive strategies to address these needs. The United States has established the National Quantum Initiative[1] (NQI), which includes dedicated funding for quantum workforce development and coordinates efforts across academia, industry, and government sectors. Investment in NQI Act-authorized activities alone exceeded $2.5 billion from 2019 to 2024.[2]

In addition to the NQI, the CHIPS and Science Act[3] included specific provisions for better evaluating quantum workforce needs and initiatives to drive quantum curriculum development and leadership.

Similarly, many other countries such as Canada[4] and Australia[5] have launched national quantum strategies specifically emphasizing workforce expansion and talent development. These initiatives recognize that developing quantum expertise should not only focus on technical training, but that it also requires creating an entire ecosystem of quantum-savvy professionals who can bridge the gap between research and practice.

> The "quantum-as-a-service model" is enabling wider access to quantum computing resources which supports relatively low cost experimentation and drives skills development in the area.

In 2025, the European Commission announced the Digital Europe work program for 2025-2027, which includes the establishment of a Quantum Digital Skills Academy with the aim of closing the talent gap and strengthening the pool of specialists. The indicative budget for the academy was announced as €10 million.[6]

In recent years the United States has created more quantum job openings than can be filled[7], with the variety of roles related to quantum expanding in academia, industry, national labs, and government. The opportunities range[8] from highly specialized jobs (e.g. error correction scientist or quantum algorithm developer) to occupations requiring a range of skills, most of which are not quantum related (e.g. business development for quantum computing firms).

The educational infrastructure supporting this growth includes the establishment of quantum hubs at universities and research institutes, specialized training programs connecting business managers with leading quantum researchers, and integration of quantum education into existing academic frameworks. The "quantum-as-a-service model" is enabling wider access to quantum computing resources that supports relatively low-cost experimentation and drives skills development in the area.

## 6.1 | Quantum skills in job postings

▸ The US labor market has shown relatively steady growth in demand for quantum skills since 2018.

US job postings requiring "quantum" skills as share of total job postings, 2011 to mid-2024

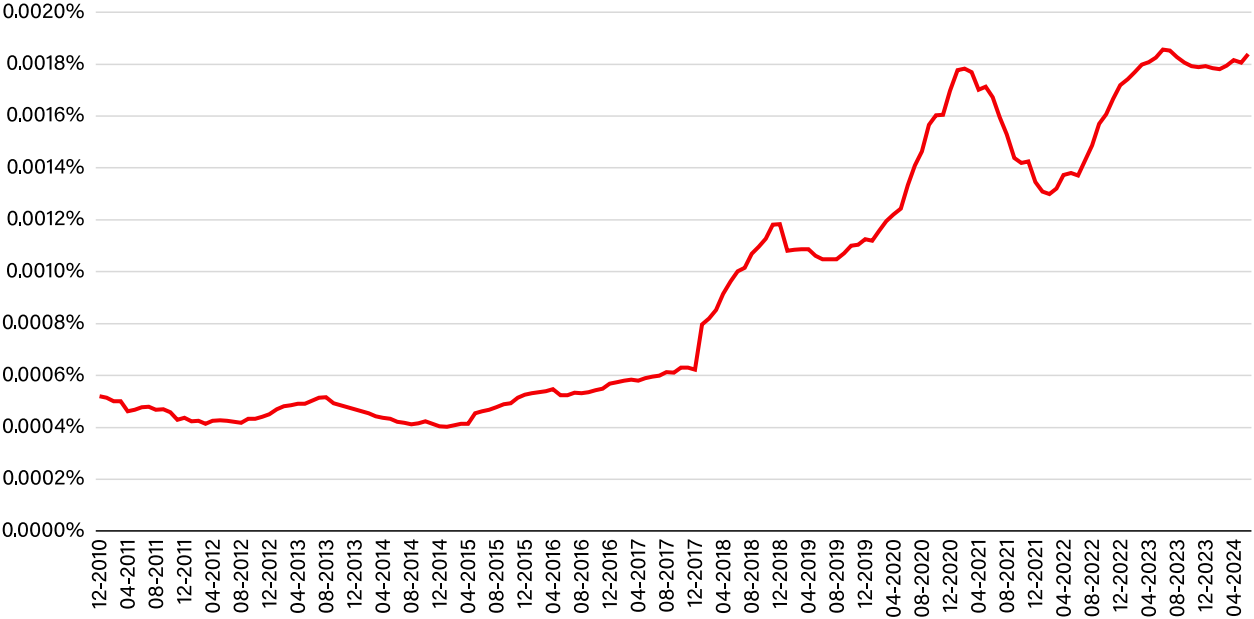

Based on the Lightcast data on US job postings requiring "quantum" skills as a share of total job postings from 2011 to mid-2024, the share of quantum skills in job postings has grown almost three times.

The data shows three distinct phases in the evolution of quantum skills demand. The initial period from 2010 to 2017 was characterized by limited growth. This was followed by an acceleration beginning in 2018, when the share almost doubled over a two-year period. Since 2021, the growth has stabilized into a more moderate but consistent upward trend, with occasional fluctuations becoming more pronounced.

Seasonal patterns seem to emerge as a significant feature of the data, with consistent quarterly variations. The highest shares of quantum skills in job postings occur during the second and third quarters of each year, while the first and fourth quarters typically show lower shares. This seasonal amplitude suggests a regular cyclical pattern in hiring demand.

The peak growth rate occurred in 2020, and while rates have since moderated, the data shows consistent upward movement, indicating sustained growth in quantum skills demand. The overall trend demonstrates the increasing importance of quantum skills in the labor market.

## 6.2 | Quantum skills in job postings

> US job postings with mentions of "quantum" began to rise
> rapidly in 2018 before peaking in 2019. There is no evidence
> of sustained growth in quantum demand versus the overall
> labor market (which was very robust in 2021-2024).

US job postings with mentions of "quantum" as share of total job postings, 2011 to mid-2024

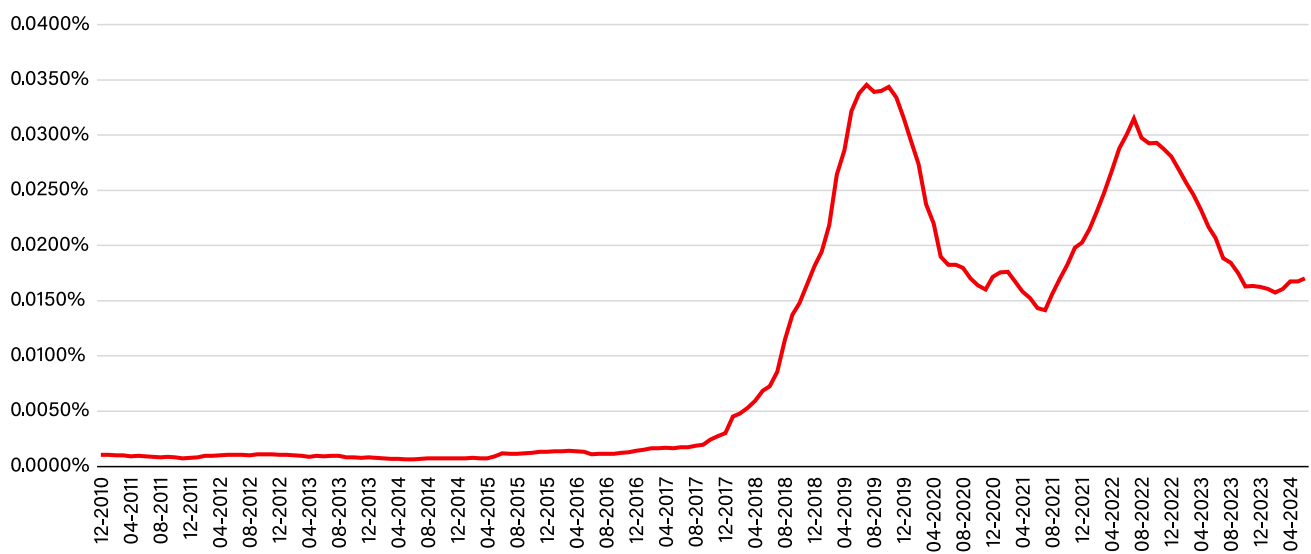

The Lightcast dataset of US job postings mentioning "quantum" spans from 2011 to mid-2024, encompassing monthly observations. The temporal pattern revealed three distinct phases in quantum workforce demand. Initially, from 2011 to 2017, the market showed remarkable stability. This early period demonstrated minimal volatility. However, beginning in 2018, the landscape underwent strong transformation, marked by increased growth that continued through 2019. During this period, quantum job postings reached their peak share in July 2019, representing a significant increase from the early period baseline.

More recently, from 2020 onward, the market has entered a phase of stabilization and moderate adjustment. While experiencing some decline from the 2019 peak, quantum-related job postings have maintained levels significantly higher than the pre-2018 era. Current figures have stabilized in the early months of 2024.

These monthly patterns suggest that quantum-related hiring typically peaks during summer months and follows a quarterly cycle with highest activity in Q3. However, it's important to note that while these trends exist, they are relatively modest compared to the overall growth trend in quantum job postings over time.

Throughout the entire period, despite fluctuations, the overall trajectory indicates sustained growth in quantum workforce demand—suggesting continued expansion in the field's employment opportunities.

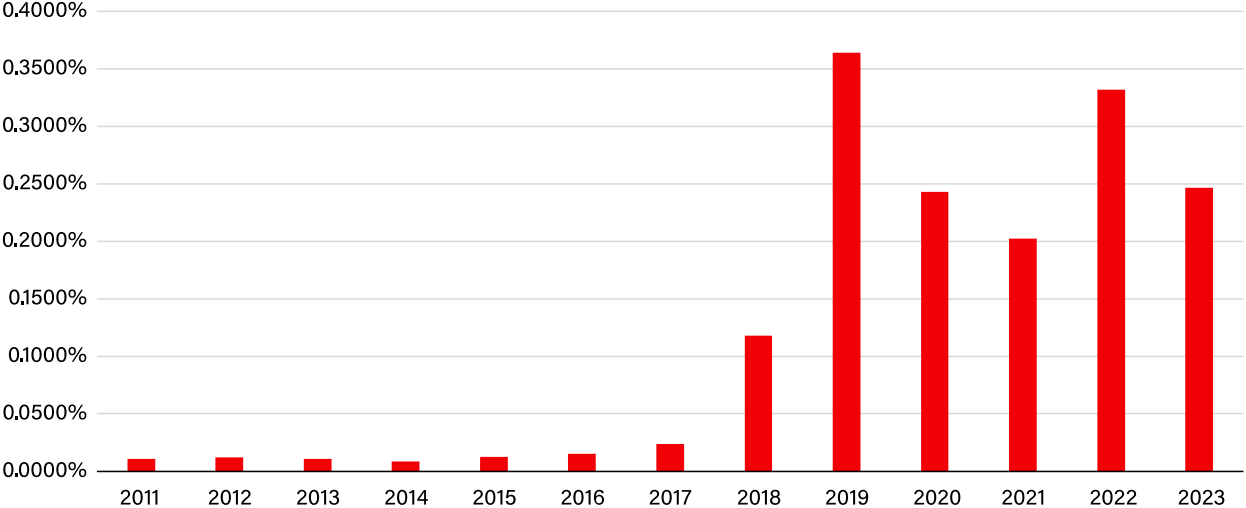

US job postings share mentioning "quantum" by year, 2011-2023

## 6.3 | Future research

We aim to continue to track this data to generate insights into the ongoing trends with the aim of better informing the community on important workforce developments. The Lightcast data we gathered suggested the quantum skills library could benefit from ongoing updates as this is a dynamic field where new job descriptions and technical requirements are continuously added. Please reach out if you are interested in collaborating on skills libraries, new resources to monitor, or the creation of additional datasets internationally.

You can reach us at contact@qir.mit.edu.

## ▸ Footnotes

[1] 115th Congress, 'H.R.6227 - National Quantum Initiative Act' (21 December 2018) <https://www.congress.gov/bill/115th-congress/house-bill/6227/text> accessed 3 April 2025.

[2] National Science and Technology Council, 'National Quantum Initiative Supplement To The President's Fy 2025 Budget'.

[3] 117th Congress, 'H.R.4346 - 117th Congress (2021-2022): CHIPS and Science Act' (9 August 2022) <https://www.congress.gov/bill/117th-congress/house-bill/4346> accessed 14 November 2024.

[4] Government of Canada, 'Canada's National Quantum Strategy' (Innovation, Science and Economic Development Canada 2025) <https://ised-isde.canada.ca/site/national-quantum-strategy/en/national-quantum-strategy-roadmap-quantum-computing> accessed 3 April 2025.

[5] Department of Industry Science and Resources, 'National Quantum Strategy' (2023) Strategy or plan <https://www.industry.gov.au/publications/national-quantum-strategy> accessed 3 April 2025.

[6] 'Commission to Invest €1.3 Billion in Artificial Intelligence, Cybersecurity and Digital Skills | Shaping Europe's Digital Future' <https://digital-strategy.ec.europa.eu/en/news/commission-invest-eu13-billion-artificial-intelligence-cybersecurity-and-digital-skills> accessed 3 April 2025.

[7] National Science and Technology Council, 'Quantum Information Science And Technology Workforce Development National Strategic Plan' (2022).

[8] Ciaran Hughes and others, 'Assessing the Needs of the Quantum Industry' (arXiv, 25 August 2021) <http://arxiv.org/abs/2109.03601> accessed 3 April 2025.

# 7 | Education

The creation of a mature quantum ecosystem depends not only on scientific breakthroughs and unlocking commercial use cases but also on the cultivation of a multidisciplinary workforce equipped to build, navigate, and govern this emergent space. However, the inherently complex nature of quantum phenomena and the reliance on advanced mathematics and physics concepts can pose a perceived barrier to education and training. Despite this there are an increasing range of global initiatives focused on providing training at all levels, from K-12, to postgraduate and professional development.

In the US, the National Q-12 Education Partnership[1] was launched in 2020 as part of the national quantum strategy and aims to increase the capabilities and number of students who are ready to engage in the quantum workforce by developing K-12 level educational materials and providing classroom tools for hands-on experiences.

There are similar examples of quantum education programs targeted at this level in China[2], and in the EU, through its Quantum Flagship's dedicated initiative to implement quantum topics in high school curricula.[3] Industry also showed an interest in filling the formal curricula lag in quantum for high school students. The Coding School, a non-profit, launched an introductory course in quantum technologies targeted at high school students in collaboration with IBM, MIT, and UC Berkeley in 2020.[4] The Coding School reports that their Introduction to Quantum Computing course was attended by over 18,000 high school students so far, and it continues to be offered in collaboration with Google Quantum AI for its September 2025 iteration.[5]

Many universities internationally offer specialized degrees in quantum technologies, at both undergraduate and graduate levels. These programs often involve interdisciplinary approaches, combining physics, computer science, and engineering to prepare students for careers in quantum research and development.

In this chapter, we present global data on master's degree programs dedicated to quantum technologies. Given the very limited number of offerings of bachelor's degree programs dedicated to quantum technologies, we present enrollment data for bachelor's degrees in the physics, computer science, and engineering fields within the US.

## 7.1 | Postgraduate education

As the quantum technology industry continues to grow, there may be increasing demand for specialized master's degrees tailored to different sectors. Some institutions have already begun to offer dedicated streams within their programs. This trend towards more specialized training reflects the growing diversity of roles in the quantum sector and the need for education to keep pace with industry demands.

▸ Germany is the leading nation in terms of master's degrees in quantum technologies, with 12 programs on offer. The UK follows closely with 10 programs, whereas the United States offers 9. France and the Netherlands are also in the top 5 countries offering master's degrees specifically referring to "quantum" in the degree title.

This distribution suggests that quantum technology is becoming increasingly important globally, with major research hubs like Germany, the United Kingdom, and the United States leading the way. The number of programs also reflects the interdisciplinary nature of quantum technology, which often involves physics, engineering, computer science, and mathematics. This diversity of possible departmental homes within universities is likely contributing to the growth of these programs. As expectations for commercial application breakthroughs in quantum computing continue to rise, we expect to see further expansion in the number of master's programs dedicated to this field globally.

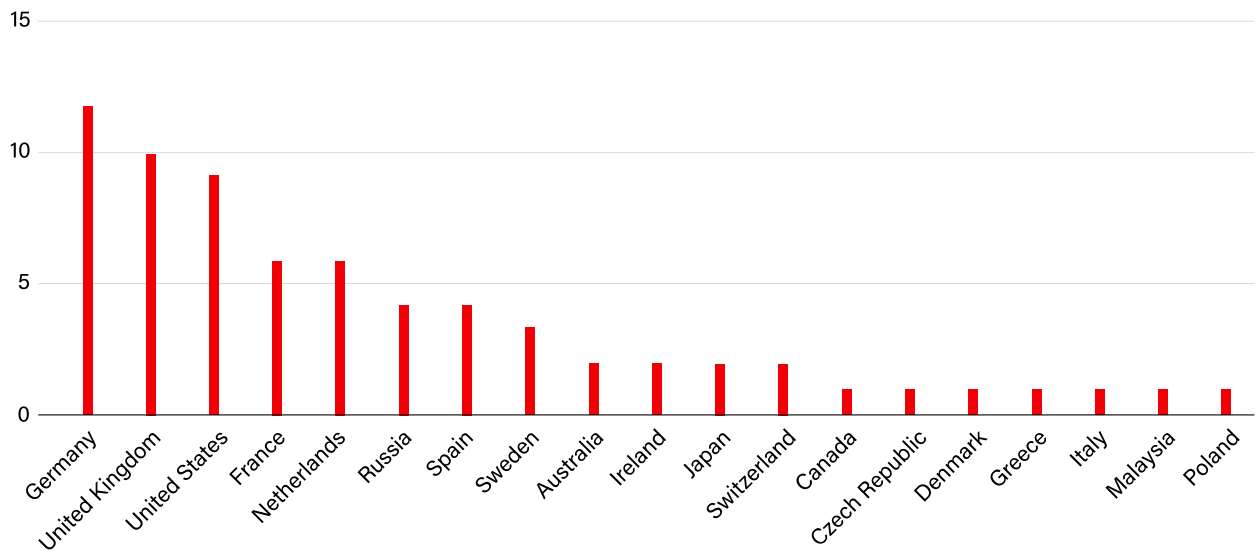

Master degrees with a specific reference to 'quantum' in the degree name

Based on Studyportals data, there are 69 master's degrees distributed across 19 countries with a specific reference to "quantum" in the degree name. Germany stands out as the leader, offering 12 quantum-related master's programs, which accounts for 17.4% of the global total. The United Kingdom follows closely behind with 10 programs (14.5%), while the United States offers 9 programs (13.0%). France and the Netherlands each contribute 6 programs (8.7% each), completing the top five countries.

The distribution pattern reveals strong concentration in European countries and the United States, with Germany, the UK, and the US together accounting for 45% of all quantum master's programs. There is a notable gap between the leading group and the majority of countries, with most offering just a single program. In this list, the Asia-Pacific region shows relatively limited representation: only Japan, Malaysia, and Australia offer programs, each contributing a single quantum-related master's degree to the total count.

The QED-C State of the Global Quantum Industry Report[6] presented a word cloud of their quantum postgraduate degrees database. The word cloud representing the data used in our report resulted in the following:

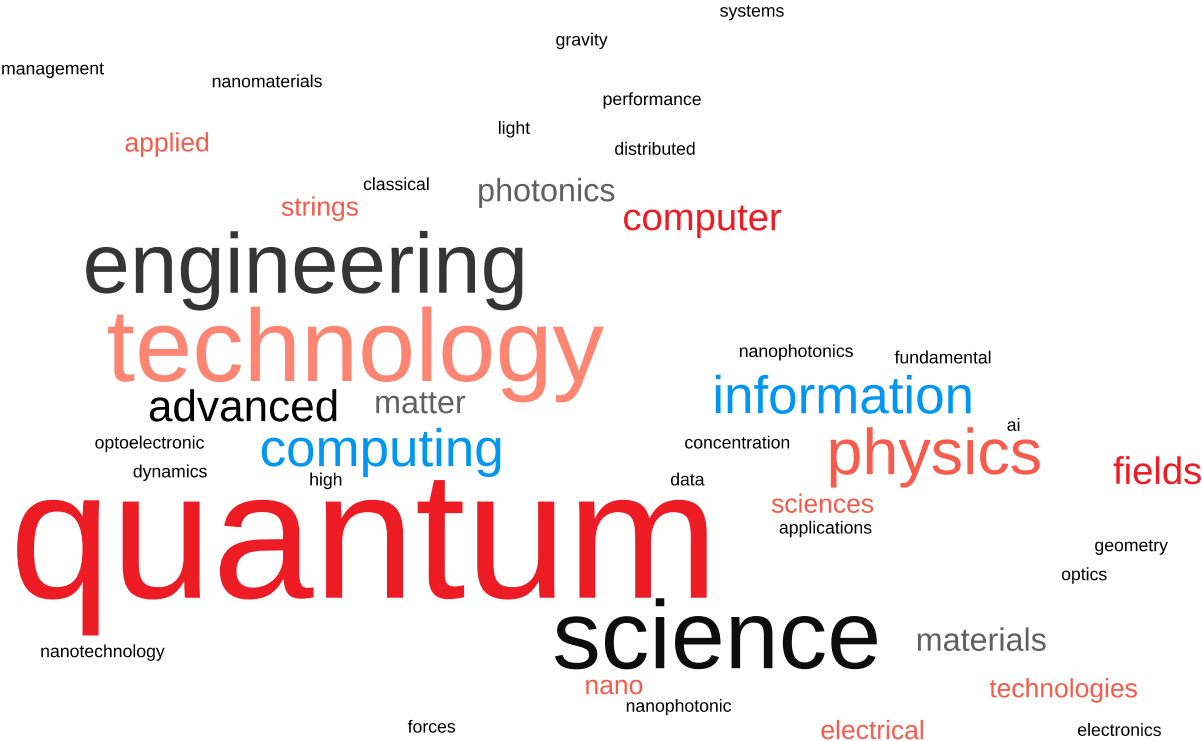

## 7.2 | Enrollment numbers

The 2021 US Report[7] "The Role of International Talent in Quantum Information Science" focuses on the future workforce needs of Quantum Information Science and Technology (QIST). The report concludes that the quantum science and technology sector faces a critical talent shortage across all major sectors, including industry, academia, and government. While the National Quantum Initiative aims to develop new workforce talent, there's an immediate need for skilled professionals and uncertainty remains about whether existing programs will sufficiently meet future demands. According to the report, international talent plays a crucial role, with foreign students comprising approximately half of US graduates in quantum-related fields.

The United States has historically benefited from retaining these international scholars, with about 70% of foreign STEM PhD graduates choosing to stay in the country as of 2017. However, developing new quantum expertise is a lengthy process requiring roughly a decade of post-secondary education and training. To address the growing workforce demands, the United States will need to pursue a dual strategy: expanding its domestic talent pipeline while maintaining its ability to attract and retain international expertise.

The report states that "the most QIST-relevant degree fields are physics, electrical engineering, and computer science" and explains that these domains were selected based on two criteria: preliminary search of keywords for online job postings and analysis of doctoral thesis titles, abstracts, and keywords.

To better understand emerging enrollment trends for physics, electrical engineering, and computer science courses, we analyzed data from the NSC Research Center (January 2025 update).

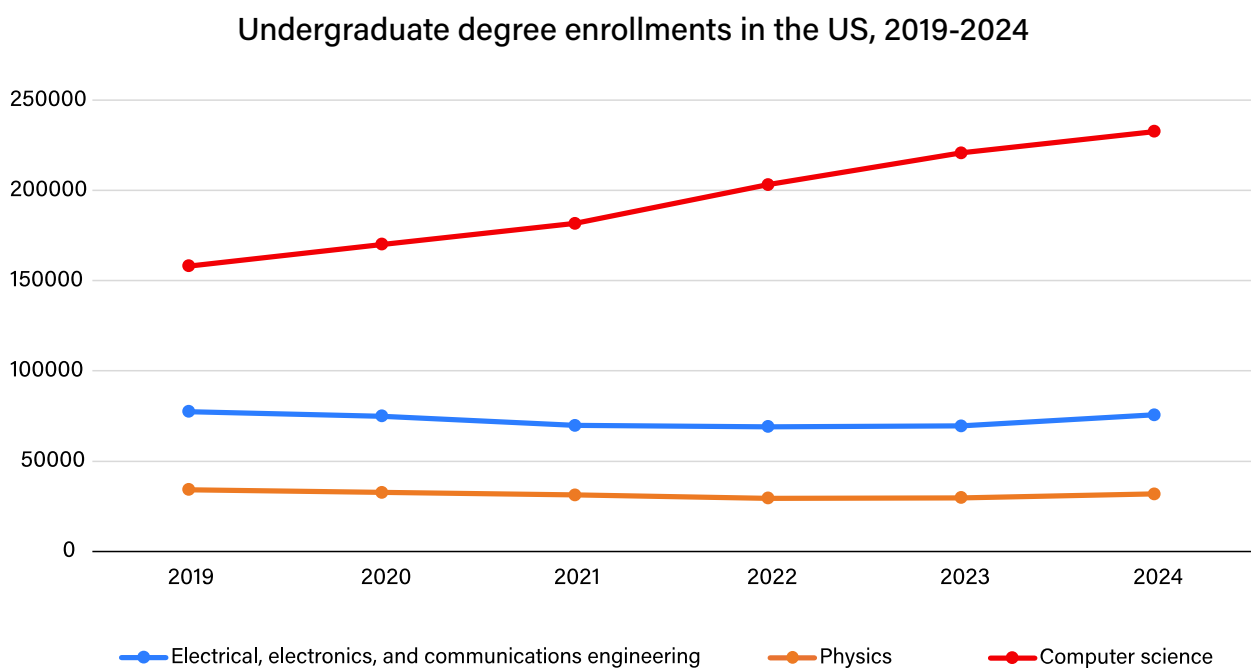

Undergraduate degree enrollments in the US, 2019-2024

Based on the enrollment data from 2019 to 2024, computer science exhibits the highest student numbers and substantial growth over the period. The electrical, electronics, and communications engineering program maintains moderate enrollment levels, whereas physics enrollments show the smallest but most consistent enrollment pattern, with a narrower range of 4,811 students between its lowest and highest enrollment figures.

The data categorization involved challenges as the major field groups at times had interconnected degrees such as "Computer and Information Science, general" and "Astronomy and Astrophysics," which are not included in the subject-level enrollment data. In order to provide a fuller picture, the report also presents the enrollment numbers for the three major field families engineering, physical sciences, and computer and information sciences and support services.

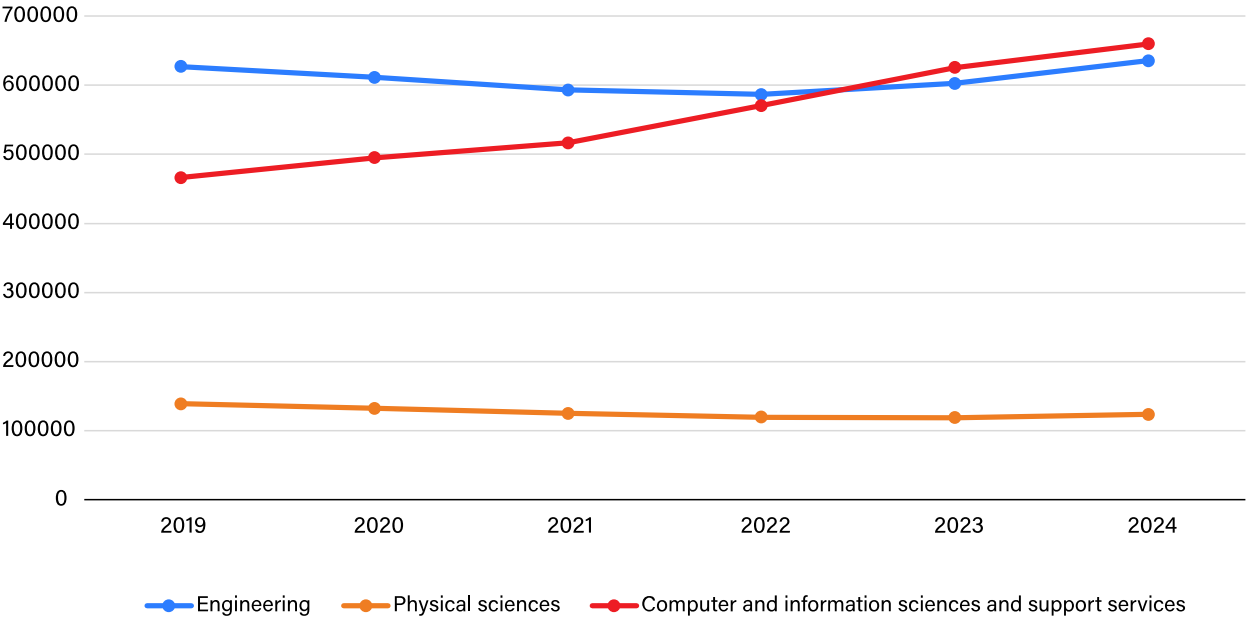

Undergraduate major field family enrollments in the US, 2019-2024

## 7.3 | Future research

The quantum education landscape is rapidly evolving, but it remains fragmented and under-documented. While anecdotal evidence points to rising interest and enrollment in quantum-related programs, there is a critical need for data on student demographics, institutional investment levels, and career outcomes. Such data is essential for identifying best practices, highlighting gaps in access and equity, and supporting evidence-based policymaking.

We invite contributions from the quantum education community to future editions of this report. The objective is to deepen and expand the insights provided.

Researchers and educators interested in sharing enrollment data, curriculum insights, or information about new programs are encouraged to contact us. We hope that our community-led approach will facilitate a comprehensive global overview of quantum education initiatives and facilitate the development of more effective educational strategies for the field.

You can reach us at contact@qir.mit.edu.

### ▸ Footnotes

[1] National Quantum Initiative, 'Enabling People' (National Quantum Initiative) <https://www.quantum.gov/workforce/> accessed 24 March 2025.

[2] SpinQ Press Release, 'Shenzhen Middle School: Building a Quantum Computing Elective Program from the Ground Up' (28 June 2024) <https://www.spinquanta.com/news-detail/shenzhen-middle-school-building-a-quantum-computing-elective-program-from20250121075716> accessed 24 March 2025.

[3] 'QTEdu- Coordination and Support Action for Quantum Technology Education' (Quantum Flagship) <https://qt.eu/projects/archive/csa-projects/qtedu> accessed 24 March 2025.

[4] 'IBM and Qubit by Qubit Offer Quantum Course | IBM Quantum Computing Blog' <https://www.ibm.com/quantum/blog/year-three-quantum-coding-school> accessed 28 March 2025.

[5] 'QubitxQubit | Course Info' <https://www.qubitbyqubit.org/course-info> accessed 28 March 2025.

[6] QED-C, 'State of the Global Quantum Industry Report' (2025) <https://quantumconsortium.org/stateofthequantumindustry2025/> accessed 24 March 2025.

[7] Subcommittee on Economic and Security Implications of Quantum Science Committee on Homeland and National Security of the National Science & Technology Council, 'The Role of International Talent in Quantum Information Science'.

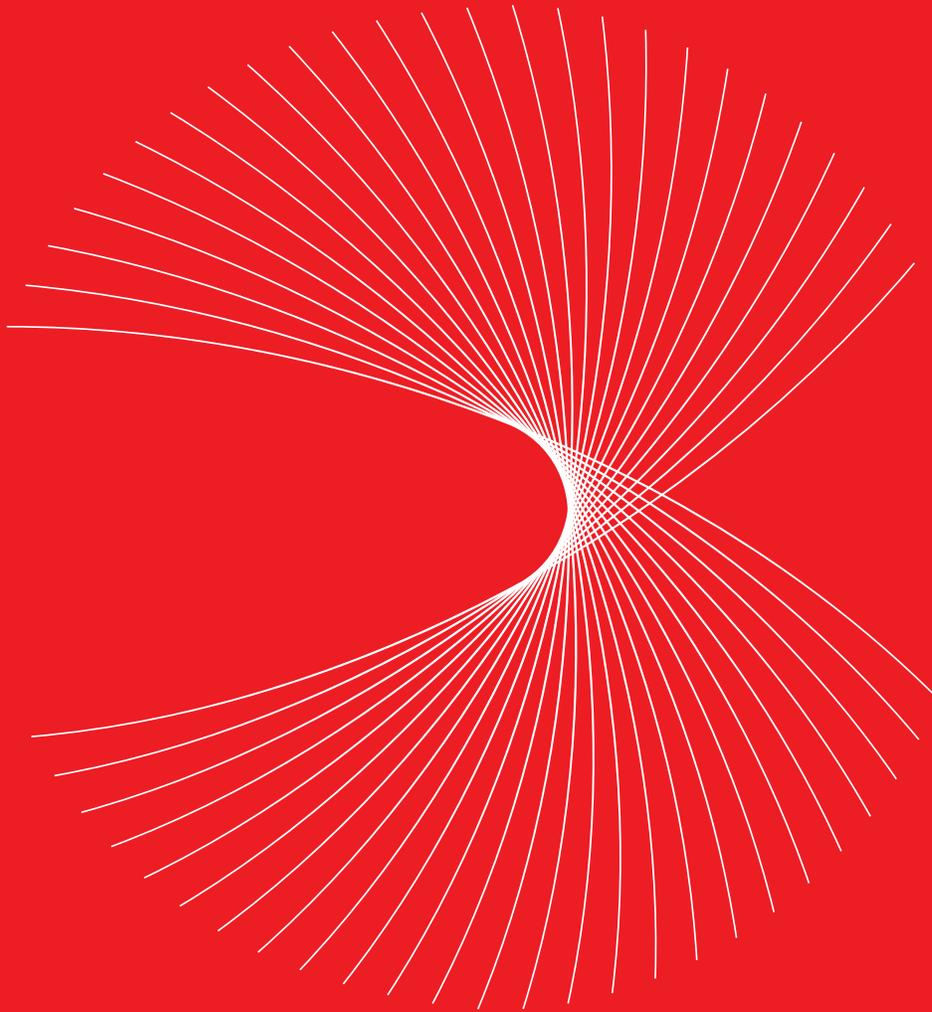

# 8 | Public Opinion

To better understand public perceptions and understanding of quantum technologies, we conducted a survey of US residents in October 2024. The survey instrument was administered to a representative panel of 1,375 US residents, with demographic sampling aligned to the US Census Bureau distributions for both gender and age groups, ensuring population representativeness. The survey was aimed at capturing attitudes, awareness levels, and expectations surrounding this emerging field. As quantum technologies transition from research laboratories to practical applications, public engagement and trust will play a critical role in shaping adoption and policy.

## 8.1 | Quantum computing

### 8.1.1. Familiarity

When asked about the familiarity levels on quantum computing, from the 1,375 survey respondents, 25% responded they were "not at all familiar," 15% said "not so familiar," 26% indicated they were "somewhat familiar," 18% claimed to be "very familiar," and 16% reported being "extremely familiar."

**How familiar are you with quantum computing?**

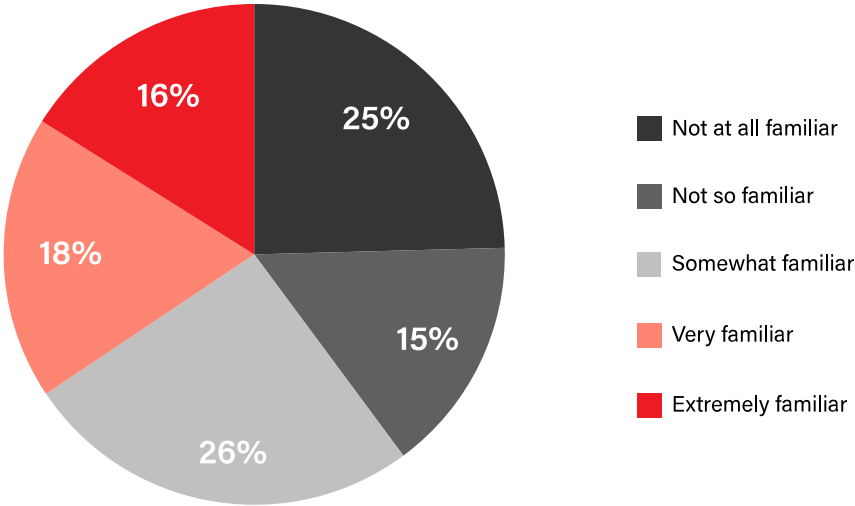

The survey reveals a diverse distribution of familiarity with quantum computing. The largest segment consists of those who are "somewhat familiar" with the topic, representing 26% of respondents. Interestingly, this moderate level of awareness is bordered by another substantial group: those with advanced familiarity represented with 34%.

> Our survey data indicates that quantum computing awareness tends to cluster at either end of the spectrum; either people have had minimal exposure or have invested significant time in understanding the technology.

Breaking down the responses further reveals that among those with limited familiarity, there's a notable distinction between those who are "not at all familiar" (25%) and those who are "not so familiar" (15%), suggesting that complete unfamiliarity is more common than partial unfamiliarity. Conversely, among those with higher levels of familiarity, there's a relatively even split between "very familiar" (18%) and "extremely familiar" (16%) respondents. This bimodal distribution indicates that quantum computing awareness tends to cluster at either end of the spectrum—people either have minimal exposure or have invested significant time in understanding the subject.

These findings might indicate an alignment with broader technological adoption patterns, particularly in emerging technologies. The presence of a large "somewhat familiar" group (26%) represents a crucial middle ground, potentially indicating recent exposure to quantum computing through media coverage or educational initiatives. This distribution suggests that quantum computing is entering mainstream discourse while indicating the importance of ongoing opportunities for education and awareness-building.[1]

### 8.1.2. Emotional repsonses

Among the 1,375 survey respondents, the emotional responses varied considerably across different potential quantum computing applications. When asked to report on to what extent they feel nervous or excited regarding potential uses of quantum computing, for materials and pharmaceuticals development, 7% felt very nervous, 11% somewhat nervous, 29% neutral, 27% somewhat excited, and 26% very excited. Regarding optimization, 7% expressed very nervous feelings, 11% somewhat nervous, 31% neutral, 24% somewhat excited, and 27% very excited. For data security and cryptography, responses showed 11% very nervous, 13% somewhat nervous, 30% neutral, 20% somewhat excited, and 26% very excited.

**Quantum computing can be potentially used for different areas. To what extent do you feel nervous or excited regarding the potential uses listed below?**

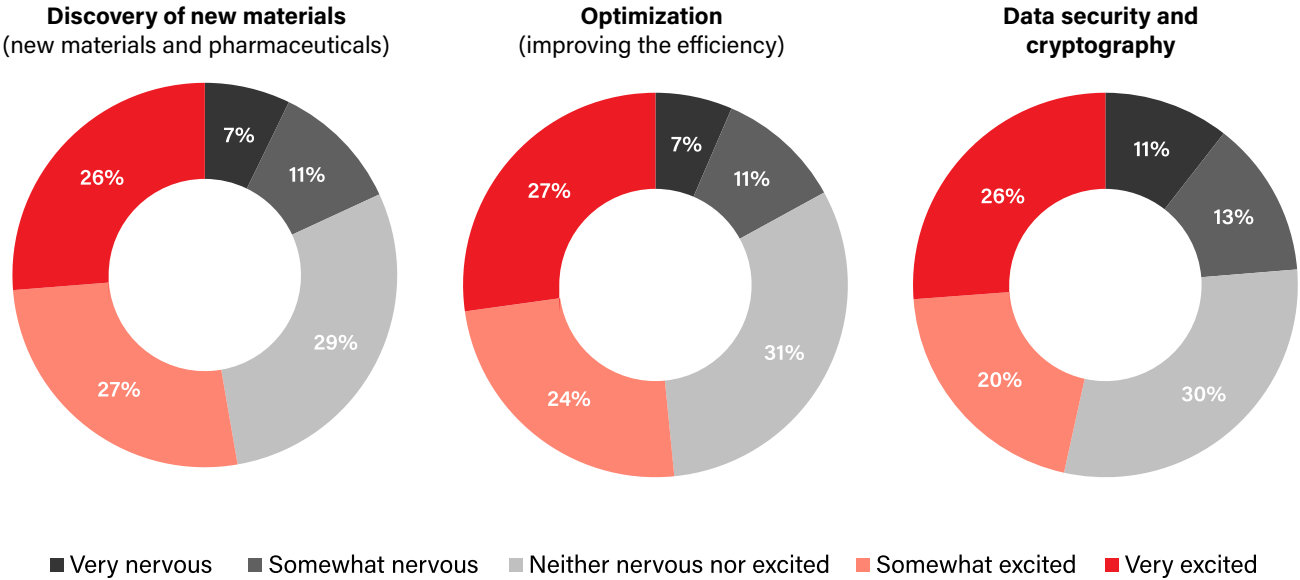

■ Very nervous ■ Somewhat nervous ■ Neither nervous nor excited ■ Somewhat excited ■ Very excited

The survey reveals the patterns in how the general public perceive different applications of quantum computing, with clear distinctions in emotional responses across various domains. Overall, across all applications, approximately half of respondents express excitement, while about one-fifth report feeling nervous, and nearly a third remain neutral.

▸ These responses suggest a generally positive outlook toward
  quantum computing's potential applications, though with
  notable variations depending on the specific use case.

Graph below compares and contrasts the number of positive and negative answers (neutral answers are not represented in this graph)

**Discovery of new materials and pharmaceuticals**

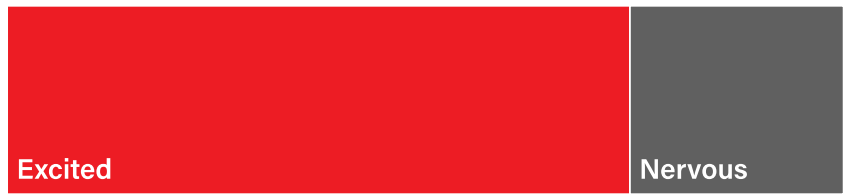

**Optimization**

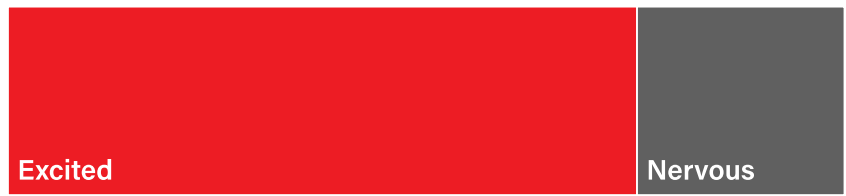

**Data security and cryptography**

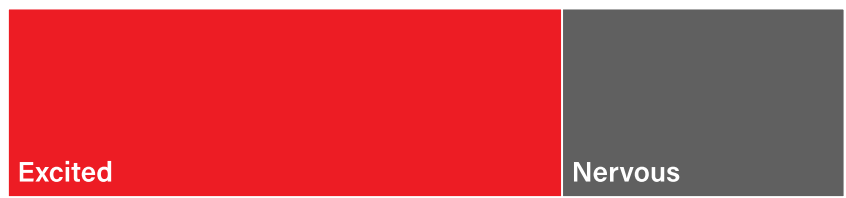

The application of quantum computing to **materials and pharmaceutical discovery** elicits the most enthusiastic response, with 53% of respondents expressing excitement and only 18% feeling nervous. This high level of enthusiasm aligns with the potential transformative impact of quantum computing in drug discovery and materials science, where breakthroughs could lead to improvements in human health and societal advancement.

In contrast, **optimization** applications show slightly lower overall excitement (51%) and similar levels of responses expressing nervousness (18%), suggesting broad acceptance of quantum computing's role in improving complex systems.

**Data security and cryptography** stands out as the most controversial application, generating significantly higher anxiety levels (24% nervous) while maintaining substantial excitement (46%). This heightened concern might reflect public awareness

81 | MIT INITIATIVE ON THE DIGITAL ECONOMY

of the dual nature of quantum computing in cryptography, such as its potential to break current encryption methods while simultaneously enabling new, quantum-resistant cryptographic solutions. The emotional responses around this application might suggest that the respondents recognize both the risks and opportunities quantum computing presents to digital security.

Across all three potential applications, approximately 30% of respondents maintain neutral positions, indicating either uncertainty about the implications or a wait-and-see attitude toward these emerging technologies. This consistent neutrality rate suggests widespread recognition that quantum computing represents a complex technology whose ultimate societal impact remains uncertain for the general public. The relatively stable neutral percentage across different applications contrasts with the varying levels of excitement and nervousness, suggesting that while people form distinct opinions about specific applications, many remain cautious about making definitive judgments.

Comparative analysis reveals that while excitement levels remain consistently high across all applications (ranging from 46% to 53%), nervousness varies from 18% to 24%. This pattern might suggest that while Americans generally welcome quantum computing's potential benefits, their comfort levels vary significantly depending on the specific domain of application.

▸ Public acceptance of quantum computing may depend heavily on how its applications are framed and communicated, with practical applications like materials discovery receiving more universal enthusiasm compared to security-related applications that raise broader societal concerns.

## 8.2 | Quantum networking

### 8.2.1. Familiarity

When asked about the familiarity levels on quantum networking, from the 1,375 survey respondents, 28% reported being "not at all familiar" with quantum networking, 18% said they were "not so familiar," 20% indicated they were "somewhat familiar," 18% claimed to be "very familiar," and 16% reported being "extremely familiar" with the technology.

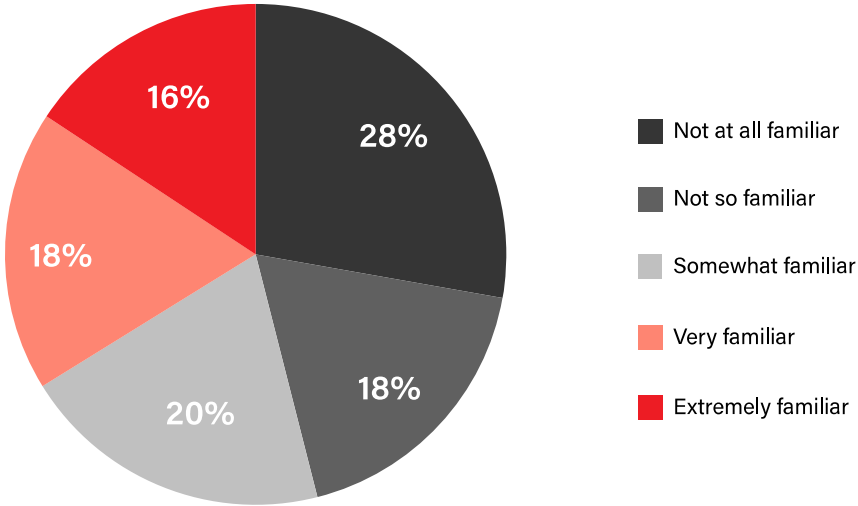

**How familiar are you with quantum networking?**

- 28% Not at all familiar
- 18% Not so familiar
- 20% Somewhat familiar
- 18% Very familiar
- 16% Extremely familiar

Nearly half (46%) of respondents reported basic or no familiarity with quantum networking. This level of basic familiarity is particularly notable, as it represents the largest single segment of responses. The distribution shows a clear progression, with 20% reporting "somewhat familiar," 18% "very familiar," and 16% "extremely familiar," resulting in a combined 34% advanced familiarity.

The relatively balanced distribution between basic and advanced familiarity levels suggests that quantum networking awareness is developing in a structured way. While the largest segment remains those with minimal familiarity, the substantial proportion of advanced familiarity (34%) might be an indication of stronger engagement from technical communities and the quantum-curious.

▸ When asked about quantum computing, 25% of the respondents reported being "not at all familiar" and 15% indicated they were "not so familiar." In contrast, quantum networking showed higher rates of unfamiliarity, with 28% reporting "not at all familiar" and 18% stating they were "not so familiar." This pattern indicates quantum networking faces greater challenges in basic public awareness than quantum computing.

### 8.2.2. Emotional responses

Among the 1,375 survey respondents, the emotional responses varied across different potential uses of quantum networking. When asked to report on to what extent they feel nervous or excited regarding potential uses of quantum networking for its relevance to secure communication: 7% of the respondents reported very nervous, 11% somewhat nervous, 29% neither nervous nor excited, 25% somewhat excited, and 28% very excited. For its relevance to scale up quantum computing by networking: 8% of the respondents reported very nervous, 12% somewhat nervous, 36% neither nervous nor excited, 20% somewhat excited, and 24% very excited.

**Quantum networking can be potentially used for different areas. To what extent do you feel nervous or excited regarding the potential uses listed below?**

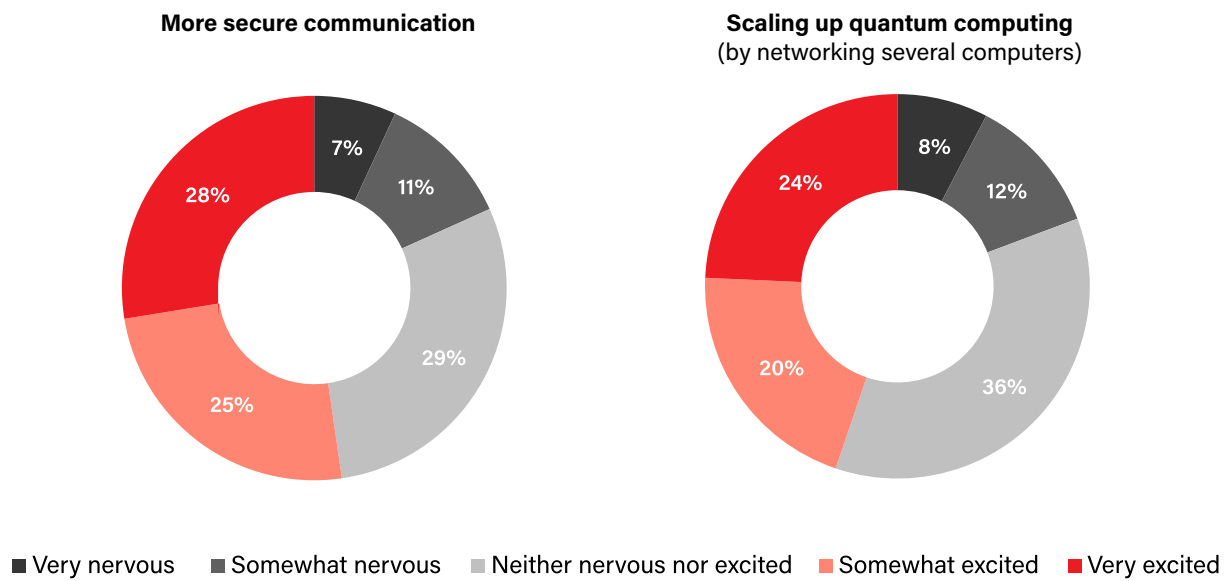

The overall sentiment analysis reveals a generally positive outlook toward both aspects of quantum networking. For secure communication, 53% of respondents expressed positive feelings (combining the answers stating "very excited" and "somewhat excited"), while 18% expressed negative feelings (combining the answers stating "very nervous" and "somewhat nervous"). For scaling quantum computing, 44% expressed positive feelings and 20% expressed negative feelings.

Graph below compares and contrasts the number of positive and negative answers (neutral answers are not represented in this graph)

**More secure communication**

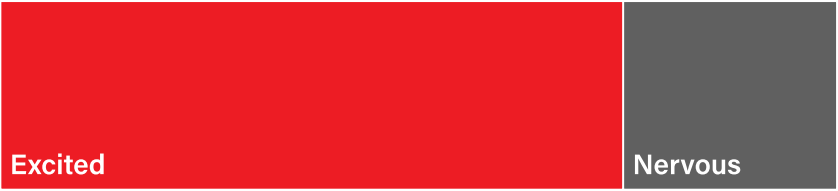

**Scaling up quantum computing**

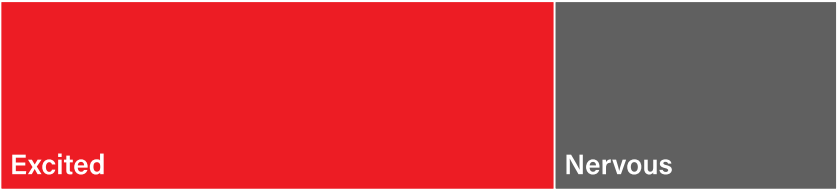

Comparing the two aspects reveals that although both show positive sentiment, secure communication generates stronger enthusiasm, with 9 percentage points more positive responses than scaling quantum computing.

> Survey data suggests that many Americans are cautiously optimistic about quantum networking, particularly regarding its potential for secure communication. The relatively high percentage of neutral responses indicates that many people are still learning about and forming opinions on this aspect of quantum technology.

## 8.3 | Governance

We asked the same 1,375 participants to report to what extent they agree or disagree with the following three statements: (1) the State can be trusted to exert effective control over organizations and companies using quantum technologies, (2) US government should fund the research and development of quantum technologies, and (3) US companies should fund the research and development of quantum technologies.

**To what extent do you agree or disagree with the following statements:**

**State can be trusted to exert effective control over organizations & companies using quantum technologies**

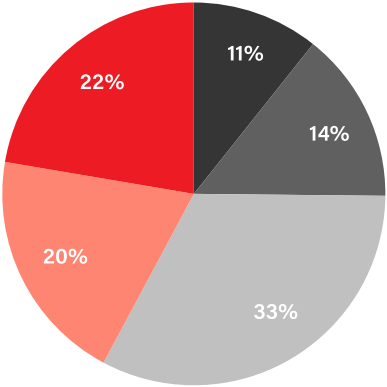

**US government should fund the research and development of quantum technologies**

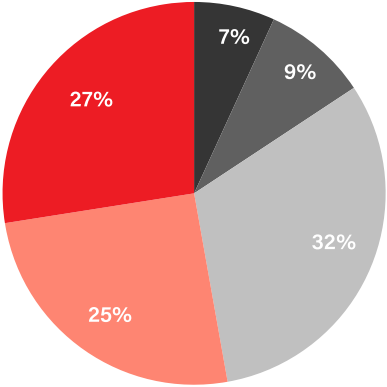

**US companies should fund the research and development of quantum technologies**

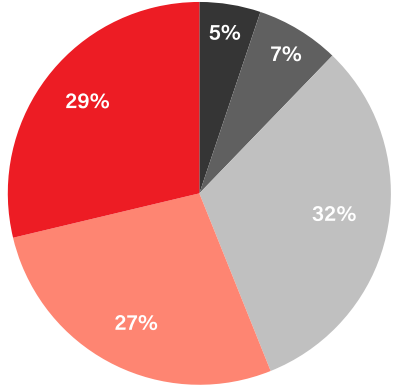

■ Strongly disagree ■ Somewhat disagree ■ Neither agree nor disagree ■ Somewhat agree ■ Strongly agree

For the statement about state control over quantum technologies: 11% strongly disagree, 14% somewhat disagree, 33% neither agree nor disagree, 20% somewhat agree, and 22% strongly agree.

For government funding of quantum technologies: 7% strongly disagree, 9% somewhat disagree, 32% neither agree nor disagree, 25% somewhat agree, and 27% strongly agree.

For company funding of quantum technologies: 5% strongly disagree, 7% somewhat disagree, 32% neither agree nor disagree, 27% somewhat agree, and 29% strongly agree.

Graph below compares and contrasts the number of supporting (agreeing) and opposing (disagreeing) answers (neutral answers are not represented in this graph)

**State can be trusted to exert effective control over organizations and companies using quantum technologies**

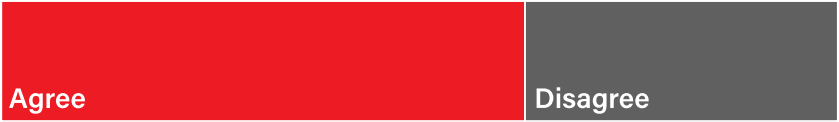

**US government should fund the research and development of quantum technologies**

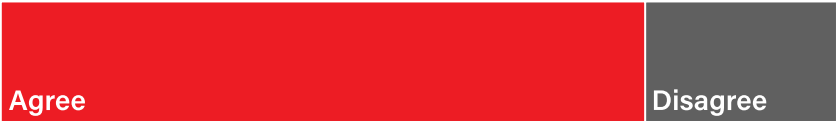

**US companies should fund the research and development of quantum technologies**

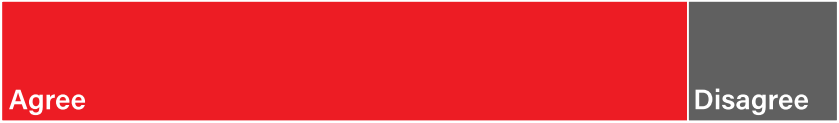

The negative responses show variations across the statements. The state control statement received the highest negative response rate at 25%, followed by government funding at 16%, and company funding at 12%. This pattern suggests that Americans are most skeptical about government control over quantum technologies, while being more comfortable with private sector involvement.

The positive responses reveal particularly high levels of enthusiasm for company funding with 56% positive responses and government funding with 52% positive responses. The state control statement received significantly lower positive responses at 42%.

▸ `The findings suggest that Americans generally support the development of quantum technologies but have nuanced views about how this development should be managed. While there is strong support for both government and private sector investment, there is more skepticism about state control over quantum technologies.`

The high percentage of neutral responses across all statements (around one third) might also indicate that many Americans are still forming their opinions about the governance of quantum technologies, highlighting the need for more public education and dialogue about these issues.

## 8.4 | Future research

Repeating the survey in future years will enable us to track shifts in public awareness, sentiment, and understanding of quantum technologies over time. Longitudinal data will help identify emerging concerns, misconceptions, or areas where targeted communication and education may be needed. It will also offer valuable insights into societal readiness and trust. We invite readers of this report—whether from the public, academia, industry, or policy communities—to share their perspectives and suggest questions or themes you believe should be included in future surveys. Your input will help ensure this effort remains relevant, inclusive, and responsive to the broader community.

You can reach us at contact@qir.mit.edu.

---

▸ **Footnotes**

[1] 'U.S. Quantum Leadership May Hinge on Public Perceptions' (Brookings) <https://www.brookings.edu/articles/u-s-quantum-leadership-may-hinge-on-public-perceptions/> accessed 12 February 2025.

# 9 | Quantum Networking

## 9.1 | Quantum networks

Quantum networks are emergent communication systems that leverage the principles of quantum mechanics to transmit information in new ways. Just as the classical internet enabled email, video calls, and online banking, quantum networks aim to enable things we can't yet do with classical networks—especially in security, computing, and sensing. Quantum networking has profound implications for national security, scientific discovery, and economic competitiveness.[1]

Quantum networking refers to the tools, protocols, and systems that enable the transmission of quantum information between different devices or locations. It incorporates fiber-optic cables, quantum repeaters to extend range, quantum routers, and the software layers needed to manage the system. The quantum internet is a closely related concept: it refers to the broader vision of what we can do once those quantum networks are built and scaled.

According to the 2024 report by the US National Quantum Initiative Advisory Committee (NQIAC), quantum networking capabilities will "play a role in US economic prosperity and national security" and continued investment in R&D of quantum networking is also necessary to clarify the magnitude of that role.[2]

In 2022, the European Commission supported the creation of the Quantum Internet Alliance (QIA) with €24 million in funding to build "a global quantum internet made in Europe."[3] In March 2025, QIA announced the creation of "the first operating system designed for quantum networks" which will facilitate program applications for quantum networks.[4] The system is planned to be made accessible for a broader audience through QIA's quantum internet demonstrator.[5]

It is critical to note that as they are understood today, quantum networks might not replace classical communications or the internet, however they have potential to offer novel functionalities such as more secure communication and the ability to connect quantum computers for enhanced computing power.[6]

## 9.2 | Quantum networking testbeds

Testbeds play a crucial role in the development of quantum networking and, by extension, the quantum internet. The National Quantum Initiative Advisory Committee defines a testbed as "a platform or facility that is accessible to multiple users to conduct replicable and rigorous testing of component technologies, protocols, and systems integration" and distinguishes it from demonstrators, prototypes and user facilities.[7]

Testbeds are essential for advancing quantum networking because they provide realistic environments in which to explore the performance, interoperability, and scalability of quantum components. According to the NQIAC, "strategically chosen and properly timed quantum networking testbeds will serve an important role in developing the theoretical underpinnings, technologies, security models, and application scenarios" for quantum networks.[8]

The importance of testbeds lies not only in technological validation but also in risk mitigation. Developing "right-sized" testbeds, those tailored in scope and cost to specific research objectives, has been a priority both in the 2021 and 2024 reports.[9,10] This strategic investment approach aims to ensure that only mature, promising technologies are scaled up for more extensive networks.

Investments in testbeds are not merely about testing hardware, they also represent a commitment to advancing the foundational science and engineering needed for a transformative quantum era.

Beyond technical development, testbeds also play a critical role in workforce training and industry engagement. They provide hands-on opportunities for students, researchers, and engineers from diverse backgrounds to develop quantum skills in a practical setting. For industry, testbeds offer a collaborative space to test products, explore market-ready solutions, and align with government and academic research. In this way, testbeds not only advance technology but also support a broader ecosystem necessary for the growth of quantum networking.

In this chapter we present data that maps quantum networking testbeds across the world from publicly available sources and in consultation with experts. Our current dataset lists 13 testbeds in the US and 15 in Europe (including UK). The distribution of these testbeds is illustrated in the maps below:

**Quantum network testbeds in the US**

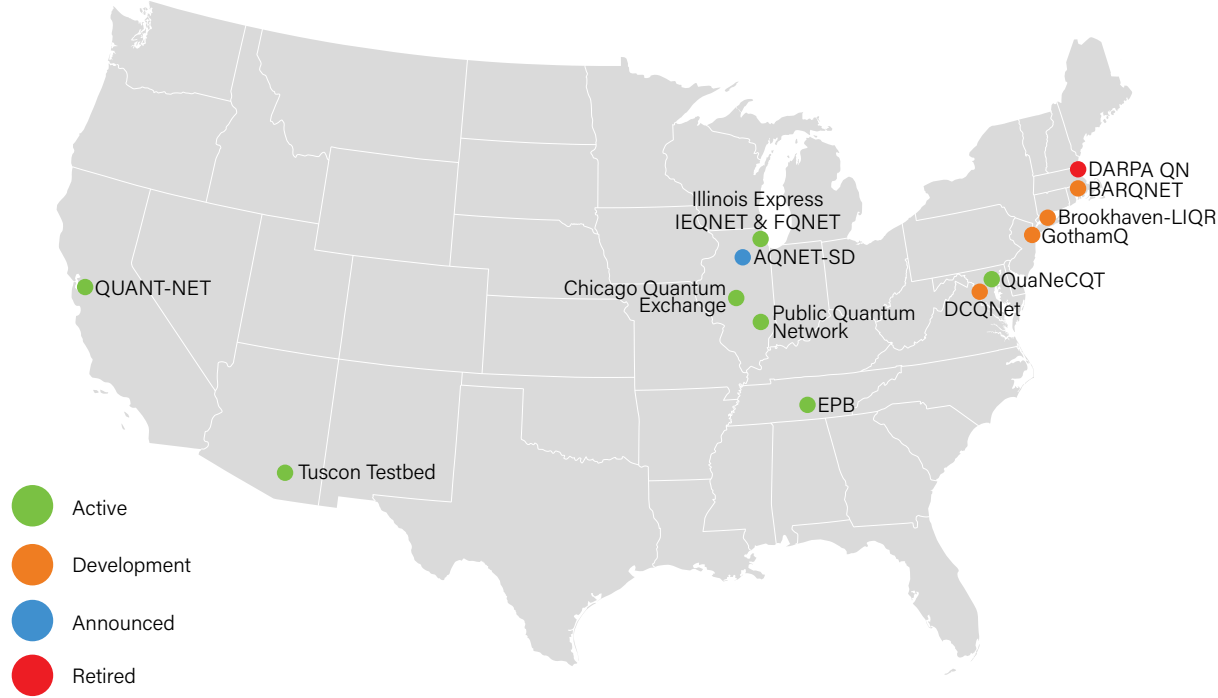

**Quantum network testbeds in Europe**

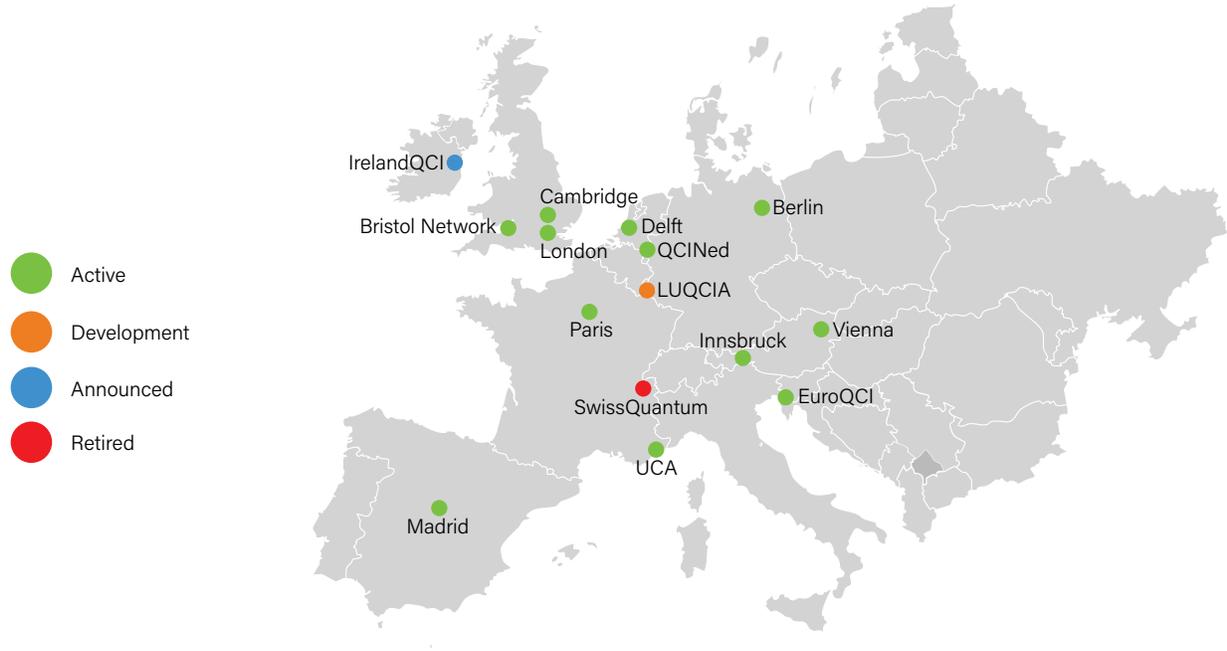

Number of quantum networking testbeds

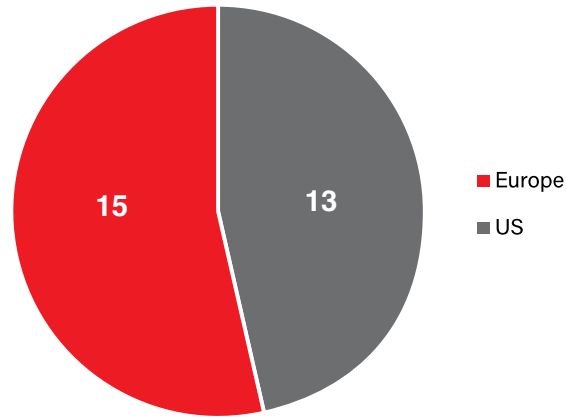

## 9.3 | Future research

We aim to systematically map and document the locations of quantum networking testbeds worldwide, creating a comprehensive open database accessible to researchers, policymakers, and industry stakeholders. We will be making this data publicly available to accelerate collaborative research, facilitate international partnerships, and inform evidence-based policy decisions regarding quantum infrastructure development. We would like to invite contributors and collaborators to join us in these efforts.

You can reach us at contact@qir.mit.edu.

### ▸ Footnotes

[1] The White House National Quantum Coordination Office, 'A Strategic Vision for America's Quantum Networks' (2020).

[2] 'Quantum Networking: Findings and Recommendations for Growing American Leadership' [2024] National Quantum Initiative Advisory Committee.

[3] Quantum Internet Alliance, 'The Quantum Internet Alliance Will Build an Advanced European Quantum Internet Ecosystem' (14 October 2022) <https://quantuminternetalliance.org/2022/10/14/the-quantum-internet-alliance-will-build-an-advanced-european-quantum-internet-ecosystem/> accessed 31 March 2025.

[4] C Delle Donne and others, 'An Operating System for Executing Applications on Quantum Network Nodes' (2025) 639 Nature 321.

[5] QIA, 'QIA Researchers Create First Operating System for Quantum Networks' (Quantum Internet Alliance, 12 March 2025) <https://quantuminternetalliance.org/2025/03/12/qia-researchers-create-first-operating-system-for-quantum-networks/> accessed 31 March 2025.

[6] 'Quantum Networking: Findings and Recommendations for Growing American Leadership' (n 6).

[7] ibid.

[8] ibid.

[9] National Science and Technology Council (n 3).

[10] 'Quantum Networking: Findings and Recommendations for Growing American Leadership' (n 6).

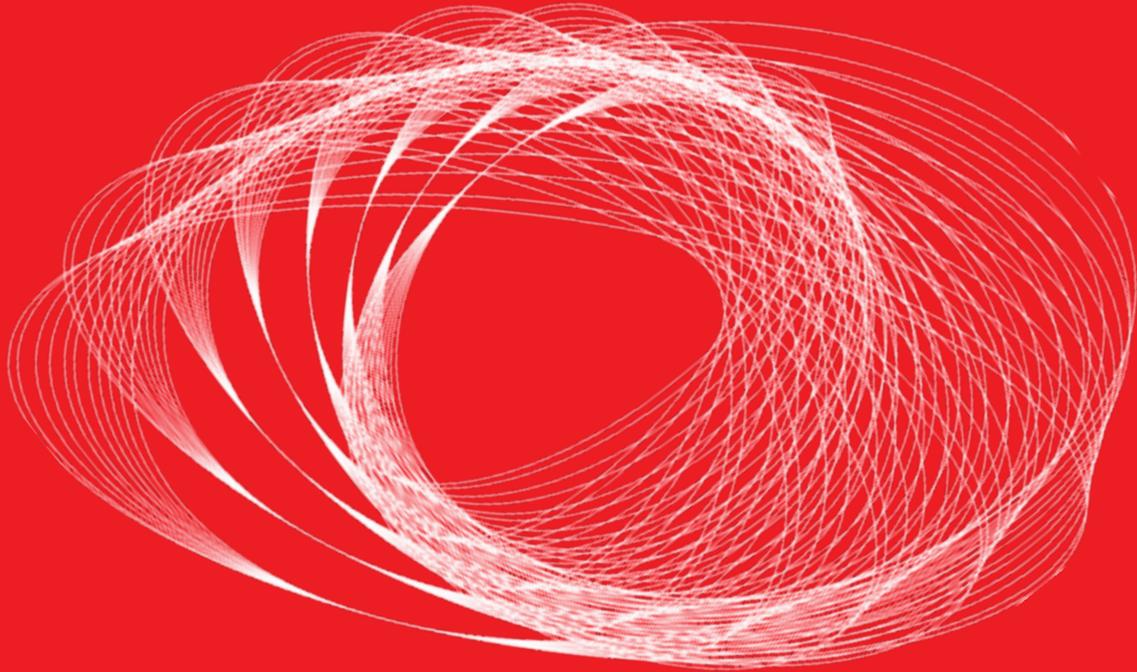

# 10 | Quantum Processor Benchmarking

It is challenging for non-experts to easily understand the performance of quantum computing today. Without this understanding, making predictions about investment, commercial deployments, use case testing, and overall strategy is prohibitive. This opacity has many drivers, including the nascent state of the technology, the existence of multiple modalities (types of quantum computers), the lack of independently verified universal performance metrics, and the context-dependent connection between hardware devices and quantum algorithms.

## Benchmarks

To enable better insight into the current state of quantum computing performance, we indexed and analyzed published data on over 200 Quantum Processing Units (QPUs) from 17 countries, including retired, prototype, current, and announced QPUs. As of April 2025, there are over 40 commercially available QPUs from at least two dozen manufacturers.[1]

It is important to preface that quantum computers remain below the performance capabilities of classical computers for all useful tasks today.

A useful analogy when considering QPU benchmarking is to consider the comparison of racecars. Racecars benefit from large horsepower and torque ratings, lightweight construction, independent suspensions, and performance aerodynamics. However, lap times on a specific racetrack are the best overall measure of how well the various elements perform together. QPUs have similar construction, design and performance idiosyncrasies. There are three primary categories of QPU benchmarks that are useful to consider:

### ▸ Physical benchmarks
e.g., number of qubits and fidelity of the qubit gates.

These are the core metrics of a QPU. They are akin to the weight and torque values of a racecar. While they are objective measures, they only provide a partial insight into the likely overall performance when described in isolation. Physical benchmarks are the category that QPU manufacturers are most likely to disclose and are a focus area for the analysis below.

### ▸ Aggregated benchmarks
e.g., Quantum Volume, CLOPS, and Logical Qubits.

These are various combinations of physical benchmarks. In the car analogy, these benchmarks are similar to power-to-weight ratios. They are more useful than singular benchmarks, but do not fully encapsulate the full performance of a QPU.

### ▸ Application-level benchmarks
e.g., Q-Score and RACBEM.

These measure the performance of QPUs when solving specific problems. They are similar to classical computing benchmarks such as LINPACK, which is used to rank classical supercomputers. These benchmarks can help compare QPUs vs other QPUs and also compare QPUs with classical computing devices. Application-level benchmarks are analogous to a racecar's lap time at a given racetrack in defined weather conditions. They allow for a limited comparison between competing cars. However, there are different benchmarks that put emphasis on different algorithmic challenges, similar to how a Formula 1 car might be set up to perform well at the Monaco Grand Prix, but would be ineffective on a NASCAR circuit. Manufacturers do not regularly publish application-level benchmarks, as today's QPUs are not capable enough to run sizable applications. As QPUs become more powerful, we expect to be able to track application-level benchmarks in future reports.

# Benchmarks

Our QPU dataset was primarily generated through a combination of manufacturer announcements, online searches, and direct queries to QPU providers. The dataset captured a variety of benchmarks. **The key ones are:**

**Qubit Counts** can be misleading as the number of qubits increases, so too can error rates. This has even prompted some manufacturers to reduce qubit counts to improve overall performance in certain situations. For example, in 2023, IBM followed the release of its 1121-qubit Condor and 433-qubit Osprey QPUs with the higher-performing 133-qubit Heron. Similarly, Quantinuum has been developing and promoting their achievements in higher overall performance on their 20-Qubit H1 QPU[2], even though their larger 56-Qubit H2 QPU has been commercially available for several years already. This underscores that qubit count alone is not a definitive measure of QPU capability and must be considered alongside other key performance benchmarks.

**Coherence** refers to how long a qubit maintains its quantum state. Due to interactions with their environment, qubits inevitably lose their quantum information, a process known as decoherence. This is characterized by two timescales: $T_1$ (energy relaxation) and $T_2$ (dephasing). $T_1$ and $T_2$ are important variables as they dictate the time that calculations can be executed. Trapped ion, and to a slightly lesser extent, neutral atom modalities exhibit $T_2$ times several orders of magnitude longer than Superconducting qubits, offering an inherent advantage for applications that require longer coherence.

**Fidelity:** Quantum computers currently experience error rates several orders of magnitude higher than classical systems. These errors arise from imperfections in many areas such as control pulses, inter-qubit couplings, and qubit state measurements, and they reflect the engineering limitations of today's quantum hardware. To characterize and compare quantum error rates, several benchmarks have been introduced. These include single-qubit gate fidelity, two-qubit gate fidelity, readout fidelity, state preparation and measurement (SPAM) error, decoherence-related errors, and crosstalk error. Manufacturers use different terminologies to describe these benchmarks, which can hinder direct comparison (e.g. mid-circuit, median, or average). Each of these benchmarks captures inaccuracies in a specific quantum operation essential to running a quantum circuit. Two-qubit gate fidelity is one of the most critical metrics as it is often the bottleneck in large circuits. It is more prevalent in such circuits and has higher error rates than one-qubit gates.

**Quantum Volume (QV)** was introduced by IBM in 2017[3] and reflects different physical-level benchmarks, such as gate fidelity, qubit count, and connectivity. Unlike the volume of a cube, QV is not computed by simple multiplication, but requires a complex set of statistical tests. QV identifies the largest square-shaped random circuits (where the number of qubits equals the circuit depth) that a quantum device can implement with high fidelity. QV has been criticized by some scholars and by IBM itself as not being useful for larger devices and also for relying on square circuits which are not typically representative of real-world quantum applications. Despite that, QV has been adopted by many hardware providers and is used in some spec sheets and marketing materials.

**Gate speed** refers to the time it takes to perform a single-qubit or two-qubit gate operation. It is directly connected to decoherence as together they determine how many operations can be conducted in a system before the qubit becomes ineffective. Systems with fast gate speeds, such as superconducting qubits and electron spins, often have shorter coherence times than systems like trapped ions or neutral atoms, which have slower gates but much longer coherence times. Current superconducting quantum computers operate at raw gate speeds[4] in the 1–100 MHz range (and 1–10 kHz when fully burdened with error correction and overhead), these speeds are significantly slower than classical CPUs, which operate at 2–5 GHz. However, such a direct comparison is only partially useful given the fundamentally different computational approaches.

### ▸ Execution time

Gate speed is a critical but often underreported metric in quantum computing—many hardware vendors do not disclose it at all. Yet it directly limits the runtime of quantum circuits. For example, consider molecular simulations using a Quantum Phase Estimation (QPE) algorithm, which can require circuits exceeding $10^{13}$ logical gates. On a Trapped-ion quantum processor, where gate speeds are typically around 10 microseconds, executing such a circuit even once would take[5] several days. Since a single molecule may require thousands of such full executions to achieve statistical confidence, and since quantum error correction dramatically increases circuit depth and gate count, total runtime could extend into years, well beyond practical limits for most applications. Businesses evaluating quantum computing should estimate execution time based on circuit size, hardware gate speed, and the overhead introduced by error correction. While gate speed imposes a fundamental limit, total runtime can be reduced by optimizing algorithms for parallelism, reducing circuit depth, and improving qubit fidelity to lower the cost of error correction.

### ▸ Error correction

Error correction is fundamental to quantum computing. Methods like Surface Codes require an increasing amount of qubits to make physical qubits into a logical one (thousands, or tens of thousands for very large circuits). Google announced an important breakthrough in 2024[6], demonstrating that their system operates below the fault-tolerance threshold, meaning that adding more qubits and correction cycles leads to a net decrease in logical error rates. This suggests a path forward for a scalable increase of Logical Qubits with a set amount of physical qubits.

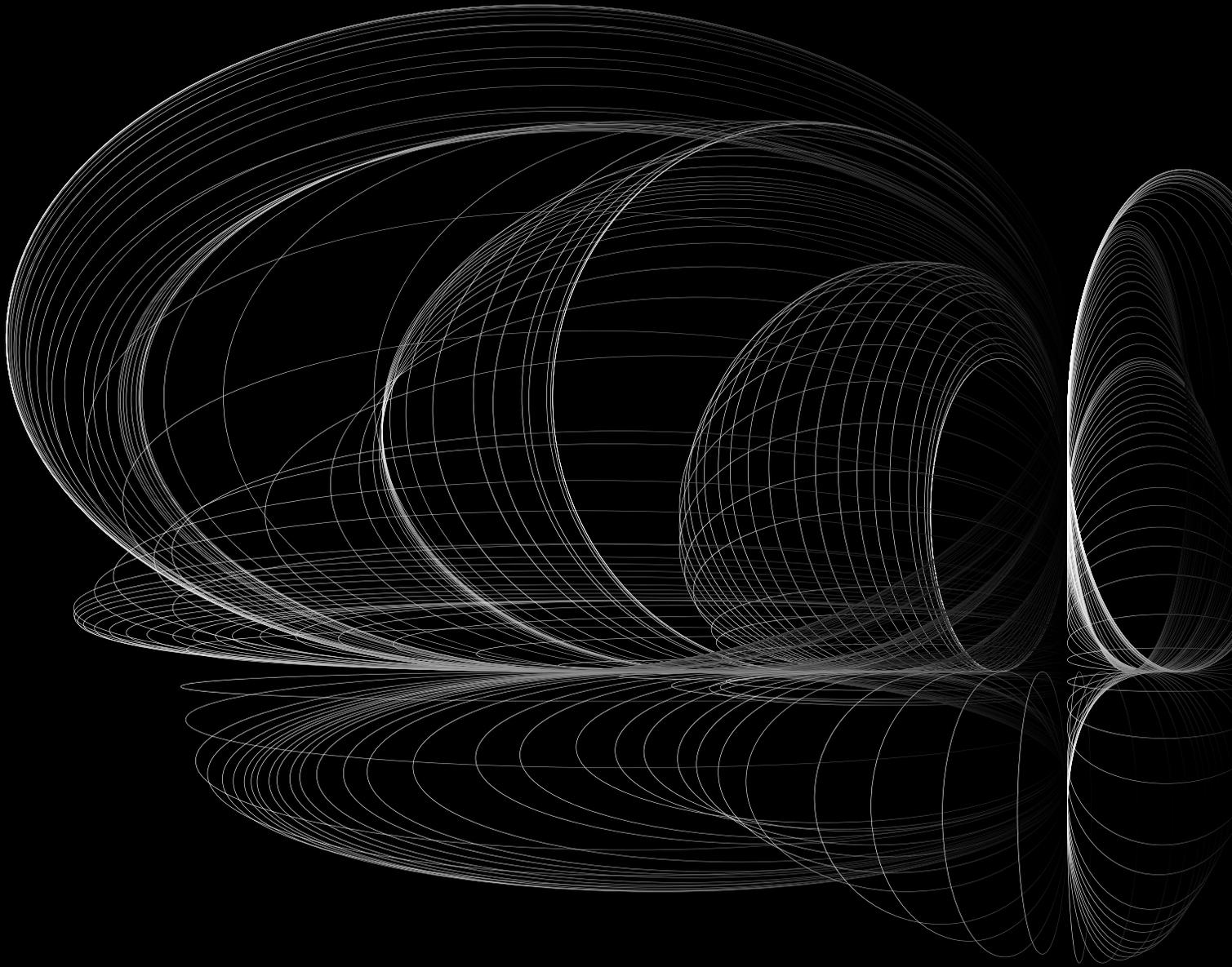

# Putting a positive spin on it

Composing a detailed QPU list is a challenging task, further complicated by quantum computing vendors often only highlighting their most favorable performance QPU benchmarks. For example, only three out of 31 Trapped Ion QPUs in our dataset reference gate speed in their publicly available specifications. Trapped ions gates are approximately 10,000 times slower than the fastest superconducting gates. IBM is among the most transparent firms when it comes to QPU benchmarking. Most of its relevant performance metrics are publicly accessible, including data for individual QPU instances and even individual qubits.

# Modalities

There are several different approaches to designing and operating a specific physical quantum computing system: these are known as modalities. Each modality uses different technological approaches to encoding, manipulating, and reading out quantum information, but result in similar functionality (also called gate-based or measurement-based). Each modality has inherent benefits and weaknesses, reflected in benchmarks such as number of qubits, fidelity, and speed. The table below illustrates the best-in-class commercial or prototyped device from each modality. No clear winner has yet emerged from these modalities.

|  | Trapped Ions | Superconducting | Neutral Atoms | Photonics | Electron Spin / NV |
|---|---|---|---|---|---|
| Qubit Count | 56 | 1121 | 1125 | 216 | 424 |
| 2-Q Gate Fidelity | 99.91% | 99.70% | 99.48% | 93.80% | 99.20% |
| 1-Q Gate Fidelity | 99.99% | 99.97% | 99.90% | 99.60% | 99.99% |
| 2Q Gate Speed (LogHz) | N/A | 7.32 | 6.18 | N/A | N/A |
| 1Q Gate Speed (LogHz) | 4.00 | 7.70 | 5.39 | N/A | N/A |
| Coherence - $T_1$ | 1,000,00 | 198 | 40,000,00 | N/A | N/A |

Best ⟶ Worst

**A quantum circuit** is a sequence of operations (quantum gates) that a QPU follows to solve a problem. It's the foundation of quantum algorithms, which use these circuits to process information.

**Superconducting QPUs** are electronic circuits created with lithography techniques used for classical computing fabrication. These circuits are cooled to millikelvin temperatures that aid in suppressing thermal noise and allow coherent quantum behavior. They excel in gate speed, qubit count, and have reasonable fidelities, but need to be cooled extensively.

**Trapped-ion QPUs** implement gate-based quantum computing using individual ions held in place by radiofrequency traps. Gate operations are performed using lasers or microwaves that manipulate the ions' internal quantum states. Trapped Ions have high fidelity, coherence, and qubit connectivity, but have slower gate speeds and have not yet scaled to large qubit counts.

**Photonic QPUs** use photons as qubits. Photons propagate through photonic integrated circuits containing linear optical elements such as beam splitters and phase shifters, which can implement certain quantum gates.

**Neutral Atom QPUs** use atoms—typically alkali or alkaline-earth metals—that are laser-cooled and confined in vacuum chambers using optical and magnetic trapping techniques. While the atoms themselves are ultracold, the hardware operates at near room temperature. Relative to other modalities, they show promise in high qubit counts, but have slower gate speed and lower fidelity although they are experiencing rapid development.[7]

**Electron Spin QPUs** leverage the quantum state of single electrons as qubits, offering relatively long coherence times and the potential for high-fidelity control with the added benefit of compatibility with existing semiconductor fabrication techniques. Scalable control and readout of large arrays remain active areas of research.

**Nitrogen-Vacancy (NV)** centers in diamond are a promising solid-state platform, where a nitrogen impurity adjacent to a lattice vacancy hosts a localized electron spin used as a qubit. NV center QPUs can operate at room temperature, unlike most other quantum computer hardware.

### ▸ Cats, Rails, and Flux

Superconducting quantum processors are among the most promising modalities. Different sub-architectures are being pursued in designing these QPUs, such as the "Cat qubits" pursued by Amazon and Alice & Bob. The Cat qubit is more difficult to fabricate but increases the fidelity of the qubit as there is only one type of error, phase flip (no bit flip error). Another interesting approach is taken by Quantum Circuits Inc., which is pursuing Dual Rail QPUs that implement a physical redundancy to their Transmon qubits to protect from errors (with a second cavity/rail). Atlantic Quantum is using Fluxonium qubits instead of more traditional Transmon qubits. The Fluxonium qubit shows longer coherence and error correction possibilities—again, at the cost of fabrication complexity.

### ▸ Annealers

A distinct class of quantum computer is the adiabatic quantum computer, also called an annealer, inspired by the metallurgical process with the same name. The principle behind quantum annealing is rooted in the adiabatic theorem, which states that a quantum system will remain in its lowest-energy state if its parameters are changed slowly enough and in the absence of significant noise. Using this phenomenon, an optimization problem can be mapped as an energy landscape of possible solutions with the lowest energy being the best solution. By annealing (i.e., adjusting the system parameters), the system is guided toward the lowest-energy state, which—if reached—yields the optimal solution. D-Wave produced the first commercial annealer in 2010, reaching 128 qubits. Today, the company produces commercial systems with 5,000 qubits. Annealers are treated separately in this report, as their architecture is not directly comparable to gate-based quantum computers. Annealers can achieve much larger qubit counts, but do not implement universal gate-based control. This limits annealers to a narrower class of problems when compared to gate-based QPUs. Only one manufacturer besides D-Wave has announced plans for releasing annealers in the future.

▸ **Majorana Qubits**

Another type of superconducting qubit is the Majorana qubit, which gained significant attention in early 2025. Microsoft has invested in this approach for over a decade and remains the primary industry player actively pursuing Majorana-based quantum computing. This design uses superconducting nanowires that host Majorana zero modes, exotic quasiparticles predicted to appear at the ends of the wire under specific conditions. Microsoft's Majorana 1 QPU represents a significant milestone toward realizing a Majorana-based quantum processor, though the announcement[8] (February 2025) has been met with skepticism from parts of the scientific community, as conclusive evidence for the topological nature of the Majorana zero modes remains under debate.

▸ **Logical Qubits**

Most quantum algorithms assume that a qubit is "perfect," i.e., that it behaves perfectly throughout the operations of the algorithm. In reality, qubits are error-prone and short-lived, so we combine many physical qubits using quantum error correction techniques (such as surface codes) to form a more stable unit known as a Logical qubit. Some manufacturers have started using this metric for their QPUs. The term can be misleading, as its practical utility depends heavily on the size and complexity of the circuit it can reliably support. To be viable for applications such as simulating complex molecules, a Logical qubit would need to support circuits millions to billions of gates long—several orders of magnitude beyond current capabilities. As such, when one is presented with a number of Logical qubits for a QPU, the key follow-up question should be: "at what circuit depth?" Only then does the number of Logical qubits convey meaningful information.

▸ **Quantum emulators**

Since quantum algorithms are inherently probabilistic, they can be emulated by classical computers to a certain level, i.e., run on a classical computer without the need of a QPU. Emulators do not physically utilize quantum effects such as entanglement or superposition. They are particularly useful for testing, debugging, and benchmarking quantum algorithms. Existing classical supercomputers can emulate circuits[9] with approximately 50 logical qubits. For classical computers, emulating additional qubits becomes exponentially more difficult, while for QPUs, this requires adding incremental logical qubits. Today's best quantum computers are orders of magnitude slower and more expensive to run than the equivalent CPUs.

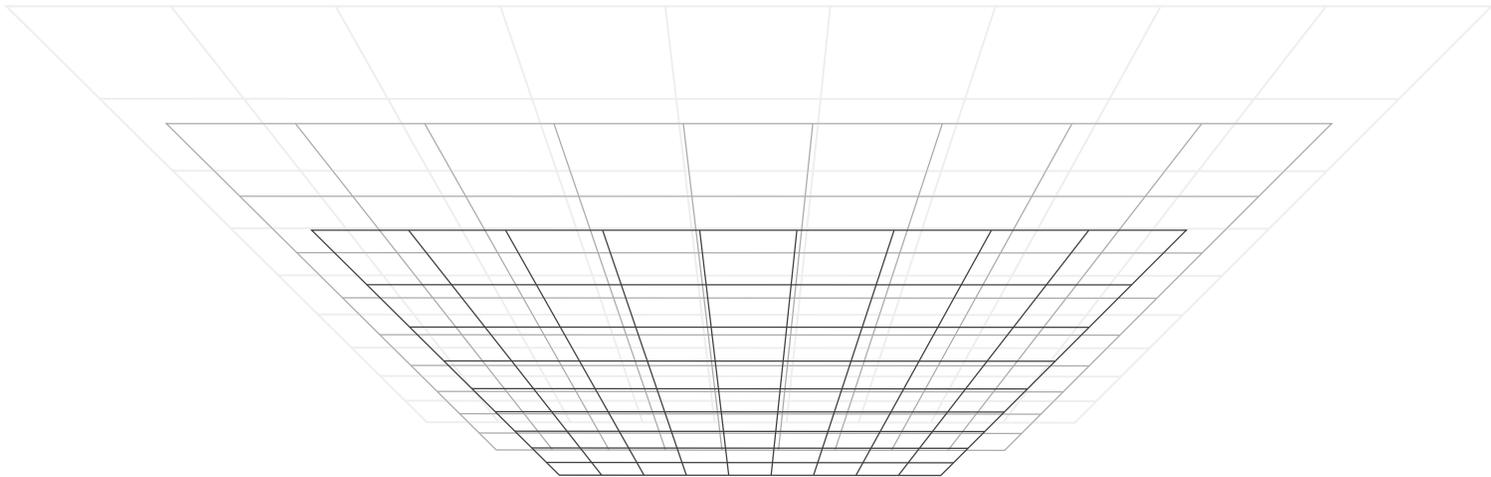

Countries are in a strategic race to achieve high-performance QPUs. The amount of commercially available QPUs globally is in the range of 40 QPUs from two dozen manufacturers. The race is led by the US.

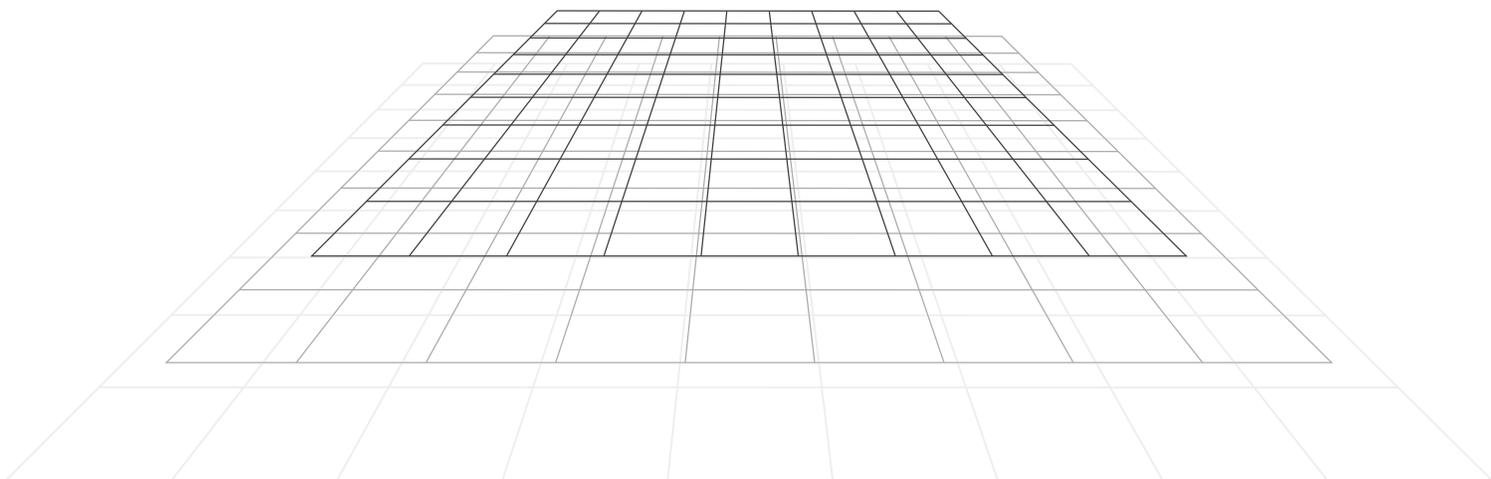

## 10.1 | QPUs per country and modality

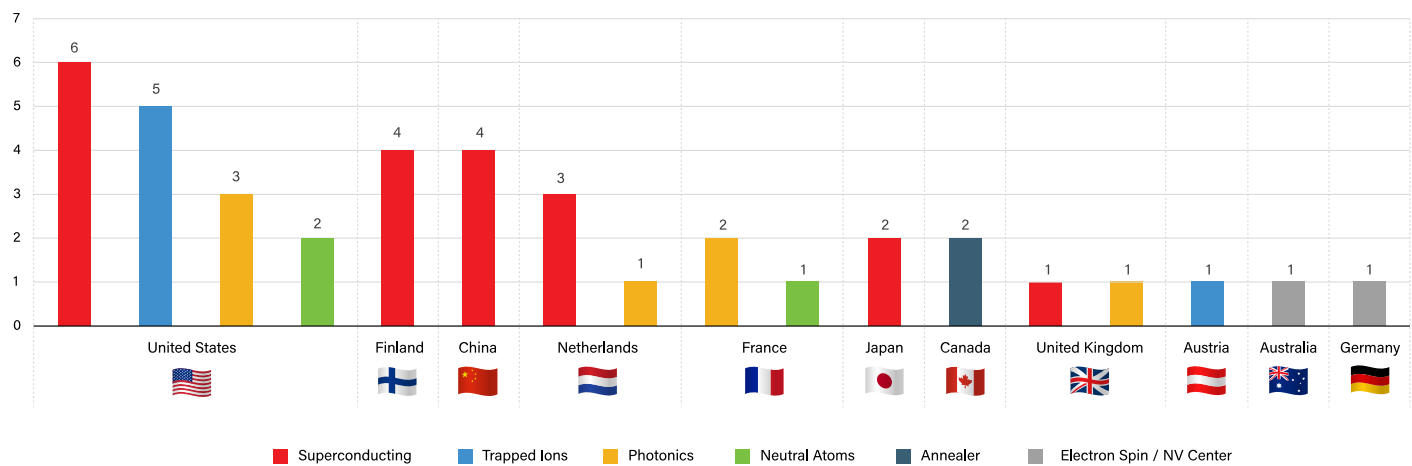

Commercially available QPU models per country

Countries are in a strategic race to achieve high-performance QPUs. The amount of commercially available QPUs globally is in the range of 40 QPUs from two dozen manufacturers. The race is led by the United States, which has the highest number of QPUs and diversity of modalities. China, Finland, and the Netherlands share the second position but their commercial QPUs are of lower performance and smaller in size than the US.

▸ **Data considerations**

We list QPUs that are commercially available per country and per quantum computing modality. Country data is allocated based on the location of the manufacturer's headquarters as described in their official materials and website. The QPUs are classified as commercially available if there is public access to the QPU either via on-premise or cloud. This also includes QPUs that may not be available on public clouds, but access is provided to specific partner companies for commercial use, e.g., Google or PsiQuantum QPUs. However, the device must be intended as a useful quantum computer for commercial use and not solely for experimental purposes. The amount of QPUs is determined as uniquely differentiated products actively provided and marketed by the provider, e.g., IBM Eagle and IBM Heron are two distinct QPUs, but Eagle r3-Brussels and Eagle r3-Sherbrooke are considered as one QPU. The amount of QPUs is not necessarily an indication of the progress of each country in quantum computing, as some manufacturers have made several very small QPUs available for basic academic research and teaching, while others have retired smaller but powerful QPUs from their offerings (e.g., IBM).

## 10.2 | QPUs per modality

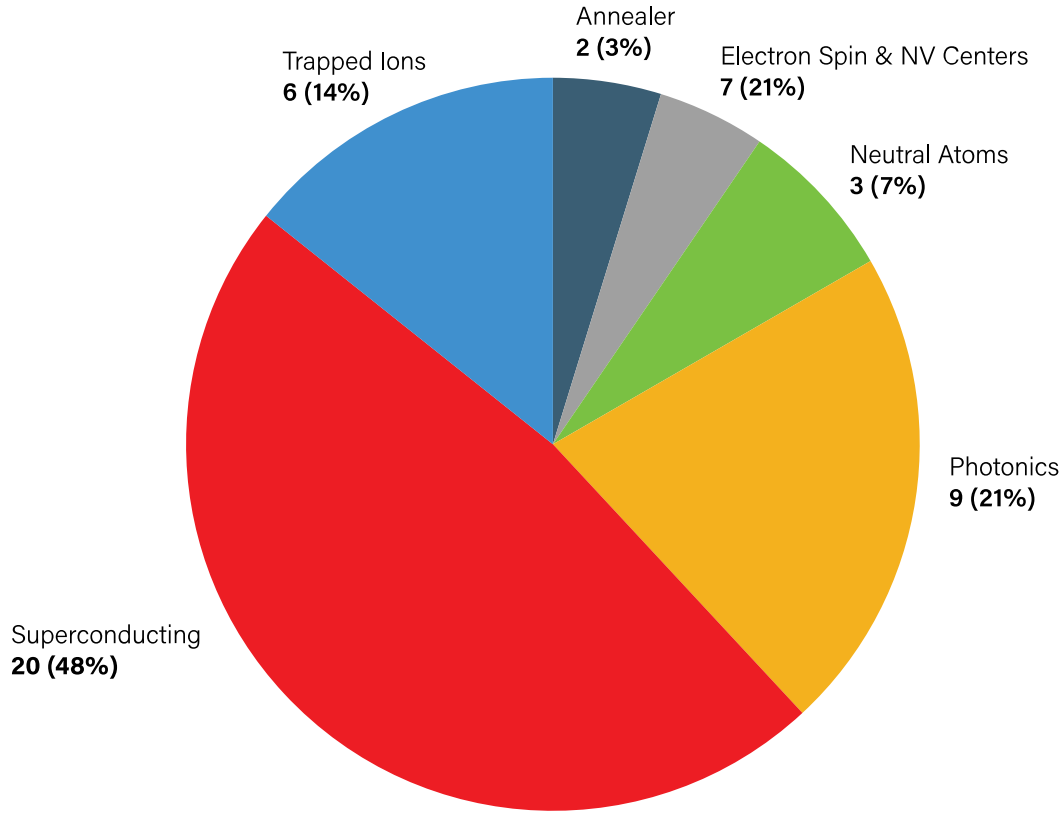

Commercially available QPU models per modality

- Annealer: 2 (3%)
- Electron Spin & NV Centers: 7 (21%)
- Neutral Atoms: 3 (7%)
- Photonics: 9 (21%)
- Superconducting: 20 (48%)
- Trapped Ions: 6 (14%)

The leading quantum computing modality is superconducting with more than 40% of commercially available QPUs. This is partially driven by the inherent manufacturing benefits and a historic head start in R&D. However, photonics, trapped ions—and especially neutral atoms and electron spins—are accelerating in quantity and it is expected this trend will continue, while Annealers are becoming increasingly marginalized and NMR QPUs are practically phased out (see Chapter 10.7.1).

## 10.3.1 | Qubit count per modality

Largest QPU released, prototyped, or planned per year

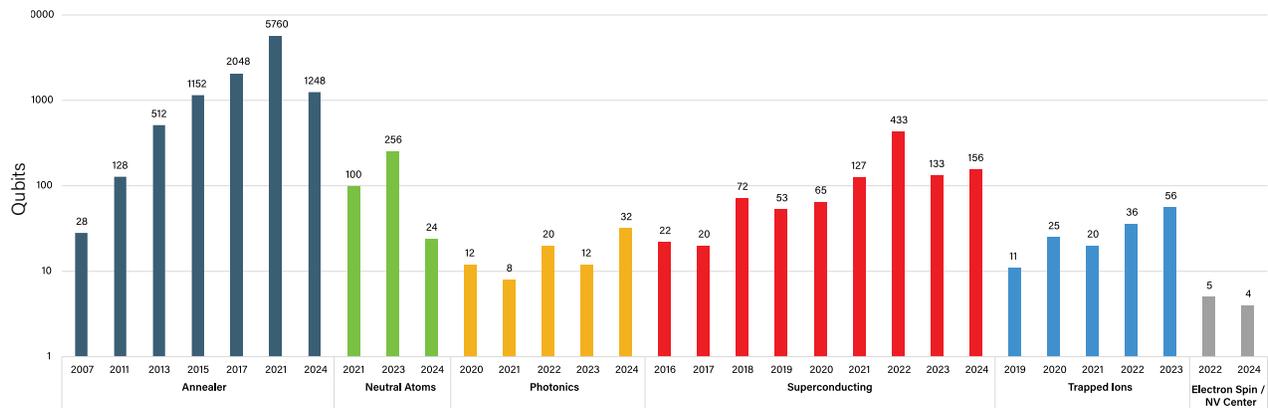

Superconducting QPUs expanded qubit count up to 2022. The more recent decline in absolute numbers reflects an increased focus on improved error correction and higher fidelity (see 10.3.2). Leading Trapped Ion devices are consistently growing their qubit count on an annual basis. The qubit count amongst leading quantum annealers grew steadily across the decade leading up to 2017 but has since stabilized.

▸ **Data considerations**

This graph shows the progression of the number of qubits in our dataset over time, considering only the largest QPU announced per modality in that year. For a QPU to be considered, it needed to be officially announced by a manufacturer and made commercially available in the given year (or expanded, e.g., Quantinuum H1 to H1-1). The data does not always show a steady increase, as some calendar years only contained new QPUs that were smaller than previously available.

# 10.3.2 | Fidelity per modality over time

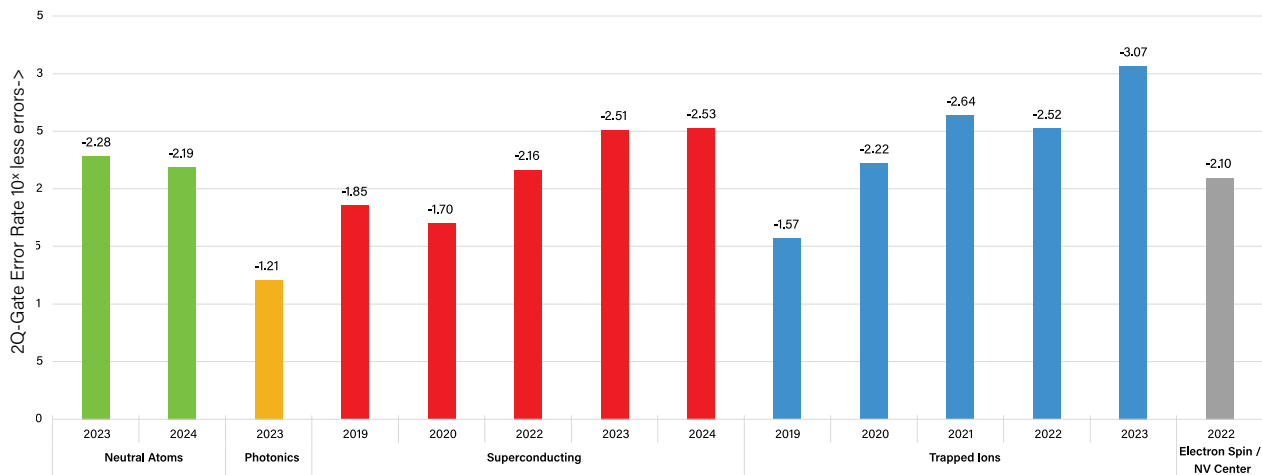

Best 2Q-gate error rate over time
(available, planned, and prototyped QPUs)

Fidelity for 2-Qubit is a key metric of performance improvement. Trapped Ions have shown consistent growth and demonstrated the highest overall fidelity. Superconducting QPUs experienced a decline in top fidelity from 2018 to 2022 until peak fidelity was achieved by the Alibaba QPU. Alibaba subsequently withdrew from the quantum computing market, and IBM and Google caught up to similar fidelity rates in 2023 and 2024. Photonics and NV Center QPUs are still relatively nascent and haven't achieved the top-performing fidelities of trapped ion and superconducting QPUs.

▸ **Data considerations**

This graph shows the progression of fidelity levels in new, commercially available QPUs in our dataset over time. We only considered the highest fidelity announced per modality per year. Error rates are given in a log10 scale, i.e. -3 translates to a 0.001 error rate which corresponds to a 99.9% or 0.999 fidelity. The noted error rates should be treated with caution as there are significant differences in the way they are measured for each QPU, e.g., mid-circuit vs first-gate-measurement, average vs median of several measurements across qubits, different gates (CZ, SWAP, etc.), and different iterations of the same QPU that give different values. In the case of conflicting values, we followed the data mismatch process detailed in the methodology chapter.

# 10.4 | Qubits versus 2Q gate fidelity

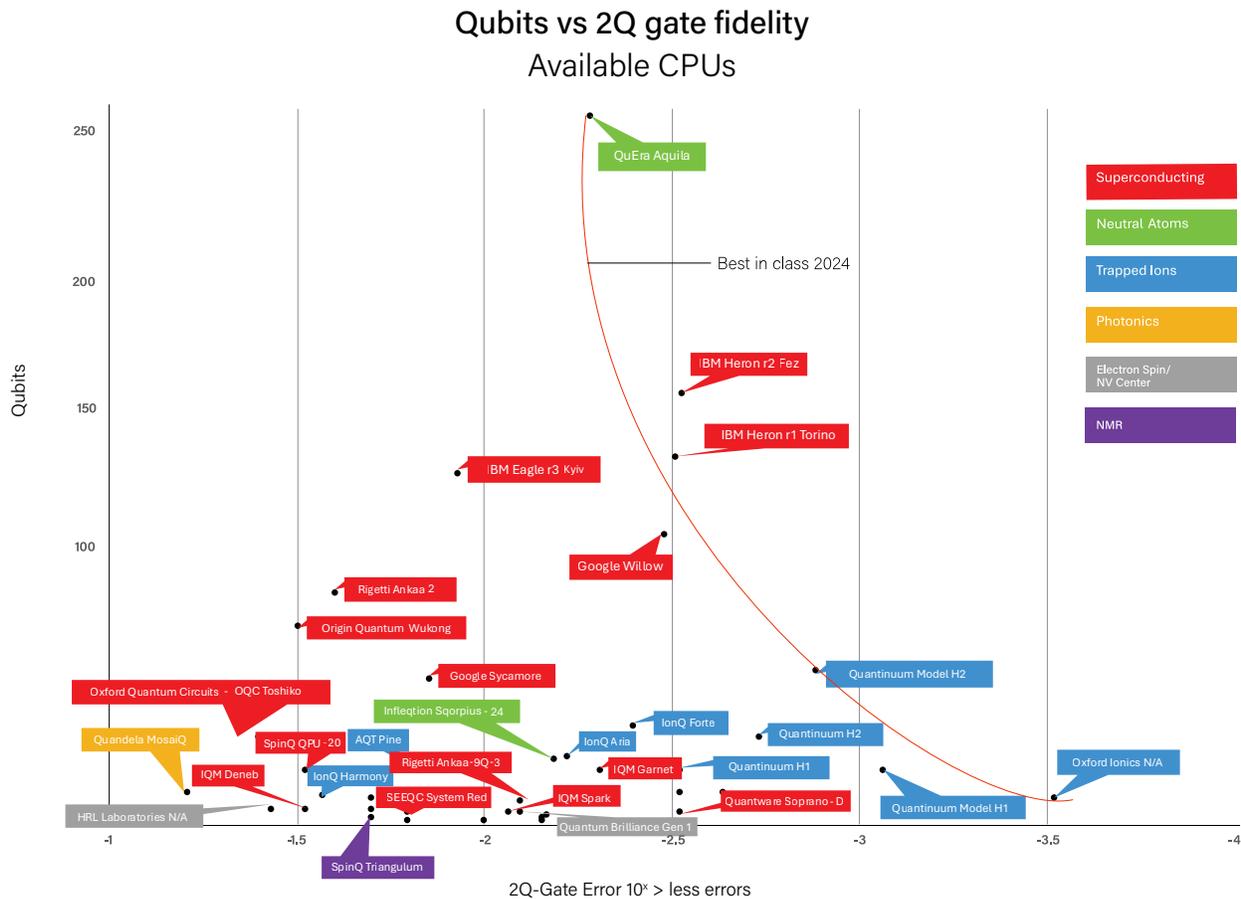

Qubits vs 2Q gate fidelity
Available CPUs

Error rates and 2-Qubit gate-errors are key metrics to benchmark QPUs. Together with the amount of qubits they indicate one of the key combined metrics indicating progress on QPUs. As such, the 2Q-gate errors are crucial to determine the performance of a QPU.

QuEra's Aquila neutral atom chips are leading in qubit count but achieve a lower fidelity. In contrast are the Trapped Ion devices from Quantinuum and Oxford Ionics which reached a 0.999 ("triple-nine") fidelity, an important rubicon. However, this was achieved with relatively smaller qubit sizes. Amongst Superconducting QPUs, Google and IBM are class leaders, with the IBM's Heron r2 achieving the highest performance across this benchmark.

▸ **Data considerations**

2-Qubit-Gates like CZ, CNOT, and SWAP are used for most quantum algorithms and make up the majority of gates for these circuits. Error rates are given in a log10 scale, i.e., 10-3 translates to a 0.001 error rate, which corresponds to a 99.9% or 0.999 fidelity. The noted error rates should be treated with caution as there are significant differences in QPU measurement approaches, e.g., mid-circuit vs first-gate-measurement, average vs median of several measurements across qubits, different gates (CZ, SWAP, etc.), and different iterations of the same QPU that give different values. In case of conflicting values, we followed the methodology of data mismatches detailed under the methodology section.

## 10.5 | Gate time versus gate fidelity

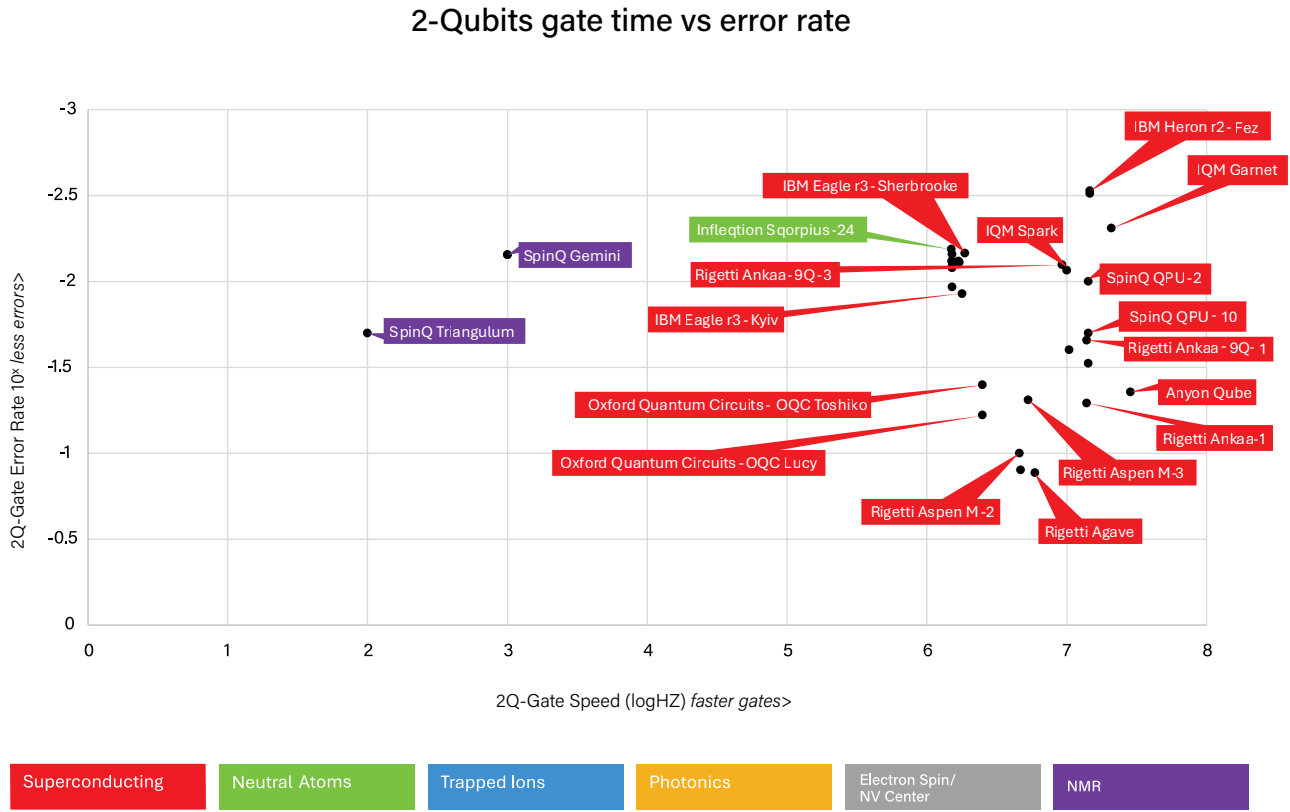

2-Qubits gate time vs error rate

To determine the maximum length of a circuit for a QPU, an important metric is the comparison of the speed of executing a single gate to how accurate this gate is (fidelity). For real-life scenarios such as Shor's Algorithm for decrypting information utilizing RSA2048, more than $10^{13}$ Logical Gates are required. Slow gate speeds at that size lead to calculation times of days or even months with some modalities.

1-Qubit and 2-Qubit Gates devices are identified above; the latter is more interesting as they are more common in large circuits for most algorithms. The superconducting IBM Heron and IQM Garnet are the class leaders. Notably absent are ion traps and neutral atoms QPUs, as manufacturers tend not to disclose exact gate speeds, which are expected to be orders of magnitude slower than superconducting QPUs (as can be seen in the 1-Qubit graph).

# 1-Qubit gate time vs error rate

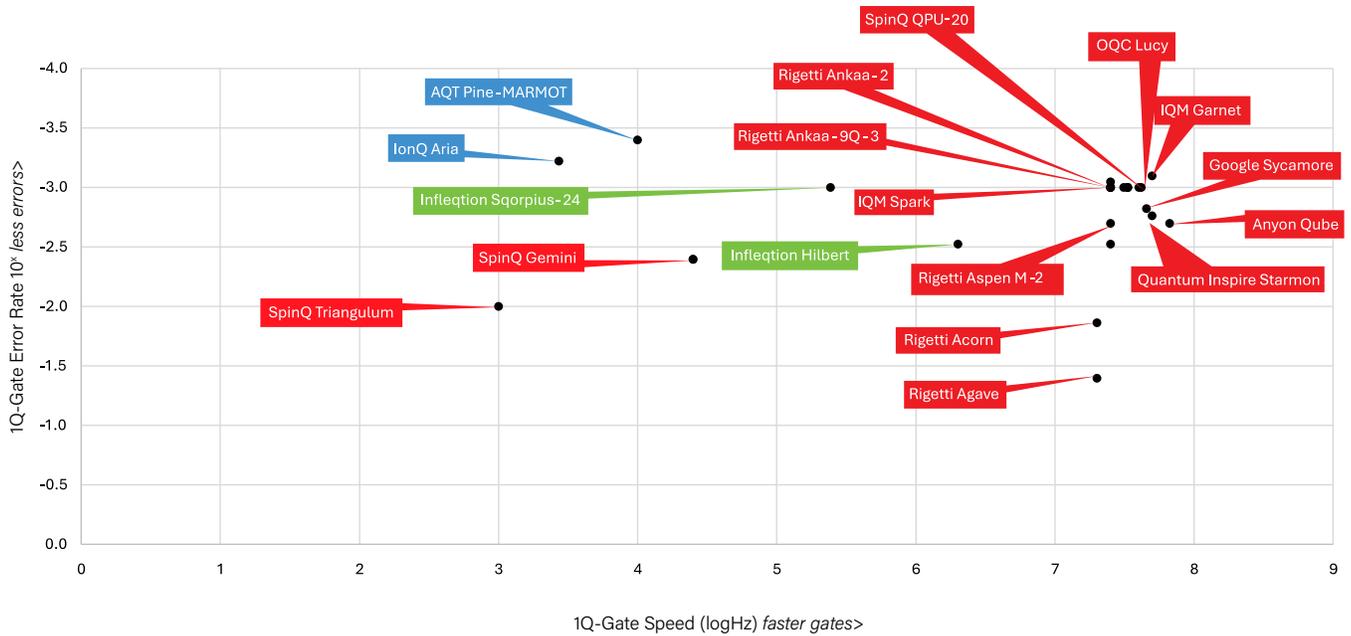

### ▸ Data considerations

We chose 2-Qubit-Gates like CZ, CNOT, and SWAP as they are used for most quantum algorithms and make up the majority of gates for these circuits. The 2Q Gate time is the time required for the execution of a 2-Qubit gate and is given in Hertz (Hz) in a logarithmic scale, where higher values mean faster gate speeds, i.e., 7 logHz corresponds to a 100ns gate speed and 3 logHz to 1,000,000ns. The error rates are given in a log10 scale where higher is better, i.e., 10-3 translates to a 0.001 error rate, which is 99.9% fidelity. The datapoints with missing labels in the graphs are closely related QPUs (e.g., different instances of IBM Eagle). To illustrate the performance comparison, we included the 1-Qubit Gate graph, which shows that these would likely land on the top left quadrant, trading high fidelity for low gate speeds.

# 10.6 | Quantum Volume

Max QPU Quantum Volume
(Log2) over time

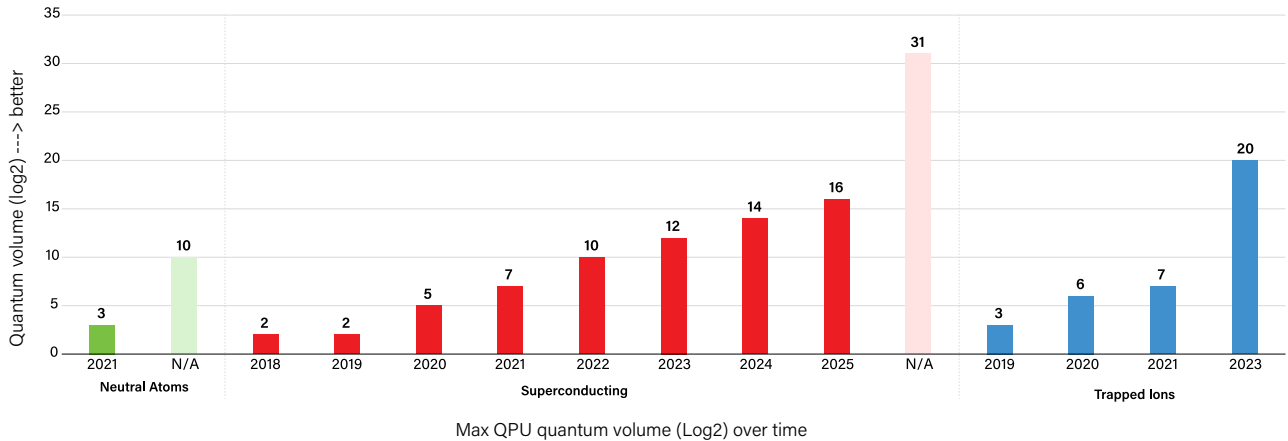

Max QPU quantum volume (Log2) over time

As described in an earlier section, Quantum Volume (QV) is an aggregated metric designed to reflect a more holistic view of overall performance of a QPU. Although its usefulness is heavily debated, manufacturers have been claiming this metric. There is a substantial and consistent increase in QV performance of the top-performing QPU across modalities. However, the stated QV values are hard to validate, and in some cases the only data has been manufacturer claims.

▸ **Data considerations**

The Quantum Volume (in log2 basis) is listed for each QPU at the given year. QV is itself a debated benchmark (even by IBM themselves), which has been used less in the last few years. Despite all these caveats, we chose to include QV, as out of all aggregated metrics (like RACBEM, Algorithmic Qubit, CLOPS, etc.) it is the one that has published values for a sizable amount of QPUs, and as such a progression on quantum computing capabilities can be roughly traced over time. To note is also that the values under 2025 and "N/A" are manufacturer plans, not yet available QPUs.

# 10.7.1 | A look into the future: QPUs per country and modality

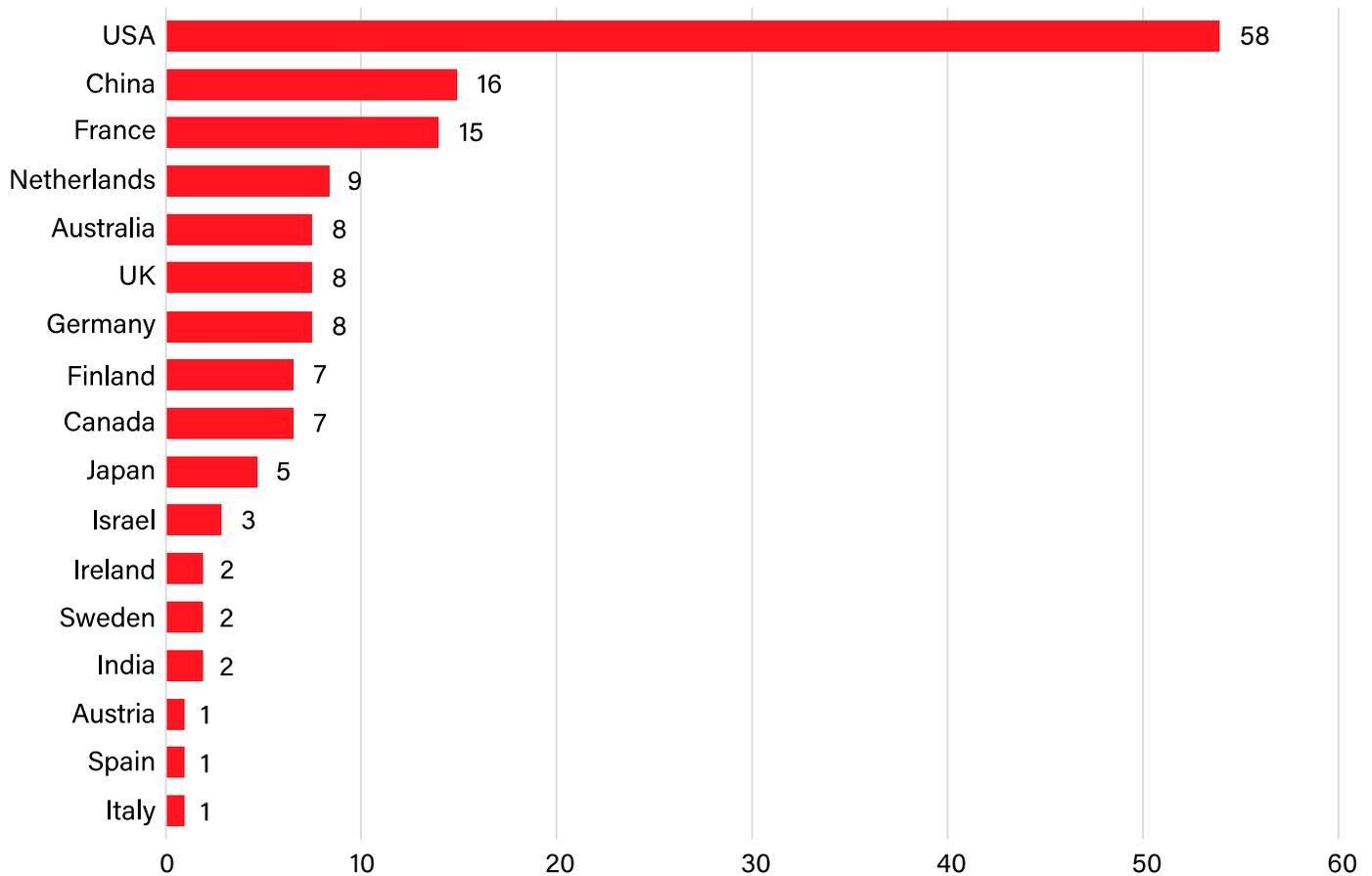

Planned, prototyped, and available QPUs per country

| Country | QPUs |
|---|---|
| USA | 58 |
| China | 16 |
| France | 15 |
| Netherlands | 9 |
| Australia | 8 |
| UK | 8 |
| Germany | 8 |
| Finland | 7 |
| Canada | 7 |
| Japan | 5 |
| Israel | 3 |
| Ireland | 2 |
| Sweden | 2 |
| India | 2 |
| Austria | 1 |
| Spain | 1 |
| Italy | 1 |

Quantum computing remains in the early stages of technical development. Numerous startups and established companies are actively developing prototypes and announcing roadmaps for future QPUs. A comprehensive overview of these efforts provides valuable insight into the evolving technological landscape and strategic directions within the field. The US is the clear leader in the number and diversity of QPUs announced. China is in second place, but is closely followed by France. The Netherlands, Germany, Australia, Canada, Finland, and the UK have announced 7-10 QPUs each. This data gives an approximate measure of country activity but is not an indication of QPU quality. It is also an artifact of QPU planned announcements and does not fully reflect the probability of their successful release.

# 10.7.1 | A look into the future: QPUs per country and modality

QPUs commercially available, prototype, or planned
per modality; change in percentage points to current distribution of QPUs

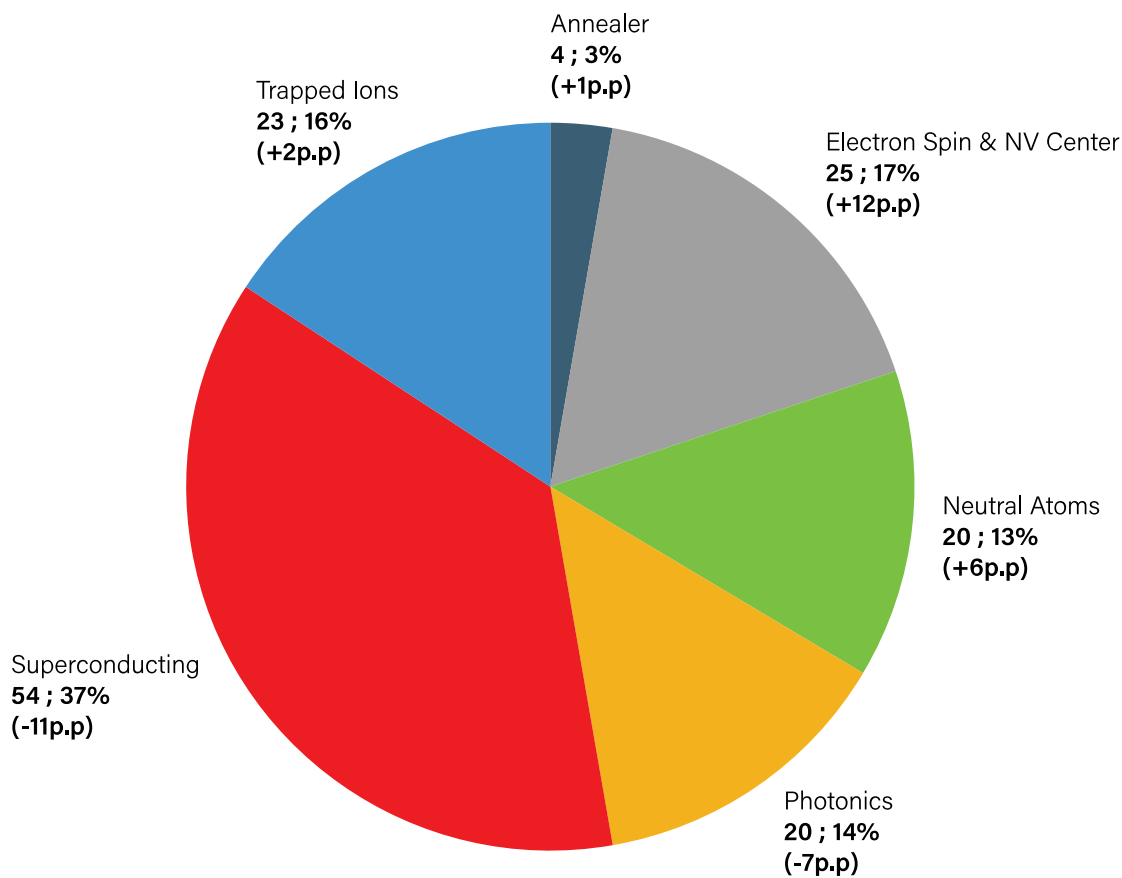

Looking at the future for different modalities, electron spin, NV Centers and neutral atoms are planned to become increasingly prevalent while NMRs and annealers are stagnant and may be phased out. Photonics, superconducting, and trapped ion QPUs may have lower overall shares in the future due to the higher growth levels of other modalities.

▸ **Data considerations**

The data in these graphs includes prototype devices, which are not intended for commercial usage and are not available to the wider community of researchers but are used by a manufacturer for research to develop a new product. It also includes Future Planned QPUs, which are announced in a manufacturer roadmap or interview. Due to the dynamic evolution of startups in the space, recent announcements and changes in QPU roadmaps may not be fully captured in our dataset. The amount of QPUs is not necessarily an indication of the progress of each country in quantum computing, as some manufacturers have made several very small QPUs available for basic academic research and teaching, while others have retired smaller but powerful QPUs from their offerings (e.g., IBM).

# 10.7.2 | A look into the future: qubit count and fidelity

QPUs vs 2Q gate fidelity
Announced QPUs

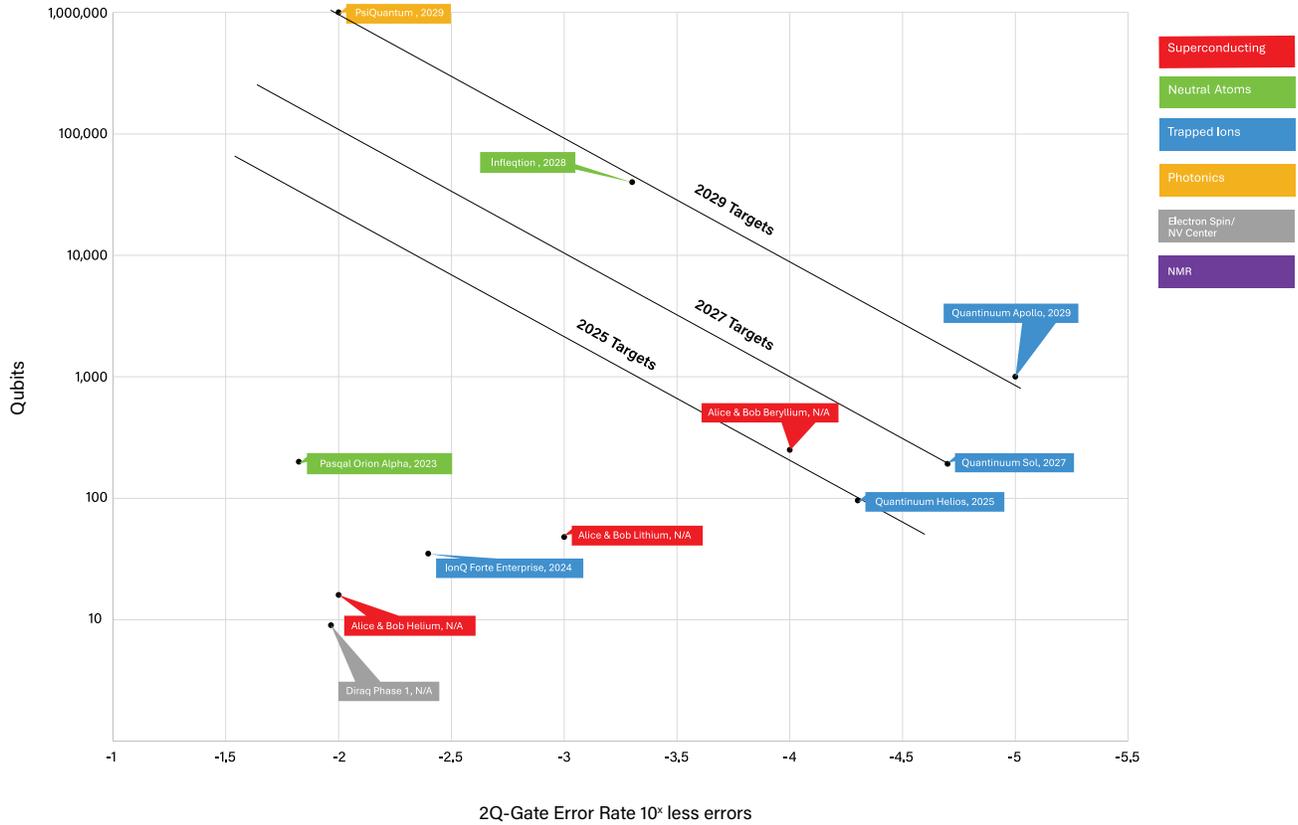

QPU manufacturers are beginning to share more forward-looking roadmaps, offering insights into how their systems might compare across benchmarks. PsiQuantum leads in projected qubit count, Quantinuum excels in error rates, and Infleqtion positions itself as a balanced performer across both metrics.

# 10.7.2 | A look into the future: qubit count and fidelity

Largest QPU released, prototyped, or planned per year

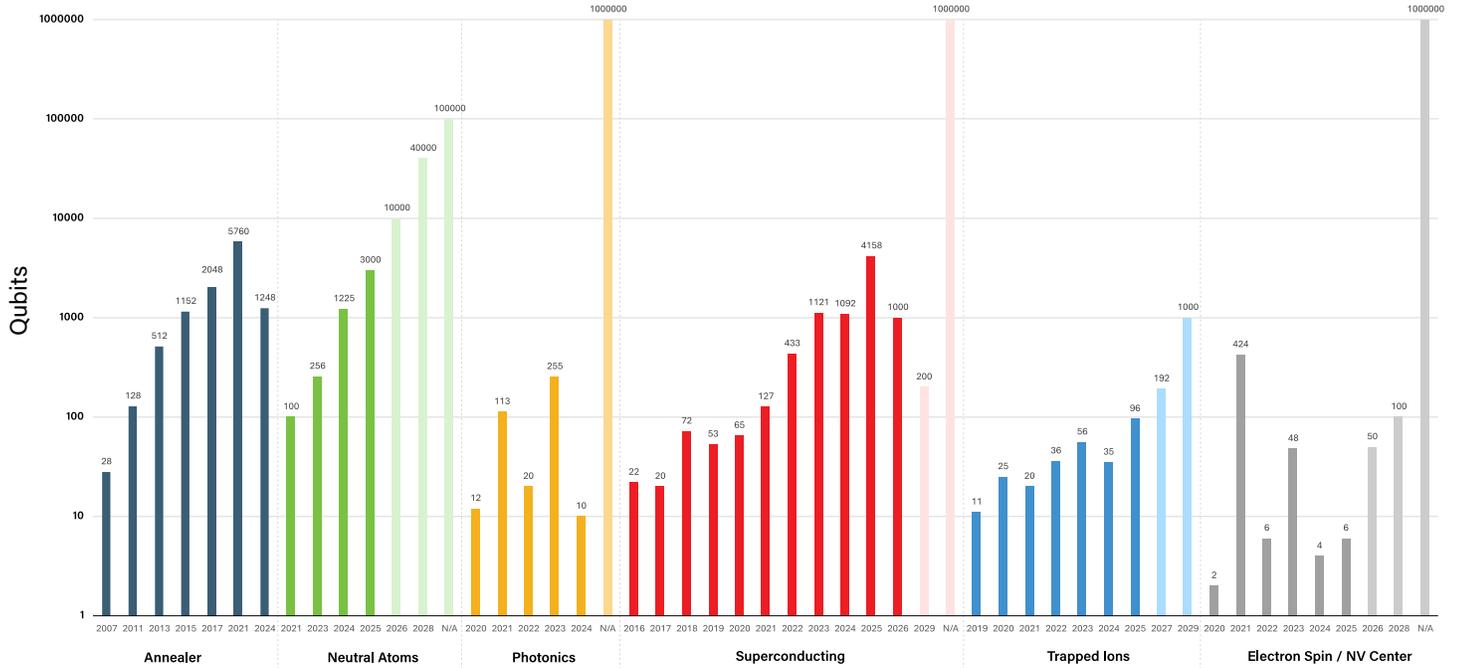

While qubit counts are expected to continue rising, the pace of growth is moderating. This reflects a shift in focus toward improving performance through better error correction and higher qubit fidelity rather than simply scaling up qubit numbers.

### Best 2Q-gate error rate over time
Available, planned, and prototyped QPUs

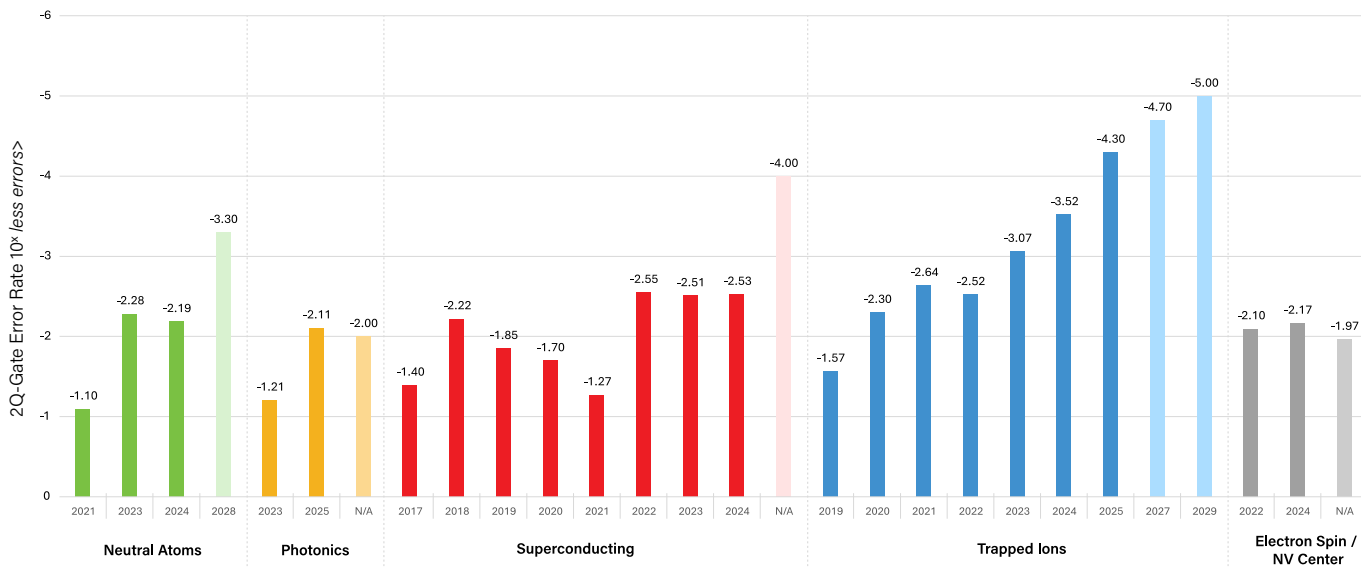

Trapped ion systems aim for exponential gains in fidelity and are on track to continue outperforming other modalities in that benchmark. Neutral atom platforms also show strong ambitions in this area, while other technologies appear more conservative in their likely fidelity trajectories.

▸ **Data considerations**

These graphs show the progression of the fidelity and qubits counts for published QPUs over time, including any future plans, considering only the largest QPU announced per modality in that year. The data does not demonstrate a constantly increasing trend, as some calendar years saw smaller QPUs than previously available. Although approximately 60 manufacturers have announced approximately 90 future QPU models, only 11 QPUs have been provided with both target qubit count and fidelities, which is why there are fewer QPUs in the first graph, QPU vs 2-Qubit Fidelity.

## 10.8 | Future research

Benchmarking is an important exercise in the advancement of our understanding of quantum computing technology as it enables informed decision-making and supports the longer-term goal of standardized comparisons. Our contribution of a publicly accessible overview aims to improve general transparency and allows researchers and community members to engage in a more detailed dialogue regarding the performance of various systems. We encourage industry members and other stakeholders to contribute to these goals by adding their data on an ongoing basis. This will help bridge the gap in this domain, where standardized datasets have been scarce. This is a rapidly and constantly evolving space. By keeping this resource updated and relevant, we are hoping to foster further collaboration and innovation.

You can reach us at contact@qir.mit.edu.

---

### ▸ Footnotes

[1] The difference of 200 indexed QPUs vs 40 commercially available in approximate numbers: 40 are retired, 30 are prototypes but not commercially accessible, and 90 are planned i.e., not released. To be commercially available, they have to be accessible via cloud or on-premise (10 out of 40 QPUs in our dataset are on-premise).

[2] 'Quantinuum Extends Its Significant Lead in Quantum Computing, Achieving Historic Milestones for Hardware Fidelity and Quantum Volume' <https://www.quantinuum.com/blog/quantinuum-extends-its-significant-lead-in-quantum-computing-achieving-historic-milestones-for-hardware-fidelity-and-quantum-volume> accessed 3 April 2025.

[3] Bishop, L. S., Bravyi, S., Cross, A., Gambetta, J. M., & Smolin, J. (2017). Quantum Volume.

[4] Olivier Ezratty, 'Understanding Quantum Technologies 2024' (Opinions Libres - Le blog d'Olivier Ezratty) <https://www.oezratty.net/wordpress/2024/understanding-quantum-technologies-2024/> accessed 3 April 2025.

[5] Raffaele Santagati and others, 'Drug Design on Quantum Computers' (2024) 20 Nature Physics 549.

[6] Rajeev Acharya and others, 'Quantum Error Correction below the Surface Code Threshold' (2025) 638 Nature 920.

[7] Simon J Evered and others, 'High-Fidelity Parallel Entangling Gates on a Neutral-Atom Quantum Computer' (2023) 622 Nature 268.

[8] 'Microsoft's Majorana 1 Chip Carves New Path for Quantum Computing' (Source) <https://news.microsoft.com/source/features/innovation/microsofts-majorana-1-chip-carves-new-path-for-quantum-computing/> accessed 3 April 2025.

[9] Thomas Häner and Damian S Steiger, '0.5 Petabyte Simulation of a 45-Qubit Quantum Circuit', Proceedings of the International Conference for High Performance Computing, Networking, Storage and Analysis (2017) <http://arxiv.org/abs/1704.01127> accessed 3 April 2025.

# 11 | Appendix

# Chapter 1 | Patents

The data in this section is based on patent families and was based on data acquired by Accenture Research from LexisNexis Patent database based on select IPC codes relevant to quantum technologies. No screening was made regarding international patent families. The list of IPC codes used for this dataset can be found here:

| IPC Code | Description | Sub-category |
|---|---|---|
| G06N0010200000 | Models of quantum computing, e.g. quantum circuits or universal quantum computers [2022.01] | Quantum Computing |
| G06N0010600000 | Quantum algorithms, e.g. based on quantum optimisation, or quantum Fourier or Hadamard transforms | Quantum Computing, Software |
| H04B0010700000 | Photonic quantum communication | Quantum Communications, Quantum Networking |
| G06N0010000000 | Quantum computing, i.e. information processing based on quantum-mechanical phenomena | Quantum Computing |
| B82Y0020000000 | Nanooptics, e.g., quantum optics or photonic crystals | Quantum Computing, Hardware |
| G02F0002020000 | Frequency-changing of light, e.g. by quantum counters | Quantum Computing, Hardware |
| G06N0010800000 | Quantum programming, e.g. interfaces, languages or software-development kits for creating or handling programs capable of running on quantum computers; Platforms for simulating or accessing quantum computers, e.g. cloud-based quantum computing | Quantum Computing, Software |
| H01L0033040000 | With a quantum effect structure or superlattice, e.g. tunnel junction | Quantum Computing, Hardware |
| H01S0005340000 | Comprising quantum well or superlattice structures, e.g. single quantum well [SQW] lasers, multiple quantum well [MQW] lasers or graded index separate confinement heterostructure [GRINSCH] lasers (H01S 5/36 takes precedence) | Quantum Computing |

| G02F0001017000 | Structures with periodic or quasi periodic potential variation, e.g. superlattices, quantum wells | Quantum Computing |
| H01L0029775000 | With one-dimensional charge carrier gas channel, e.g. quantum wire FET | Quantum Computing |
| H01L0033060000 | Within the light emitting region, e.g. quantum confinement structure or tunnel barrier | Quantum Computing |
| H10K0050115000 | Comprising active inorganic nanostructures, e.g. luminescent quantum dot | Quantum Computing |
| B82Y0010000000 | Nanotechnology for information processing, storage or transmission, e.g. quantum computing or single electron logic | Quantum Computing, Quantum Networking |
| B82Y0015000000 | Nanotechnology for interacting, sensing or actuating, e.g. quantum dots as markers in protein assays or molecular motors | Quantum Computing |
| G06N0010400000 | Physical realisations or architectures of quantum processors or components for manipulating qubits, e.g. qubit coupling or qubit | Quantum Computing |
| G06N0010700000 | Quantum error correction, detection or prevention, e.g. surface codes or magic state distillation | Quantum Computing |
| G16C0010000000 | Computational theoretical chemistry, i.e. ICT specially adapted for theoretical aspects of quantum chemistry, molecular mechanics, molecular dynamics or the like | Quantum Computing |
| H01L0029150000 | Structures with periodic or quasi periodic potential variation, e.g. multiple quantum wells, superlattices (such structures applied for the control of light G02F 1/017; applied in semiconductor lasers H01S 5/34) | Quantum Computing |
| H04B0010000000 | Transmission systems employing electromagnetic waves other than radio-waves, e.g. infrared, visible or ultraviolet light, or employing corpuscular radiation, e.g. quantum communication | Quantum Computing |

Regarding the categorization of the patent data presented by origin, Accenture researchers applied the following methodology:

**Step 1:** For applicants with name but without type (e.g. university, corporate, etc.), search was filtered using keywords such as specific origin entity specific keywords (such as hospital, etc.) to categorize some of the applicants. The filtered results are stored as Result 1.

**Step 2:** Researchers identified those without keywords (over 1,000 names) using am AI LLM (Gemini-1.5-flash-001) to automatically detect their type. This result is stored as Result 2.

**Step 3:** Researchers merged the applicants with identified types (from both initial filtering and LLM results) with the applicants who originally had type information provided by the vendor.

**Step 4:** Merged data was manually controlled to ensure non-duplication and accuracy.

The data presented in patent applications by country was provided by Accenture in collaboration with The Quantum Insider (TQI). TQI acquired the patent data directly from patent offices. For national patents, national patent office data for each country (e.g. in the US patents are sourced from the United States Patent and Trademark Office (USPTO); in China patents are sourced from the China National Intellectual Property Administration). WO = World Intellectual Property Organization (WIPO), EP = European Patent Office. For "Other Countries" it is typically a longer tail of nations which have been aggregated. For some countries it is typically a longer tail of other countries which have been aggregated.

The TQI patent data was collected based on the following keywords:

Adiabatic Theorem, Bosonic Creutz Ladder, Dicke Model, Distributed quantum computation, Fault-tolerant quantum computation, gray zone assault, Hadamard Gate, Harrow Hassidim Lloyd, Harrow Hassidim Lloyd, HHL algorithm, ion traps, Josephson junctions, neutral atoms, Noisy Intermediate-Scale Quantum era, Open Quantum Systems, Photonic Quantum Computing, QAOA, qbits, qbytes, QEC, QNLP, QSVM, qtrits, Quantum accelerators, quantum annealing, Quantum Advantage, Quantum accelerators, quantum annealing, Quantum Advantage, quantum algorithms, Quantum applications, quantum approaches, quantum approximate optimization, quantum approximate optimization algorithms, quantum arithmetic, Quantum artificial intelligence, quantum backtracking, quantum bits, Quantum Bosonic Systems, quantum bytes, quantum chaos, quantum chaos, quantum chemistry, Quantum circuits, quantum classifier, quantum communication, quantum compiler, Quantum complexity, Quantum component, Quantum computation, Quantum computational, Quantum computer, Quantum Computing Architectures, Quantum Control, quantum correlation, Quantum cryptanalysis, Quantum cryptoalgorithm, Quantum cryptog, Quantum cryptographic, Quantum cryptology, Quantum cryptosystem, Quantum Cryptology, Quantum Cryptology, Quantum cryptosystem, Quantum decoding, quantum devices, quantum distillation, quantum dots, quantum dynamics, quantum dynamics, quantum eigensolver, quantum encryption, Quantum Entanglement, quantum entanglement distillation, quantum error correction, Quantum Error Detection, Quantum Field theory, quantum Fourier transform, Quantum gas sensors, Quantum gate, Quantum gate fidelity, Quantum gate-based, Quantum gates, quantum Grover, Quantum hardware, Quantum hardware security, quantum image sensor, quantum imager array, quantum information, Quantum information processing, Quantum information science, Quantum information systems, Quantum information theory, Quantum interference, Quantum ions, quantum Josza, Quantum Kernel, Quantum key, Quantum key distribution, Quantum key distribution network, Quantum key distribution protocol, Quantum key distribution systems, quantum key exchange, Quantum LDPC Codes, Quantum Linear Optics, quantum logic, quantum machine learning, quantum machines, quantum magic states, Quantum Maps, Quantum Measurement, Quantum Memristors, Quantum metrological, Quantum metrology, Quantum metrology standards, Quantum Monte Carlo, Quantum Natural Language Processing, Quantum Networks, quantum neural networks, Quantum Oscillator, quantum phase amplifiers, Quantum precision measurement sensors, Quantum process, Quantum processing, Quantum Programs, Quantum proof, quantum public key, Quantum Quantile Mechanics, Quantum Quantizer, Quantum random number, Quantum random number generation, Quantum random number generation device, Quantum random number generator, Quantum random number sequences, Quantum safe network, Quantum sensing, Quantum sensing technology, Quantum sensor, Quantum sensor networks, quantum sensors, Quantum shared key, quantum Shor, Quantum Signal Processing, quantum simulation, Quantum Simulator, quantum single photon, quantum software, Quantum Speedup, quantum spin, quantum spintronics, quantum state, Quantum Subroutines, Quantum superposition, quantum supremacy, quantum supremacy, quantum switches, quantum systems, Quantum Technologies, quantum teleportation, quantum tensor, quantum toffoli gate, Quantum transmons, quantum variational, quantum video, Quantum Week, Quantum-enhanced, Quantum-enhanced, Quantum-resistance, qubits, qubytes, qudits, qumodes, qutrits, silicon qubits, VQE.

# Chapter 2 | Academic research

Accenture provided the data for this chapter which was gathered from the Critical Technology Tracker Project by the Australian Strategic Policy Institute[1] (ASPI).

**ASPI methodology summary:**

The Web of Science (WoS) Core Collection database was used as the primary source of research publication data. WoS Core Collection was chosen because it has well-understood performance characteristics and is used extensively by researchers who study scientific trends. The dataset includes conference and journal publications and excluded bibliographic records that were deemed to not reflect research advances, such as book reviews, retracted publications and letters submitted to academic journals. In addition, data from the Research Organization Registry (ROR) was used to clean institution names, and data from the Open Researcher and Contributor ID (ORCID) database was used to build career profiles for the researchers plotted in the ASPI Talent Tracker.

**Country-level quality metrics:**

The methodology includes "both the top 10% and the H-index as neither is perfect and both add a unique insight." In technologies in which first and second place flip depending on which quality metric is used, the race really is too close to call. However, more often, the lead is large and unambiguous, and both metrics are consistent regarding who is leading."

ASPI methodology states that "The top 10% of the most highly cited papers were analyzed to generate insights into which countries are publishing the greatest share of high-quality, innovative and high-impact research. Credit for each publication was divided among authors and their affiliations and not assigned only to the first author (for example, if there were two authors, they would each be assigned half the allocation). Fractional allocation of credit is a better prediction of individuals who go on to win Nobel Prizes or fellowship of prestigious societies. Fractional allocation of credit was used for all metrics.'

'The number of institutions that a country has in the world's top 10 institutions is used to illustrate research concentration and dominance. This list is based on the number of papers that the institutions have in the top 10% of highly cited papers."

**Definitions as explained in the ASPI methodology:**

**Quality Metrics:** "Distinguishing innovative and high-impact research papers from low-quality papers is critical when estimating the current and future technical capability of nations. Not all the millions of research papers published each year are high quality."

**Citation:** "When a scientific paper references another paper, that's known as a citation. The number of times a paper is cited reflects the impact of the paper. As time goes by, there are more opportunities for a paper to be cited, so only papers of a similar age should be compared using citation counts (as was done in this report)." This data was

# Chapter 3 | Venture funding

This data was gathered by Accenture in collaboration with The Quantum Insider (TQI) using The Quantum Insider Funding Database. The methodology and limitations are explained below:

Funding numbers are obtained from open media sources (press releases, articles, etc). For example, Riverlane funding round: https://www.riverlane.com/press-release/riverlane-raises-75-million-to-meet-surging-global-demand-for-quantum-error-correction-technology.

Where possible TQI emails the companies to validate if they are missing investors or details. Not all companies disclose the size of funding rounds (e.g. QEDMA shows as $4.7 million seed but they haven't publicly disclosed their top up round so they have asked not to be included in the dataset). Based on this, there will be gaps in reporting and the data should be viewed as indicative rather than complete.

# Chapter 4 | Quantum in corporate communications

The data was collected by Accenture through AlphaSense on 10th March 2025 using keyword search term "quantum computing." The documents presented in this section include five categories, (1) Company Documents consisting of US Filings, Global Filings, Company House Filings, Private Company Filings, Event Transcripts, ESG, Thought Leadership, Other Company Publications; (2) Research Documents consisting of Broker Research, IDC Research, Consultancy Research, Broker Feed; (3) Transcript Documents consisting of Event Transcripts; (4) News Documents consisting of Financial Times, Market News, General News, Trade Publications, RSS Feeds, LexisNexis, (5) Expert Call Documents consisting of Expert interviews.

# Chapter 5 | Policy

The policy research was completed through comprehensive desk research specifically designed to capture the rapidly evolving landscape of quantum technology initiatives across multiple countries, including detecting and analyzing the national strategy documents and implementation plans, which involved cross-referencing multiple official sources and analysis of policy implementation progress.

# Chapter 6 | Workforce

This data was gathered by Accenture and sourced from Lightcast. Lightcast integrates economic, labor market, demographic, education, profile, and job posting data from dozens of government and private-sector sources, creating a comprehensive and current dataset that includes both published data and detailed estimates with full United States coverage. Further information on Lightcast data sources available [here](here).

The following keywords were used in our Lightcast database searches:

**Quantum skills keywords:** Quantum Gates, Heteronuclear Single Quantum Coherence Spectroscopy,Heteronuclear Multiple Quantum Coherence,Quantum Link,Quantum Mechanics/Molecular Mechanics (QM/MM),Quantum Point Contact,Quantum Phase Transition,Quantum Dynamics,Quantum Imaging,Quantum Technology,Superconducting Quantum Interference Device (SQUID),Amazon Quantum Ledger Database (QLDB),Quantum Cryptography,Quantum GIS (QGIS),Quantum Scalar Servers,Quantum Chemistry,Quantum Mechanics,Quantum Physics,Quantum Information Sciences,Quantum Computing,Quantum Dots,Quantum Information,Quantum ESPRESSO.

**Quantum mentions keywords:** quantum computer  quantum computing quantum supremacy quantum entanglement quantum superposition quantum bit qubit topological qubit silicon spin qubit quantum advantage quantum simulation quantum machine learning quantum optimization quantum chemistry optical quantum computing gate model quantum computing photonic quantum computer quantum dots superconducting quantum computer trapped ion quantum computer quantum annealing quantum as a service quantum cloud post-quantum cryptography PQC quantum algorithm Shor's algorithm Grover's algorithm quantum encryption quantum research quantum use case

# Chapter 7 | Education

The education data represented in the "Postgraduate Education" section was collected from the StudyPortals resource and presents the master's degree programs that make a specific reference to "quantum" in the degree name found in the named resource.

Education enrollment data was collected from the publicly available data set "Current Term Enrollment Estimates" with the January 2025 updates of the NSC Research Center. The NSC states in their methodology for compiling the relevant dataset that the data is based on administrative data directly derived from college and university registrars. NSC declares that since the fall of 2021, "institutions actively submitting enrollment data to the Clearinghouse account for 97 percent of all enrollments at Title IV, degree-granting institutions in the US."

The 2021 US Report "The Role of International Talent in Quantum Information Science" states that "the most QIST-relevant degree fields are physics, electrical engineering, and computer science" and explains that these domains were selected based on two criteria:

preliminary search of keywords for online job postings and, analysis of doctoral thesis titles, abstracts, and keywords. Therefore we retrieved the student enrollment data for the relevant three degrees.

We used Electrical, Electronics, and Communications Engineering (141000 – Major Field Group CIP) enrollment data as a subcategory of Engineering; Physics degree (400800 – Major Field Group CIP) enrollment data as a subcategory of Physical Sciences; and Computer Science degree (110700 – Major Field Group CIP) enrollment data as a subcategory of Computer and Information Sciences and Support Services. The data categorization had challenges as the major field groups at times had interconnected degrees such as "Computer and Information Science, general", "Astronomy and Astrophysics" which are not included in the subject-level enrollment data. In order to provide a fuller picture, the report also presents the enrollment numbers for the three major field families Engineering, Physical Sciences, and Computer and Information Sciences and Support Services.

# Chapter 8 | Public opinion

"Survey on Public Opinion" refers to a general population survey conducted online in October 2024 with US participants. The survey instrument was administered to a representative panel of 1,375 US residents, with demographic sampling aligned to US Census Bureau distributions for both gender and age groups, ensuring population representativeness. Data collection procedures followed established survey research protocols, with items grouped by thematic content to enhance respondent engagement and reduce cognitive load. The five-point scale format was selected to balance response sensitivity with participant comprehension, avoiding the potential ambiguity associated with finer-grained scales while still capturing meaningful variations in opinion intensity.

Each item was measured using a symmetric response format anchored by "Strongly Disagree" and "Strongly Agree," with intermediate positions of "Somewhat Disagree," "Neither Agree nor Disagree," and "Somewhat Agree." Regarding emotional response questions, the response format was anchored by "Very Nervous" and "Very Excited," with intermediate positions of "Somewhat Nervous," "Neither Nervous nor Excited" and "Somewhat Excited." This approach enabled precise quantification of attitudinal responses while maintaining respondent comprehension through clear, distinct categories. To better demonstrate the difference in responses, in the relevant chapter graphs were prepared to represent the ratio of "Agree" and "Disagree" responses as well as "Positive" and "Negative" emotional responses where the neutral answers were not represented. For the referred visualizations "Strongly Disagree" and "Somewhat Disagree" responses were grouped into the category "Disagree", and "Strongly Agree" and "Somewhat Agree" responses were grouped into the category "Agree". Utilizing the same approach, for the visualizations of positive and negative responses, "Very Nervous" and "Somewhat Nervous" responses were grouped into the category "Negative," and "Very Excited" and "Somewhat Excited" responses were grouped into the category "Positive."

# Chapter 9 | Quantum networking testbeds

The dataset was created by merging input from the Center for Quantum Networks (CQN) researchers, QIR professional network and publicly available information by GQI Quantum Computing Report (accessed in June 2024).

# Chapter 10 | Quantum processor benchmarking

The dataset of QPUs was composed by a combination of a keyword-based online search and official announcements, references to QPU lists made available to us, and direct query to QPU manufacturers. The data was collected from January 2024 to April 2025.

In particular, a list of known manufacturers was created based on the sources of The Quantum Insider, Olivier Ezratty, and Wikipedia. For each manufacturer, the official website was interrogated to retrieve the indicated benchmarks. For datasets not on manufacturer's websites, we utilized web searches (Google) for official announcements from manufacturers and related news articles.

Additionally, scholarly articles were identified via Arxiv and Google Scholar using the following keywords for benchmarks: Quantum Volume, CLOPS, EPLG, Q-Score, benchmarking.

During this process, additional manufacturers/QPUs were identified and added to the QPU list. Lastly, each manufacturer was contacted for verification of records—either to an existing contact of the QIR team, or to the communications address listed on manufacturer's website. The final list was reviewed by the QIR team and experts in their professional network.

## ▸ Footnotes

[1] Gaida, J., Wong-Leung, J., & Robin, S. (2023). Critical technology tracker. Who Is Leading the Critical Technology Race. A Project by the Australian Strategic Policy Institute. https://techtracker.aspi.org.au

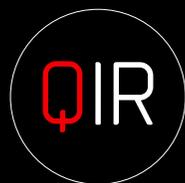

Quantum Index Report 2025
MIT Initiative on the Digital Economy

245 First Street, Room E94-1521, Cambridge, MA 02142  USA

qir.mit.edu  |  ide.mit.edu